\documentclass[superscriptaddress,
 reprint,
 amssymb, amsmath,
 aip, cha 
]{revtex4-1}
\usepackage[utf8]{inputenc}
\usepackage[T1]{fontenc}
\usepackage{amsmath}
\usepackage{amsfonts}
\usepackage{amssymb}
\usepackage{graphicx}
\usepackage{bm}
\usepackage{comment}
\usepackage{grffile} 
\usepackage{dcolumn}
\usepackage{lipsum}
\usepackage{multirow}
\usepackage{tabularx}
\usepackage{booktabs}
\usepackage{subcaption}
\usepackage{tikz}
\usepackage{soul}

\DeclareMathAlphabet{\mathmybb}{U}{bbold}{m}{n}
\newcommand{\1}{\mathmybb{1}}

\usepackage{pifont}
\usepackage{url}
\usepackage{listings}
\newcommand{\var}{{\operatorname{var}}}

\newcommand*{\Id}{\operatorname{I}}

\renewcommand*{\Xi}{{\boldsymbol{\xi}}}

\newcommand*{\R}{{\mathbb{R}}}

\newcommand*{\mypop}{{(\alpha)}}

\DeclareFontFamily{U}{stix2bb}{}
\DeclareFontShape{U}{stix2bb}{m}{n} {<-> stix2-mathbb}{}

\graphicspath{{}}

\usepackage{mathtools}

\definecolor{darkgreen}{rgb}{.15,.55,0}

\begin{document}


\title{Finite-size induced random switching of chimeras in a deterministic two-population Kuramoto-Sakaguchi model}

\author{Henry Irvine}
 \email{H.Irvine@maths.usyd.edu.au}
 \affiliation{\mbox{School of Mathematics and Statistics, The University of Sydney, Sydney, NSW 2006, Australia}}
\author{Georg~A. Gottwald}
 \email{georg.gottwald@sydney.edu.au}
 \affiliation{\mbox{School of Mathematics and Statistics, The University of Sydney, Sydney, NSW 2006, Australia}}



\begin{abstract}
The two-population Kuramoto-Sakaguchi model for interacting populations of phase oscillators exhibits chimera states whereby one population is synchronised, and the other is desynchronised. Which of the two populations is synchronised depends on the initial conditions. We show that this deterministic model exhibits random switches of their chimera states, alternating between which of the two populations is synchronised and which is not. We show that these random switches are induced by the finite size of the network. We provide numerical evidence that the switches are governed by a Poisson process and that the time between switches grows exponentially with the system size, rendering switches unobservable for all practical purposes in sufficiently large networks. We develop a reduced stochastic model for the synchronised population, based on a central limit theorem controlling the collective effect of the desynchronised population on the synchronised one, and show that this stochastic model well reproduces the statistical behaviour of the full deterministic model. We further determine critical fluctuation sizes capable of inducing switches and provide estimates for the mean switching times from an associated Kramers problem. 
\end{abstract}

\maketitle


\begin{quotation}

\end{quotation}

\section{Introduction}
Coupled oscillators describe systems ranging from neurons in the brain \cite{CrookEtAl97, SheebaEtAl08, BhowmikShanahan12} and synchronous firefly flashing \cite{MirolloStrogatz90}, to Josephson junction arrays \cite{WatanabeStrogatz94, WiesenfeldEtAl98} and to power grids \cite{FilatrellaEtAl08, NishikawaMotter15}. The celebrated Kuramoto model and its numerous variants \cite{Kuramoto84, SakaguchiKuramoto86, Strogatz00, PikovskyEtAl01, AcebronEtAl05, OsipovEtAl07, ArenasEtAl08, DorflerBullo14, RodriguesEtAl16} constitute a paradigmatic model to describe the temporal evolution of such coupled oscillators and their often rich dynamical behaviour. Much progress has been made by employing mean-field theory, valid in the thermodynamic limit of infinitely many oscillators \cite{SakaguchiKuramoto86, OttAntonsen08, DelgadinoEtAl21}. However, a number of finite-size effects have recently attracted attention, including finite-size-induced stochastic drift of oscillators that vanishes in the thermodynamic limit \cite{GiacominPoquet15, Lucon15, Gottwald15, Gottwald17}. Such finite-size effects are not amenable to classical mean-field theory and require new analytical approaches~\cite{HildebrandtEtAl07, BuiceChow07, HongEtAl07, Tang11, GiacominPoquet15, Lucon15, Gottwald15, Gottwald17, OmelchenkoGottwald25}. 

We present a dynamical phenomenon that is due entirely to the finite size of the number of oscillators. We describe the emergence of stochastic switching of chimera states in a deterministic system of a two-population Kuramoto-Sakaguchi (KS) model \cite{OkudaKuramoto91, MontbrioEtAl04}. Chimeras denote a symmetry-breaking state in which a group of oscillators is synchronised while the rest of the oscillators are desynchronised, and where the partition into a synchronised and desynchronised group depends on the initial condition \cite{KuramotoBattogkh02, OmelchenkoEtAl08, OmelchenkoEtAl10, Omelchenko18}. Recently, chimeras were also found in an Ising model with long-range diffusion \cite{DeHaroGarciaTorres25}.  The existence of chimera states in a two-population KS model is well known and has been studied in either very small networks \cite{PanaggioEtAl16, BurylkoEtAl22} or in the thermodynamic limit \cite{AbramsEtAl08, Laing09, MartensEtAl16}. Chimeras can undergo complex dynamical behaviour. In switching (or alternating) chimeras, the synchronised and desynchronised regions alternate in time.
To date, such switching has been attributed to either the existence of metastable states \cite{MaEtAl10, Laing12, BuscarinoEtAl15, SemenovaEtAl16, ZhangEtAl20} or to heteroclinic cycles \cite{Bick18, Bick19, BickLohse19, EbrahimzadehEtAl20, GoldschmidtEtAl19, BrezetskyEtAl21, LeeKrischer23}.  In a more complex generalised Kuramoto-Sakaguchi model aperiodic alternating chimeras were observed and analysed that exist in the thermodynamic limit \cite{LeeKrischer23b}. In contrast, here, we report on effectively random switching of chimeras in a deterministic two-population KS model that is non-transient and mediated entirely by finite-size effects, and that disappears in sufficiently large networks.  Numerical simulations will show that the mean switching time increases exponentially with the number of oscillators $N$ and that the variance of the order parameter scales as $1/N$, suggesting an underlying central limit theorem which captures fluctuations around the well-known deterministic thermodynamic limit. We will describe the fluctuations by means of the recently proposed stochastic model reduction for deterministic KS models \cite{YueGottwald24}. We adopt this stochastic modelling framework to parametrise the interaction term that couples the phases of the desynchronised population to those of the synchronised population via a Gaussian process. Gaussian processes are entirely determined by their mean and their covariance function. The mean, represented by the thermodynamic limit of infinitely many oscillators, can be estimated by classical mean field theory. The covariance function will be estimated by first approximating the Gaussian process by a complex Ornstein-Uhlenbeck process, and then numerically determining the parameters of this complex Ornstein-Uhlenbeck process to fit the observed covariance function of the finite-size fluctuations. This modelling strategy captures the statistical features of the full two-population KS model very well, including the distribution of the order parameter and the statistics of the phases of the synchronised oscillators. We will further determine the critical fluctuation size which is sufficient to induce chimera switching, and set up a classical Kramers problem to approximate the mean switching time with reasonable accuracy~\cite{Kramers40}. Our modelling strategy requires the system size to be sufficiently large to allow for the underlying central limit theorem to be valid and sufficiently small to prevent the law of large numbers described by the thermodynamic limit and classical mean-field theory. 

The paper is organised as follows. Section~\ref{sec:model} presents the deterministic two-population KS model. Section~\ref{sec:mft} reviews the mean-field theory. Section~\ref{sec:obs} presents results of numerical simulations of the two-population KS model showing that switches of the chimera states occur in an effective stochastic way in finite-size networks. Section~\ref{sec:detred} presents a deterministic model reduction which will inform the estimation of the mean switching time. Section~\ref{sec:stochred} presents the stochastic model reduction for a synchronised population. Section~\ref{sec:results} provides results of numerical simulations showing that the reduced stochastic model well captures the dynamics of the full system and quantitatively describes the statistics of the observed random switching of chimeras. We conclude with a discussion and an outlook in Section~\ref{sec:discussion}.


\section{Kuramoto-Sakaguchi model for two interacting populations}
\label{sec:model}
To account for different populations of oscillators and their mutual interaction, extensions of the KS model were proposed \cite{OkudaKuramoto91, MontbrioEtAl04}. The intricate balance between the interactions of oscillators belonging to different populations gives rise to rich dynamical behaviour. In particular, these systems support chimera states in which one population is synchronised while the other is desynchronised \cite{AbramsEtAl08, Laing09, MartensEtAl16, PanaggioEtAl16, BurylkoEtAl22}. Which of the two populations is synchronised and which is not depends on the initial condition.  We consider here the dynamics of equally sized populations $P_1$ and $P_2$ of size $N_1=N_2=N$, and where each population is governed by the same dynamics in the sense that their native frequencies are drawn from the same distribution and their respective intra- and interpopulation interactions are of equal strength. The dynamics of the phase oscillators are given by
\begin{align}
\frac{d}{dt}\theta^{(1)}_i &= \omega^{(1)}_i + \frac{K}{N}\sum_{j\in P_1}\sin(\theta^{(1)}_j-\theta^{(1)}_i-\lambda)
\nonumber
\\ 
&\hphantom{= \omega^{(1)}_i}
+ \frac{\kappa}{N}\sum_{j\in P_2}\sin(\theta^{(2)}_j-\theta^{(1)}_i-\lambda),
\label{eq:abrams_1}    \\
\frac{d}{dt}\theta_{i}^{(2)} &= \omega_{i}^{(2)} + \frac{K}{N}\sum_{j\in P_2}\sin(\theta^{(2)}_j-\theta^{(2)}_i-\lambda) 
\nonumber
\\
&\hphantom{= \omega^{(1)}_i}
+ \frac{\kappa}{N}\sum_{j\in P_1}\sin(\theta^{(1)}_j-\theta^{(2)}_i-\lambda).
\label{eq:abrams_2}
\end{align}
Here $\theta_i^{(\alpha)}$ denotes the phase of oscillator $i=1,\ldots,N$ in population $\alpha\in\{1,2\}$. The intrapopulation and interpopulation coupling strengths are denoted by $K$ and $\kappa$, respectively. We assume attractive forces and a stronger interaction between oscillators belonging to the same population with $K>\kappa>0$. The parameter $\lambda\in [0,\frac{\pi}{2}]$ accounts for a phase lag. Each oscillator is equipped with a native frequency drawn from a distribution $g(\omega)$. Note that through a change of coordinates into a rotating reference frame, $\theta_i^{(\alpha)}(t) \to \theta^{(\alpha)}_i(t) - \bar \omega \, t$ with mean native frequency $\bar \omega = \frac{1}{N} \sum \omega_i$, we may assume without loss of generality that the mean native frequency $\bar\omega$ is zero. Here we consider normally distributed natural frequencies $g(\omega) \sim \mathcal{N}(0,\sigma_{\omega}^2)$ for both populations. In all computations, we set $\sigma_{\omega}^2 = 1$,  and sample the native frequencies as equiprobable draws from $g(\omega)$. This ensures that both populations have exactly the same set of native frequencies and that no heterogeneity is introduced via random sampling. For independent random draws see Appendix~\ref{sec:appNR_R}. We remark that this also best approximates the thermodynamic limit.  In the following we shall use greek sub- or superscripts to label populations and roman subscripts to label indices of oscillators.

The two-population KS model \eqref{eq:abrams_1}--\eqref{eq:abrams_2} undergoes a transition from desynchronised to synchronised dynamics as the coupling strengths $K$ and $\kappa$ are gradually increased. At lower coupling strengths, the oscillators exhibit desynchronised dynamics and a lack of coherent phase alignment. With increasing coupling strengths, the oscillators align their phases and oscillate with a common rotation frequency $\Omega$. Notably, this synchronisation typically involves both populations achieving coherent phase locking. However, depending on the particular choice of the intrapopulation coupling strength $K$ and the interpopulation coupling strength $\kappa$, the system may exhibit symmetry breaking between the two populations and chimera states appear wherein one population achieves synchronisation while the other remains in a (relatively) desynchronised state \cite{AbramsEtAl08, Laing09, MartensEtAl16, BurylkoEtAl22}. 

The collective behaviour of each of the populations is captured by the complex mean-field variables $z_\alpha$, or equivalently by the real mean-field variables $r_\alpha$ and $\psi_\alpha$ for $\alpha=1,2$, with
\begin{align} 
	z_\alpha \coloneq r_\alpha(t)e^{i\psi_\alpha(t)} = \frac{1}{N}\sum_{j\in P_\alpha}e^{i\theta^{(\alpha)}_j(t)}.
\label{eq:z}
\end{align}
In particular, the order parameters ${\bar{r}}_\alpha$ with
\begin{align} 
	{\bar{r}}_\alpha=\lim_{T\rightarrow\infty}\frac{1}{T}\int_{0}^{T}r_\alpha(t)dt,
\end{align}
quantifies the degree of synchronisation within a population $P_\alpha$; complete phase synchronisation with $\theta^{(\alpha)}_1=\theta^{(\alpha)}_2=\cdots=\theta^{(\alpha)}_N$ implies $\bar r_\alpha=1$, whereas a fully desynchronised population with $N$ independently and uniformly spread out phases implies $\bar r_\alpha\sim \mathcal{O}(1/\sqrt{N})$ as a standard Monte Carlo error.


\subsection{Mean-field theory}
\label{sec:mft}

Sakaguchi and Kuramoto \cite{SakaguchiKuramoto86} developed a mean-field theory for the case of a single population with uniform coupling for the order parameter $r$ and the mean frequency $\Omega$. Mean-field theory has been extended for the two-population KS model \eqref{eq:abrams_1}--\eqref{eq:abrams_2}  \cite{AbramsEtAl08, Laing12, MartensEtAl16, BickEtAl20}, which is presented here for completeness. In the two-population KS model, we have mean-field variables $z_\alpha$ for each population and the common mean frequency $\Omega$. The two-population KS model \eqref{eq:abrams_1}--\eqref{eq:abrams_2} can be written in terms of the complex mean field variables \eqref{eq:z} as 
\begin{align}
\label{eq:abramsz_1}
\frac{d}{dt}\theta^{(1)}_i &= \omega_i^{(1)} - \Omega + \textrm{Im}\left[ (K z_1 + \kappa z_2)e^{-i\lambda}e^{-i\theta^{(1)}_i}\right],\\
\label{eq:abramsz_2}
\frac{d}{dt}\theta^{(2)}_i &= \omega_i^{(2)} - \Omega + \textrm{Im}\left[ (K z_2 + \kappa z_1)e^{-i\lambda}e^{-i\theta^{(2)}_i}\right],
\end{align}
where we moved into the frame of reference rotating with constant mean frequency $\Omega$. Each oscillator $\theta^\mypop_i$ only couples to the other oscillators through the mean-field variables $z_\alpha$ and $\Omega$.

In view of describing chimera states in which one of the two populations is synchronised, whereas the other one is not, we shall without loss of generality assume that population $P_1$ is (partially) synchronised and $P_2$ is desynchronised, and we move into the frame of reference of the synchronised cluster with mean phase $\psi_1$ and mean frequency $\Omega$. This implies $z_1e^{-i\psi_1}=r_1$ and motivates introducing the phase-shifted order parameter of the desynchronised population
\begin{align}
\label{eq:Z} 
Z \coloneq z_2=r_2 e^{i\Delta \psi}
\end{align}
with $\Delta \psi = \psi_2-\psi_1$.

The dynamics of the phases in \eqref{eq:abramsz_1}--\eqref{eq:abramsz_2} can now be expressed in terms of $Z$, $r_1$ and $\Omega$ as 
\begin{align}
\label{eq:babyadler1}
 \frac{d}{dt}\theta_j^{(\alpha)}
& = \omega_j^{(\alpha)} - \Omega + \text{Im}(v_{\alpha}e^{-i\theta_j^{(\alpha)}})
\end{align}
for $\alpha=1,2$, where, for clarity of exposition, we introduce
\begin{align}
v_{1} &= (K r_{1}+ \kappa Z)e^{-i\lambda},
\label{eq:babyadler2_1}
\\
v_{2} &= (K Z+ \kappa r_{1})e^{-i\lambda}.
\label{eq:babyadler2_2}
\end{align}

We focus on stationary chimeras for which, in the thermodynamic limit $\Omega$, $r_1$ and $Z$ are constant, corresponding to a stationary distribution of phases. 
For constant mean-field parameters stationary solutions of \eqref{eq:babyadler1}--\eqref{eq:babyadler2_2} exist, provided the coupling strengths are sufficiently large ensuring
\begin{align}
\label{eq:cond_stat}
|\omega^{(\alpha)}_j-\Omega|<|v_{\alpha}|,
\end{align}
and are given by
\begin{align}
\label{eq:statsol}
\theta^{(\alpha)}_j&=\arcsin\left(\frac{\omega^{(\alpha)}_j-\Omega}{|v_{\alpha}|}\right)+\arg(v_{\alpha}).
\end{align} 
If, on the other hand, \eqref{eq:cond_stat} is not satisfied, there are no stationary solutions, and instead the phases evolve according to \eqref{eq:babyadler1}--\eqref{eq:babyadler2_2}.   

In the thermodynamic limit $ N\to \infty$, the phases of oscillators $\theta\in[0,2\pi]$ within population $P_\alpha$ with a native frequency of $\omega$ can be described by a probability density function $\rho^{(\alpha)}(\theta,t; \omega)$ for $\alpha=1,2$, satisfying the continuity equation
\begin{align}
\frac{\partial}{\partial t}\rho^{(\alpha)} + \frac{\partial}{\partial \theta}(\nu^{(\alpha)} \rho^{(\alpha)}) =0,
\end{align}
where the vector fields  $\nu^{(\alpha)}(\theta;\omega)$ are given by the right-hand-side of \eqref{eq:babyadler1} with 
\begin{align}
 \nu^{(\alpha)}(\theta;\omega) &=  \omega - \Omega + \text{Im}(v_{\alpha}e^{-i\theta}),
\label{eq:nu_nonen}
\end{align}
where $v_\alpha$ is given by \eqref{eq:babyadler2_1}--\eqref{eq:babyadler2_2}. 
Stationary solutions satisfying $\partial_t \rho^{(\alpha)}=0$ depend on the equation parameters. Entrained oscillators with frequencies satisfying \eqref{eq:cond_stat} are distributed according to the stationary density 
\begin{align}
\rho^{(\alpha)}(\theta; \omega) &= \delta \left(\theta - \arcsin(\frac{\omega - \Omega}{|v_{\alpha}|})-\arg(v_{\alpha})\right),
\label{eq:rhostat_en}
\end{align}
whereas non-entrained oscillators with frequencies not satisfying \eqref{eq:cond_stat} are distributed according to the stationary density
\begin{align}
\rho^{(\alpha)}(\theta; \omega) &= C^{(\alpha)}(\omega) \frac{1}{\left|\nu^{(\alpha)}(\theta;\omega)\right|}
\label{eq:rhostat_nonen}
\end{align}
with normalisation constant $C^{(\alpha)}(\omega)=\sqrt{(\omega-\Omega)^2-|v^{(\alpha)}|^2}/2\pi$, such that the density is high in regions where the phases spend much of their time (cf. \eqref{eq:babyadler1}--\eqref{eq:babyadler2_2}). 
The order parameters  $z_1$ and $z_2$  are given in the thermodynamic limit as 
\begin{align}
z_1 &= \int_{-\infty}^\infty \int_0^{2\pi} e^{i\theta}\rho^{(1)}(\theta;\omega) g(\omega) d\theta d\omega, \label{eq:general_selconst_r}\\
z_2    &= \int_{-\infty}^\infty \int_0^{2\pi} e^{i\theta}\rho^{(2)}(\theta;\omega) g(\omega) d\theta d\omega \label{eq:general_selconst_Z}.
\end{align}

The integration over the native frequencies $\omega$ has to be split into the entrained and non-entrained ranges (cf. \eqref{eq:cond_stat} for the entrained range), each with their respective stationary density \eqref{eq:rhostat_en} or \eqref{eq:rhostat_nonen}. In particular, upon Fourier decomposition of \eqref{eq:rhostat_nonen} we obtain 

\begin{align}
        z_1&= i\frac{v_{1}}{|v_{1}|}\int_{-\infty}^{\infty}h(\frac{\omega-\Omega}{|v_1|}) g(\omega)d\omega,
         \label{eq:selconst_r} \\
        z_2&= i\frac{v_{2}}{|v_{2}|}\int_{-\infty}^{\infty}h(\frac{\omega-\Omega}{|v_2|}) g(\omega)d\omega ,
        \label{eq:selconst_Z}
\end{align}
with
\begin{align}
    h(s) &= 
\begin{cases}
(1-\sqrt{1-s^{-2}})s & \text{for } |s| > 0, \\
s-i\sqrt{1-s^2} & \text{for } |s| \leq 1,
\end{cases}
\label{eq:hs}
\end{align}
where we adopt the convenient complex formulation introduced in \cite{Omelchenko12}. 
Note that $v_\alpha=v_\alpha(z_{1,2})$. Hence, the complex equations \eqref{eq:selconst_r}--\eqref{eq:selconst_Z} constitute self-consistency relations which allow to determine $z_1$, $z_2$  and $\Omega$, at least numerically~\cite{AbramsEtAl08, Laing12, MartensEtAl16, BickEtAl20}.  
We note that the self-consistency relationships \eqref{eq:selconst_r}--\eqref{eq:selconst_Z} are phase-invariant; therefore, in practice, we additionally impose the restrictions of $z_1=r_1$ and $z_2=Z$ when solving them.


\section{Numerical results for the two-population Kuramoto-Sakaguchi model}
\label{sec:obs}
To numerically integrate the two-population KS model \eqref{eq:abrams_1}--\eqref{eq:abrams_2} we prepare the initial conditions such that population $P_1$ is synchronised with initial phase $\theta_1^{(1)}=\theta_2^{(1)}=\dots=\theta_N^{(1)}=\psi_1=0$ and population $P_2$ is desynchronised with initial phases uniformly sampled from $[-\pi,\pi]$. We employ MATLAB's Runge-Kutta method \texttt{ode45} to integrate \eqref{eq:abrams_1}--\eqref{eq:abrams_2} with relative and absolute tolerances of $2.2 \, 10^{-8}$. To prevent an eventual formation of small clusters due to finite-size sample errors in the native frequencies \cite{PeterPikovsky18, FialkowskiEtAl23} and to ensure that both populations are exactly the same , we draw the two sets of $N$ native frequencies $\omega^\mypop$ equiprobably. We show results for chimera states with $K=100$, $\kappa=60$ and $\lambda=\pi/2-0.075$. For these parameters, the evolution equations \eqref{eq:babyadler1}--\eqref{eq:babyadler2_2} support a stable synchronised solution with constant mean-field variables.

\begin{figure}[htb]
     \centering
     \begin{subfigure}{0.48\linewidth}
         \centering
         \includegraphics[width=\linewidth]{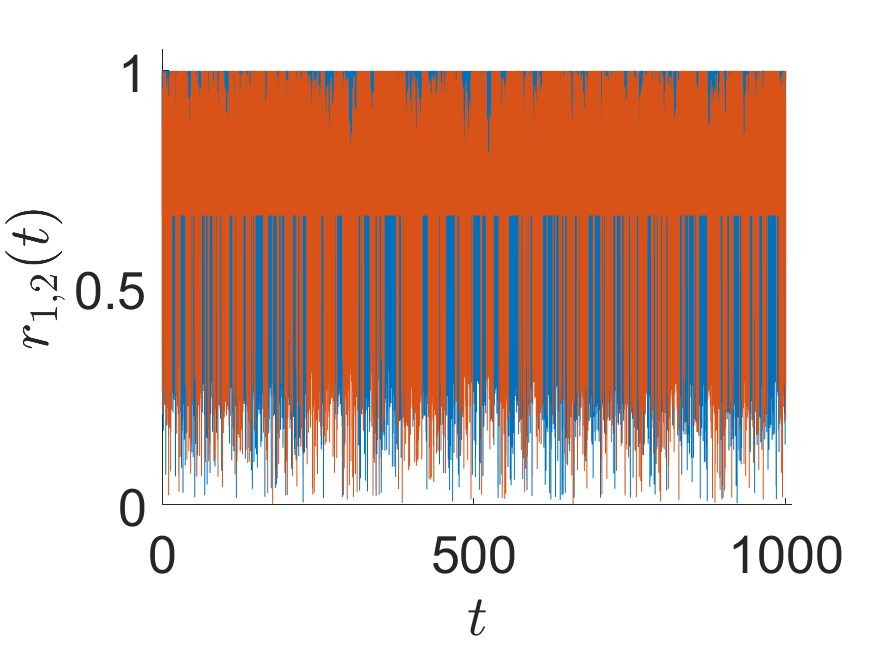}
         \caption{$N=6$}
         \label{fig:rN6}
     \end{subfigure}
     \hfill
     \begin{subfigure}{0.48\linewidth}
         \centering
         \includegraphics[width=\linewidth]{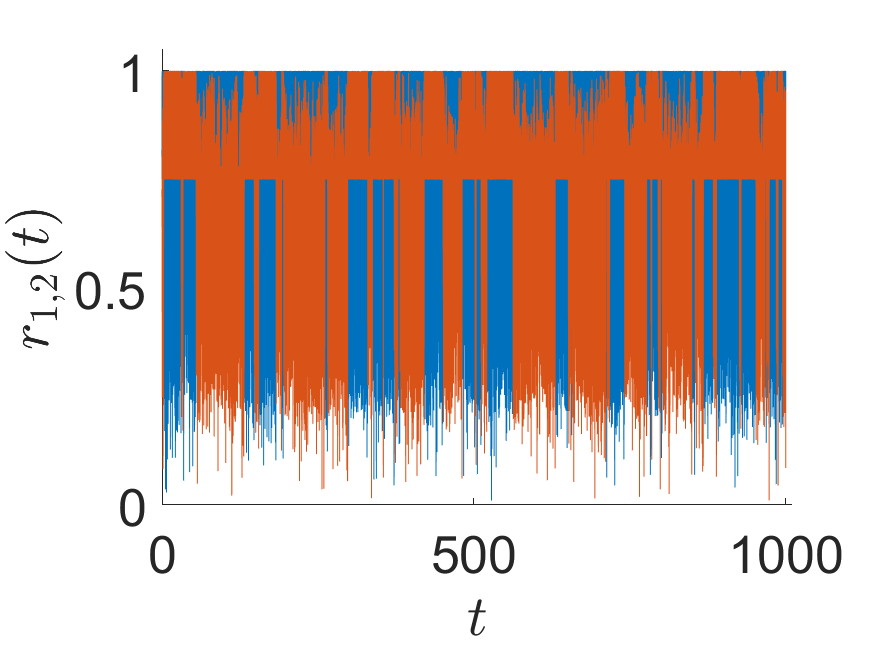}
         \caption{$N=8$}
         \label{fig:rN8}
     \end{subfigure}
     \hfill
     \begin{subfigure}{0.48\linewidth}
         \centering
         \includegraphics[width=\linewidth]{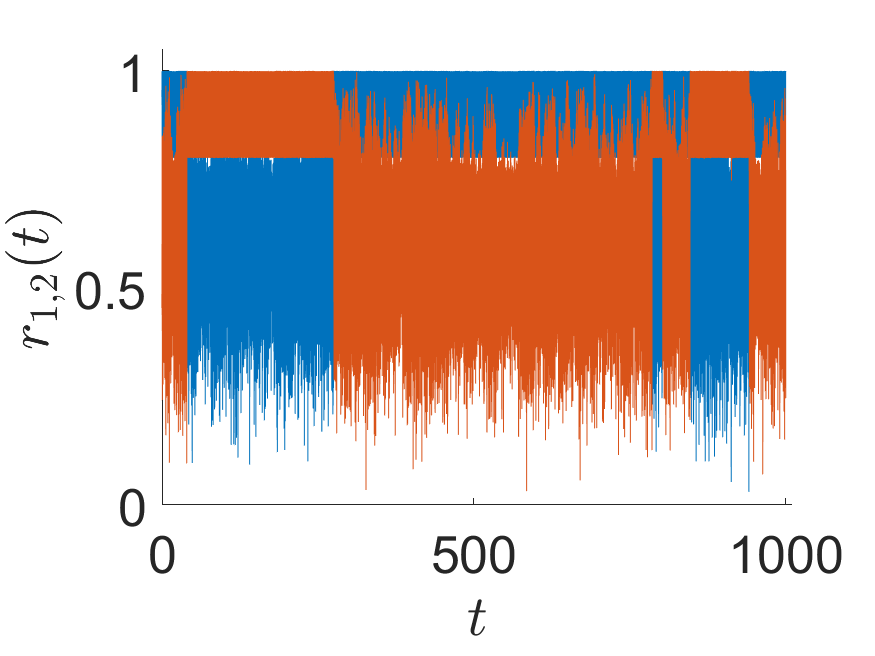}
         \caption{$N=10$}
         \label{fig:rN10}
     \end{subfigure}
     \hfill
     \begin{subfigure}{0.48\linewidth}
         \centering
         \includegraphics[width=\linewidth]{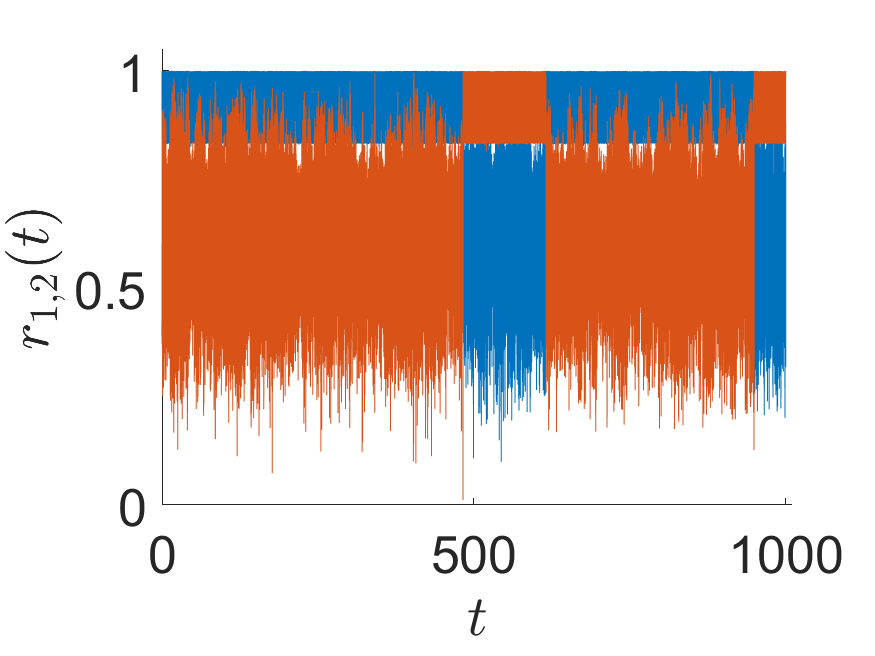}
         \caption{$N=12$}
         \label{fig:rN12}
     \end{subfigure}
     \hfill
     \begin{subfigure}{0.48\linewidth}
         \centering
         \includegraphics[width=\linewidth]{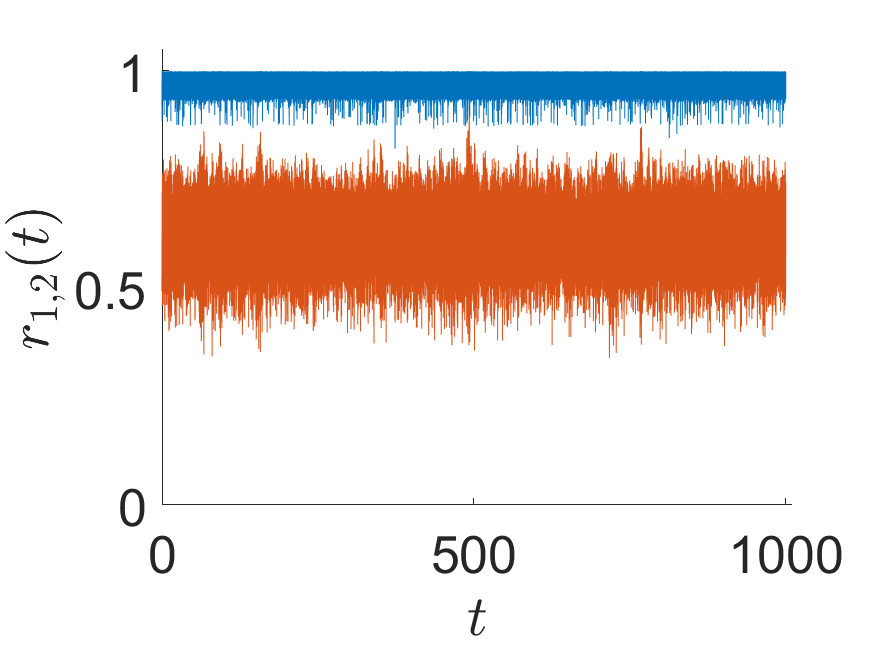}
         \caption{$N=32$}
         \label{fig:rN32}
     \end{subfigure}
     \hfill
     \begin{subfigure}{0.48\linewidth}
         \centering
         \includegraphics[width=\linewidth]{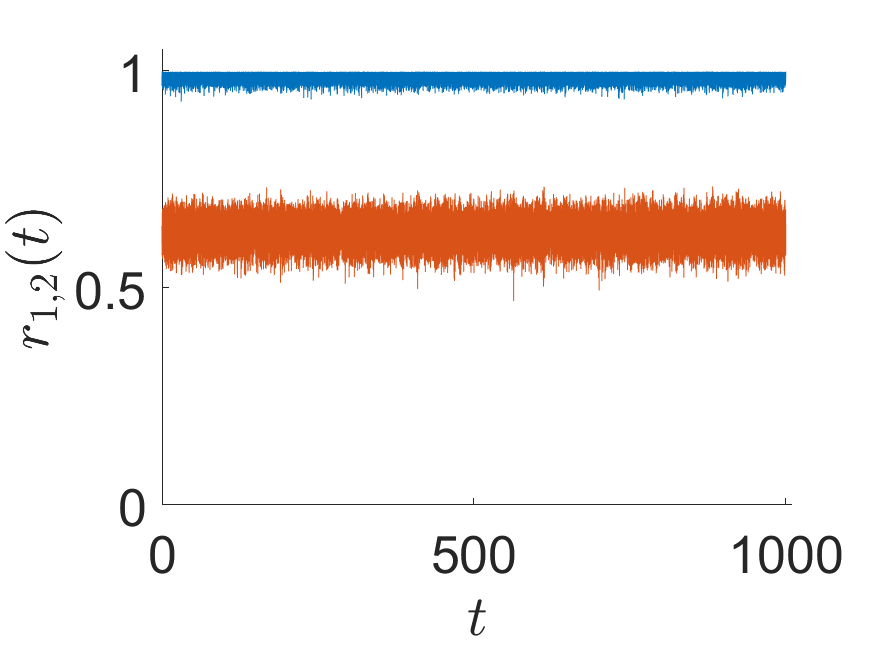}
         \caption{$N=128$}
         \label{fig:rN128}
     \end{subfigure}
        \caption{Order parameters $r_1$ (blue) and $r_2$ (red) for the two-population KS model \eqref{eq:abrams_1}--\eqref{eq:abrams_2} exhibiting switching chimeras as a function of time for various numbers of oscillators $N$. Equation parameters are $K=100$, $\kappa=60$ and $\lambda=\frac{\pi}{2}-0.075$ and native frequencies $\omega$ are drawn equiprobably from a standard normal distribution $\mathcal{N}(0,1)$  such that both populations have the same set of native frequencies.}
        \label{fig:r}
\end{figure}

\begin{figure}[htb]
    \centering
     \begin{subfigure}{0.49\linewidth}
         \centering
         \includegraphics[width=\linewidth]{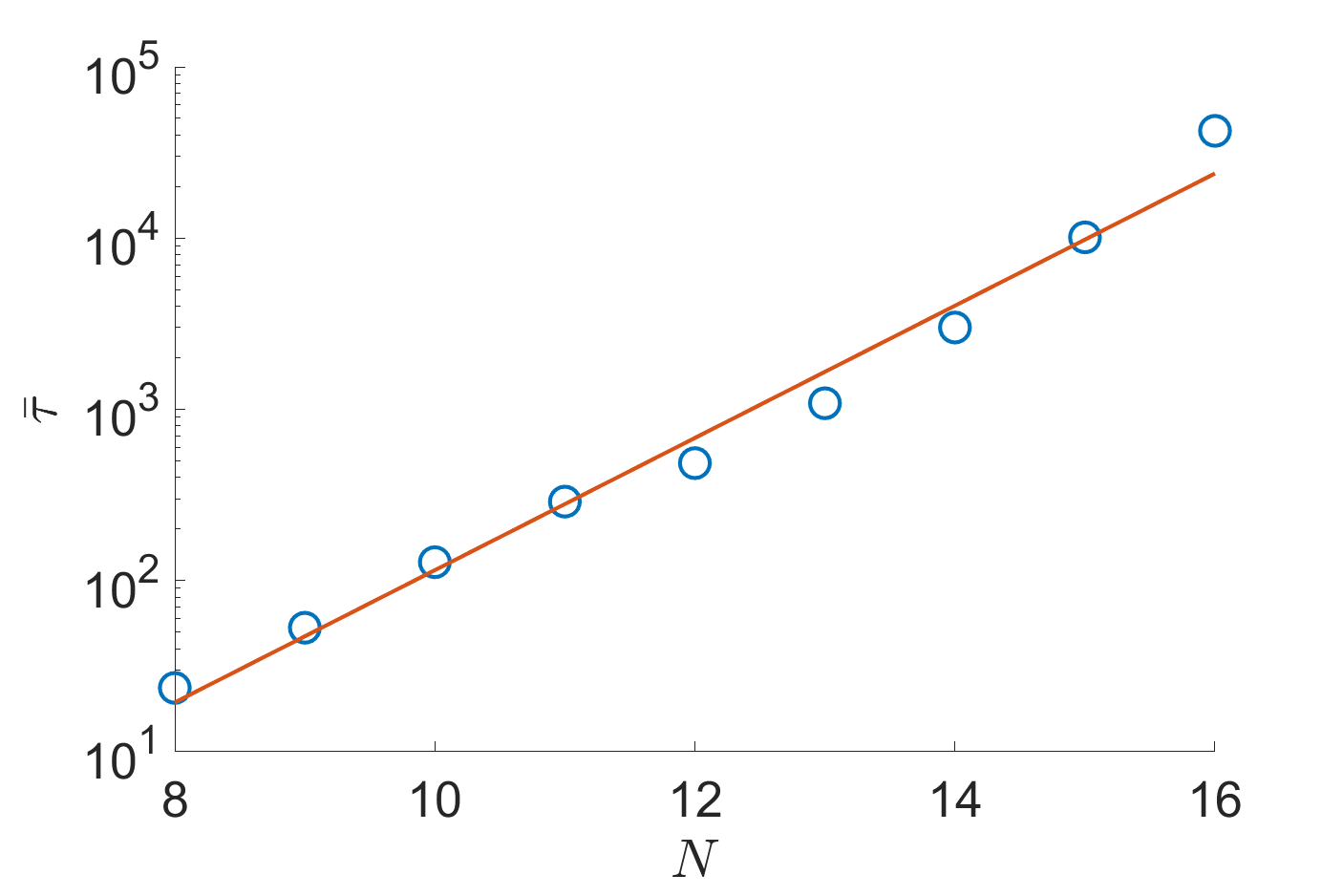}
         \caption{}
     \end{subfigure}
     \begin{subfigure}{0.49\linewidth}
         \centering
         \includegraphics[width=\linewidth]{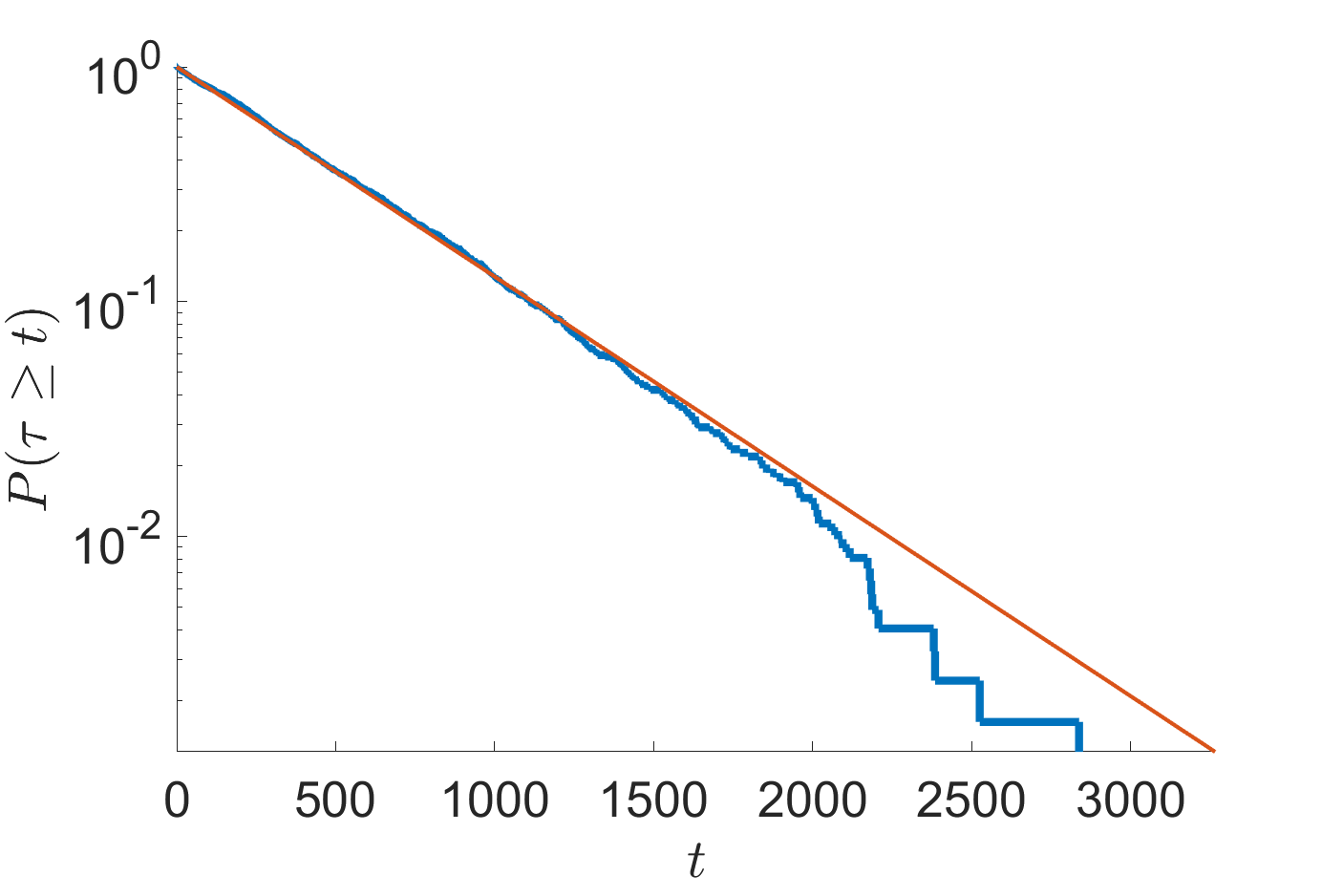}
         \caption{}
     \end{subfigure}
    \caption{Statistics of the switching time. (a) Mean switching time $\bar{\tau}$ as a function of the number of oscillators $N$ (blue circles). The red line shows a best-fit exponential regression model with $\bar{\tau}= 0.0158 \times \exp(0.889N)$.
    (b): 
    Empirical complementary cumulative distribution function $P(\tau\geq t)= \frac{1}{n}\sum_{i=1}^{n}\mathbf{1}_{\tau_i\geq t}$ for a network of size $N=12$, estimated from $n=1234$ switching events $\tau_i$ on a $\log$-scale (blue line). The red line shows a best-fit regression model corresponding to  $P(\tau\geq t)=\exp(-t/\bar\tau)$ with $\bar \tau\approx 486.0$. 
    Equation parameters are as in Figure~\ref{fig:r}.}
    \label{fig:tau}
\end{figure}

To illustrate switching chimeras and how their existence depends on the number of oscillators $N$, we show in Figure~\ref{fig:r} the temporal evolution of the order parameters $r_1$ and $r_2$ for $N$ ranging from $N=6$ to $N=128$. Switching between synchronised and desynchronised states is clearly seen. Switching events are not cyclic but appear randomly, and the time between switches increases with $N$. We checked that for $N=21$, no chimera switching occurred for our initial condition when integrating for $10^6$ time units.  
It is further seen that the synchronised population exhibits less variation in the order parameter for larger numbers of oscillators. Figure~\ref{fig:tau}(a) shows the mean switching time $\bar\tau$, estimated from $1,000$ switching events, as a function of the system size $N$. The mean switching time $\bar\tau$ increases exponentially with the network size $N$. A best-fit line suggests $\bar{\tau}\sim \exp(0.889 N)$. This suggests that  chimera switching is a finite-size effect.  In Figure~\ref{fig:tau}(b) we show the empirical probability $P(\tau\geq t)$ of switching times $\tau$ which are greater than or equal to some $t$ for fixed $N=12$, revealing an exponential distribution $P(\tau\geq t)\approx \exp(-t/\bar{\tau})$ with $\bar{\tau}\approx 486.0$. This suggests that switching events occur independently and follow a Poisson process.  We shall see that switching as well as the variance of the order parameter of the synchronised population is caused by an effective stochastic driving of the synchronised population by the desynchronised population.


\begin{figure}[htb]
     \centering
     \includegraphics[width=0.70 \linewidth]{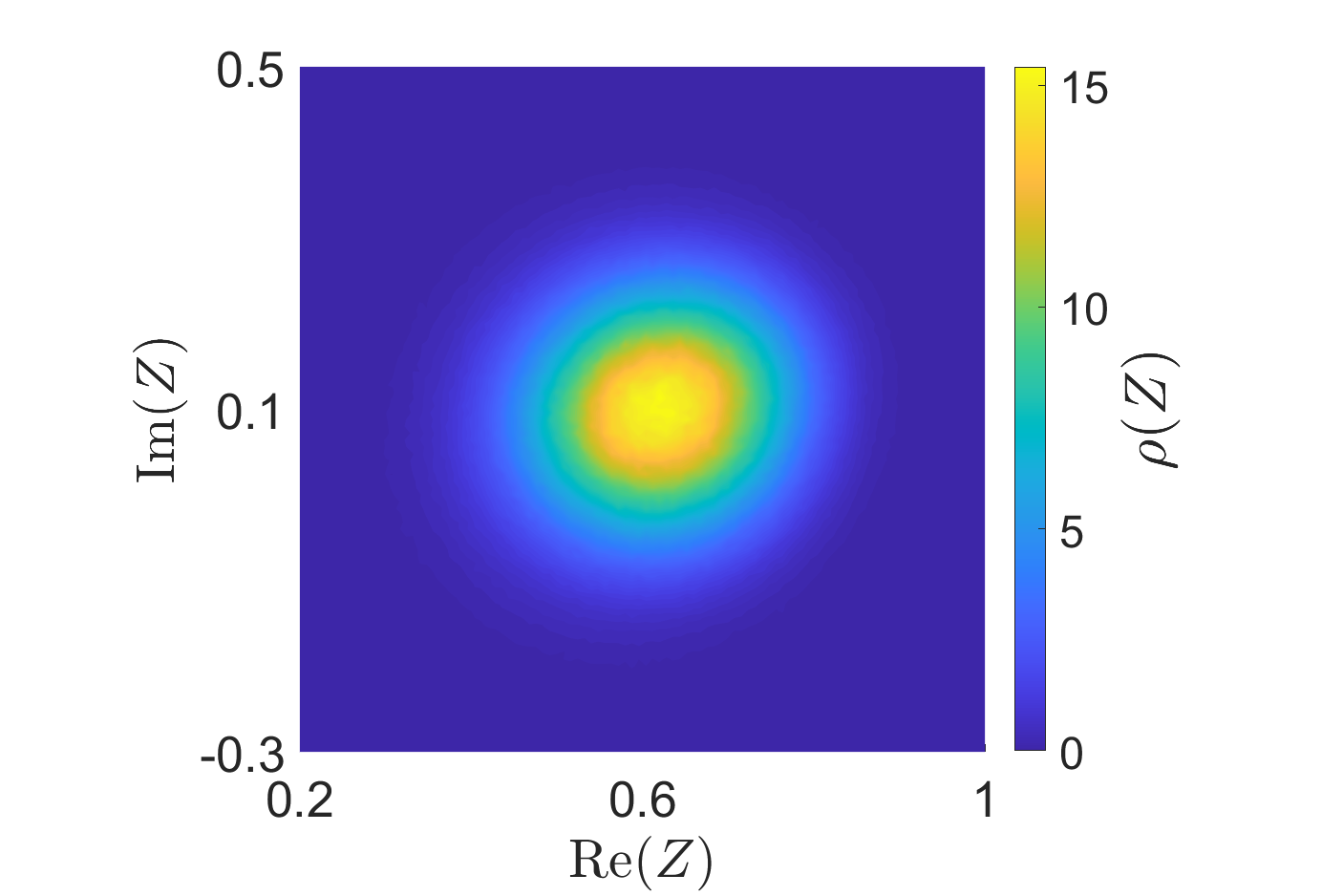}\\
     \includegraphics[width=0.49\linewidth]{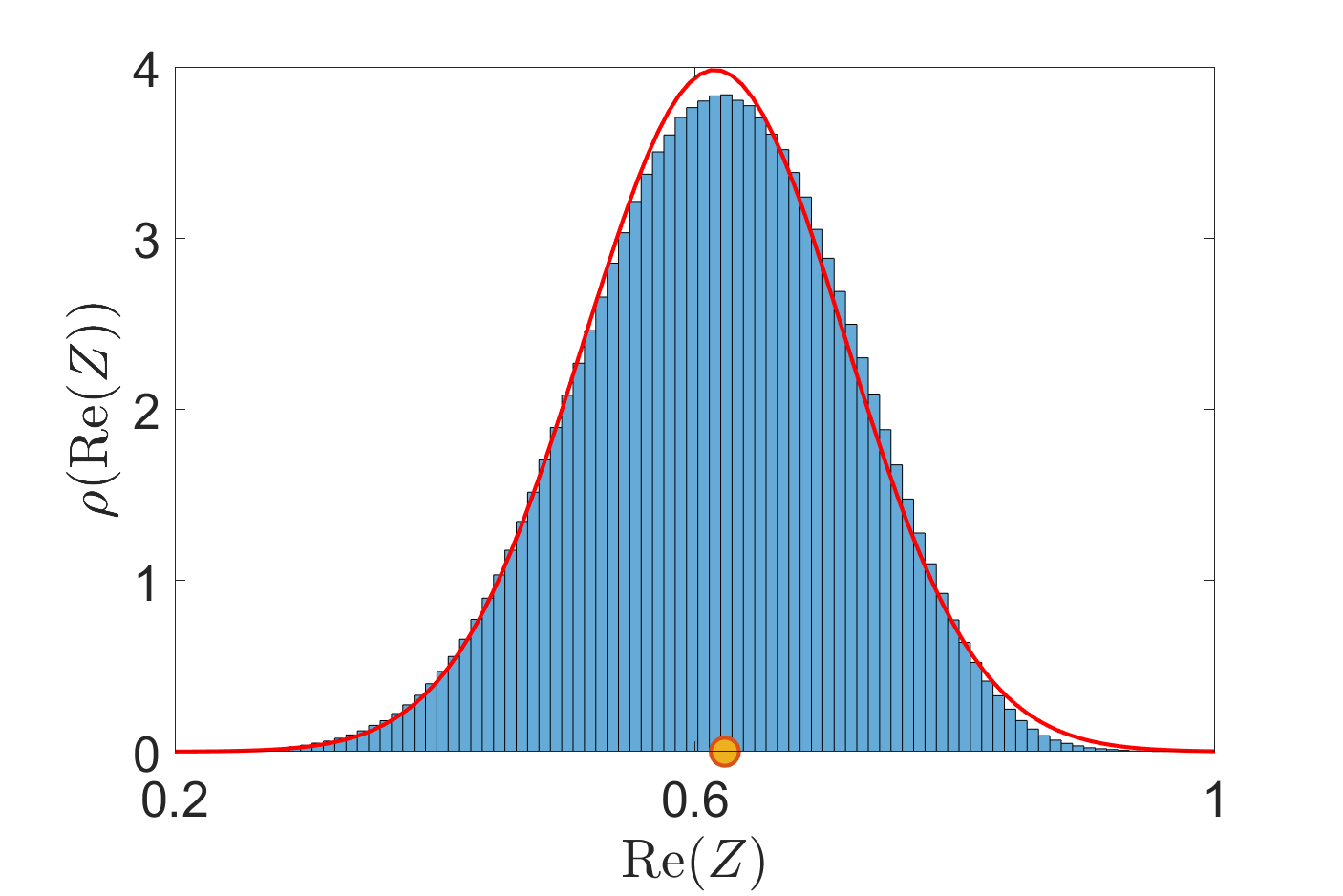}
     \includegraphics[width=0.49 \linewidth]{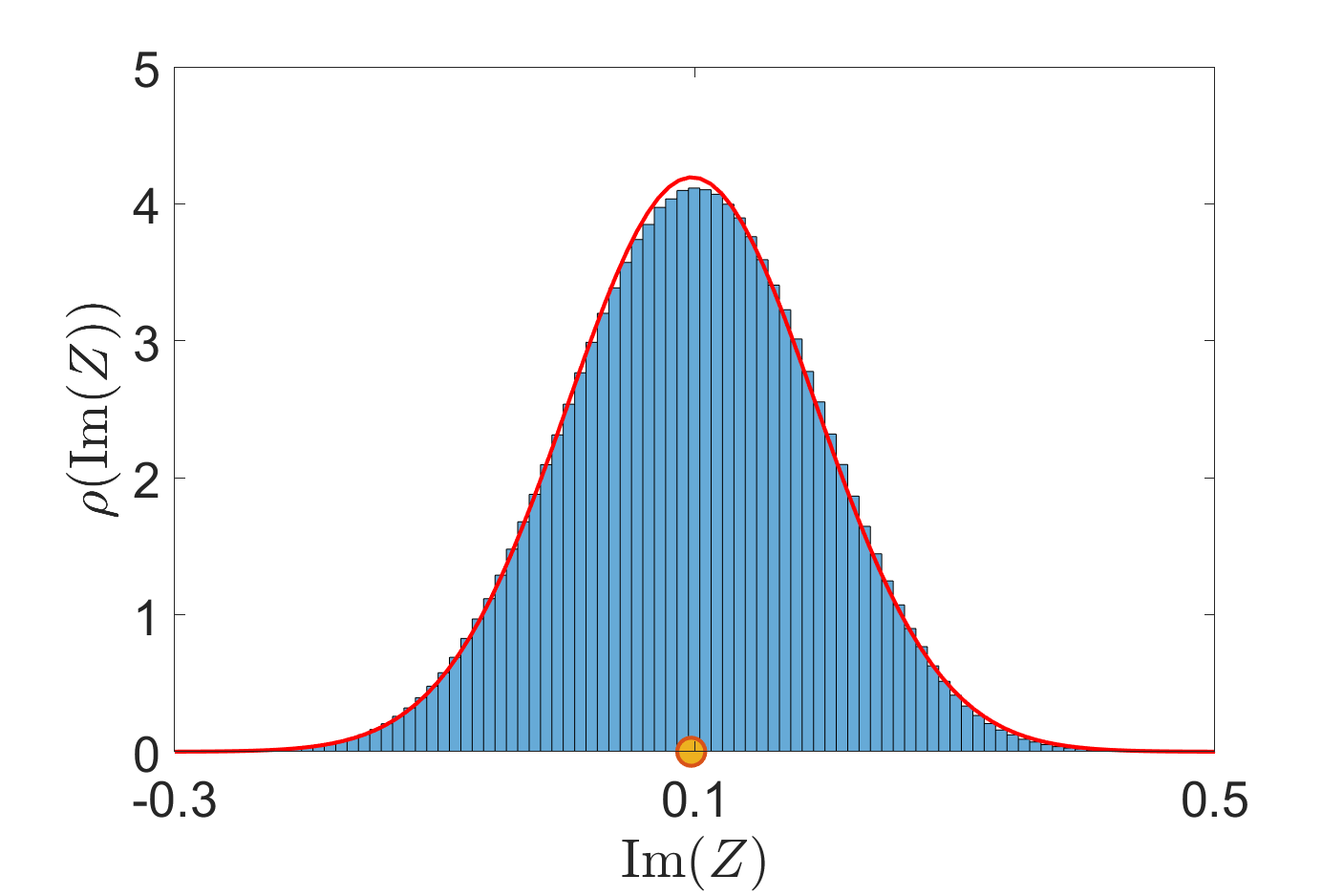}
\caption{Empirical joint (top) and marginal (bottom) distributions of $Z$ obtained from simulating \eqref{eq:abrams_1}--\eqref{eq:abrams_2} with $N=16$. The filled orange circles denote the thermodynamic value of the real and imaginary part of $Z$ as calculated from the self-consistency relations \eqref{eq:selconst_r}--\eqref{eq:selconst_Z}  of the mean-field theory. The red lines are best-fit Gaussian distributions. Equation parameters are as in Figure~\ref{fig:r}.}
        \label{fig:heat}
\end{figure}

To investigate how such an effective stochastic dynamics is generated in the deterministic two-population KS model, we examine the statistical behaviour of $Z$, which captures the driving interaction term of the synchronised population $P_1$ with the desynchronised population $P_2$ (cf. \eqref{eq:babyadler1}). Figure~\ref{fig:heat} shows the empirical distribution of the driving term $Z$ for $N=16$ oscillators. To generate the empirical distribution, we simulated \eqref{eq:abrams_1}--\eqref{eq:abrams_2} for $10,000$ time units, and determined $Z$ from the desynchronised population, employing the symmetry of the model, irrespective of whether that is population $P_1$ or $P_2$. The real and imaginary parts of $Z$ are well approximated by a Gaussian with an empirical mean of $\langle Z\rangle=0.6243 + 0.1007\, i$. The angular brackets $\langle\cdot \rangle$ denote a temporal average. 

\begin{figure}[htb]
     \centering
     \includegraphics[width=0.49\linewidth]{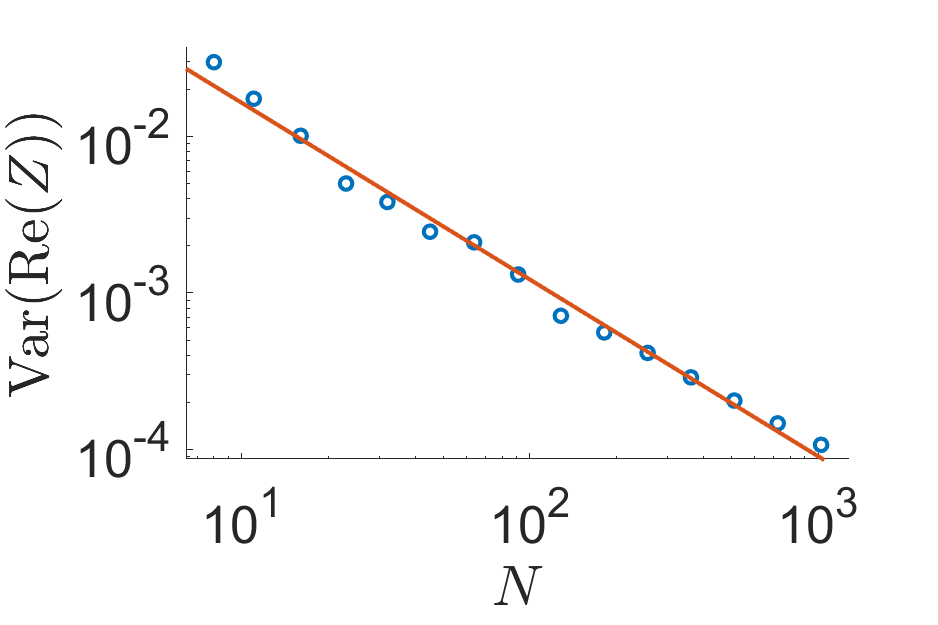}
     \includegraphics[width=0.49\linewidth]{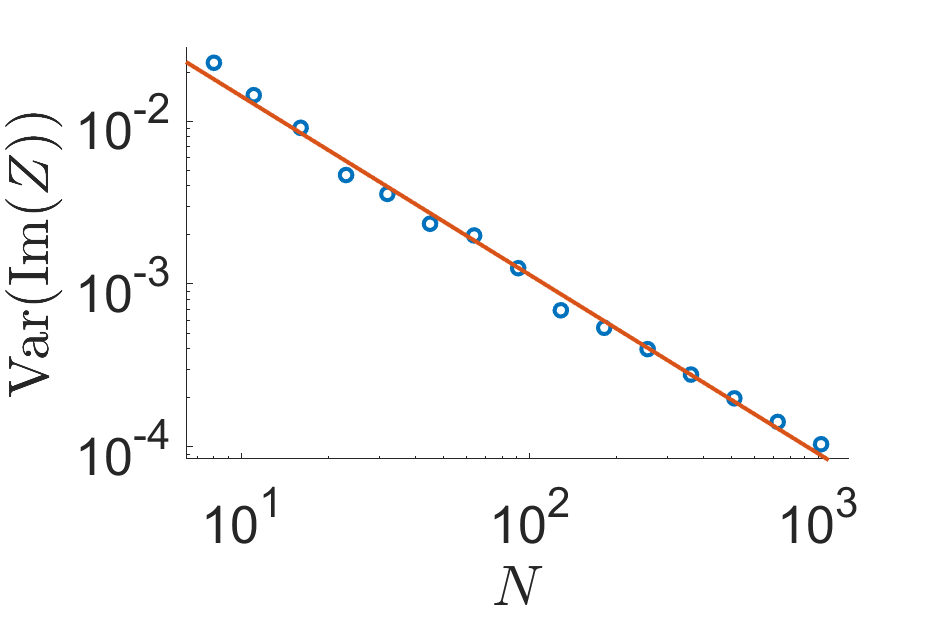}
\caption{Variance of the real (left) and imaginary (right) component of $Z$ as a function of the number of oscillators $N$. The red lines show a power law best fit indicating a scaling with $N^{-1.13}$ and $N^{-1.10}$ for the real and imaginary parts, respectively. Equation parameters are as in Figure~\ref{fig:r}.}
\label{fig:varN}
\end{figure}

\begin{figure}[htb]
     \centering
     \begin{subfigure}{0.48\linewidth}
         \centering
         \includegraphics[width=\linewidth]{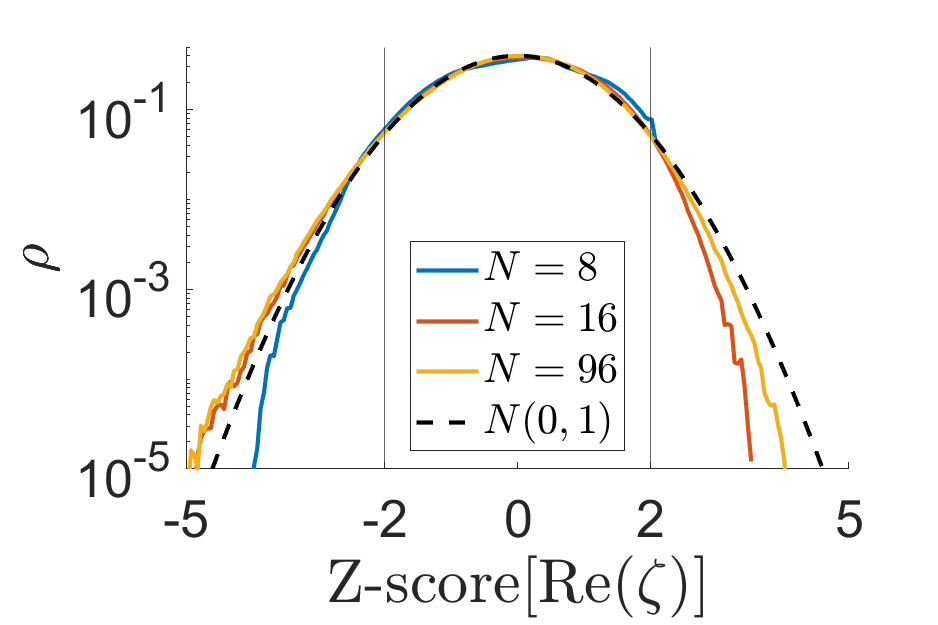}
         \caption{}
         \label{fig:zeta_marginal_real}
     \end{subfigure}
     \hfill
     \begin{subfigure}{0.48\linewidth}
         \centering
         \includegraphics[width=\linewidth]{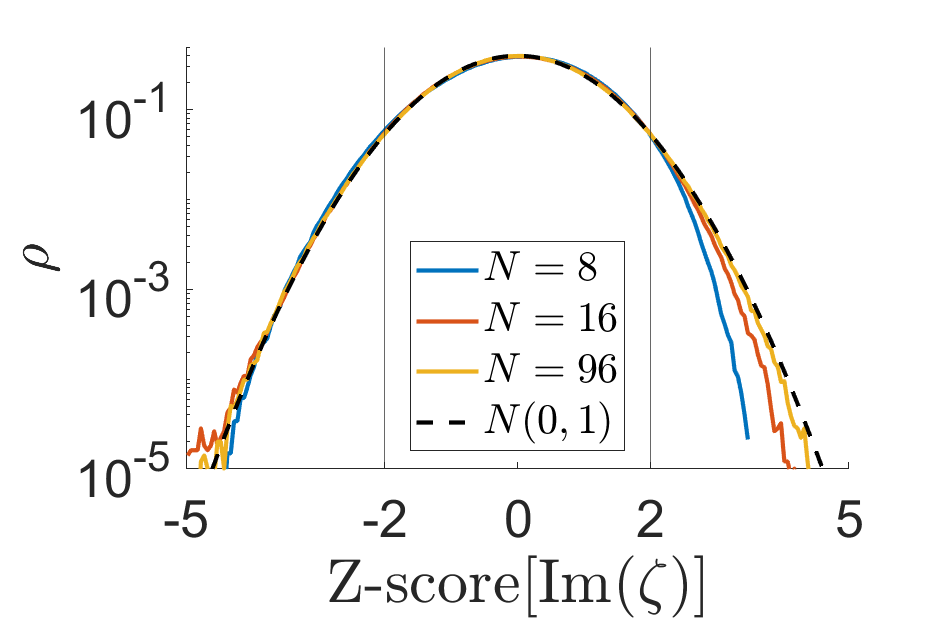}
         \caption{}
         \label{fig:zeta_marginal_imag}
     \end{subfigure}
     \hfill
     \begin{subfigure}{0.70\linewidth}
         \centering
         \includegraphics[width=\linewidth]{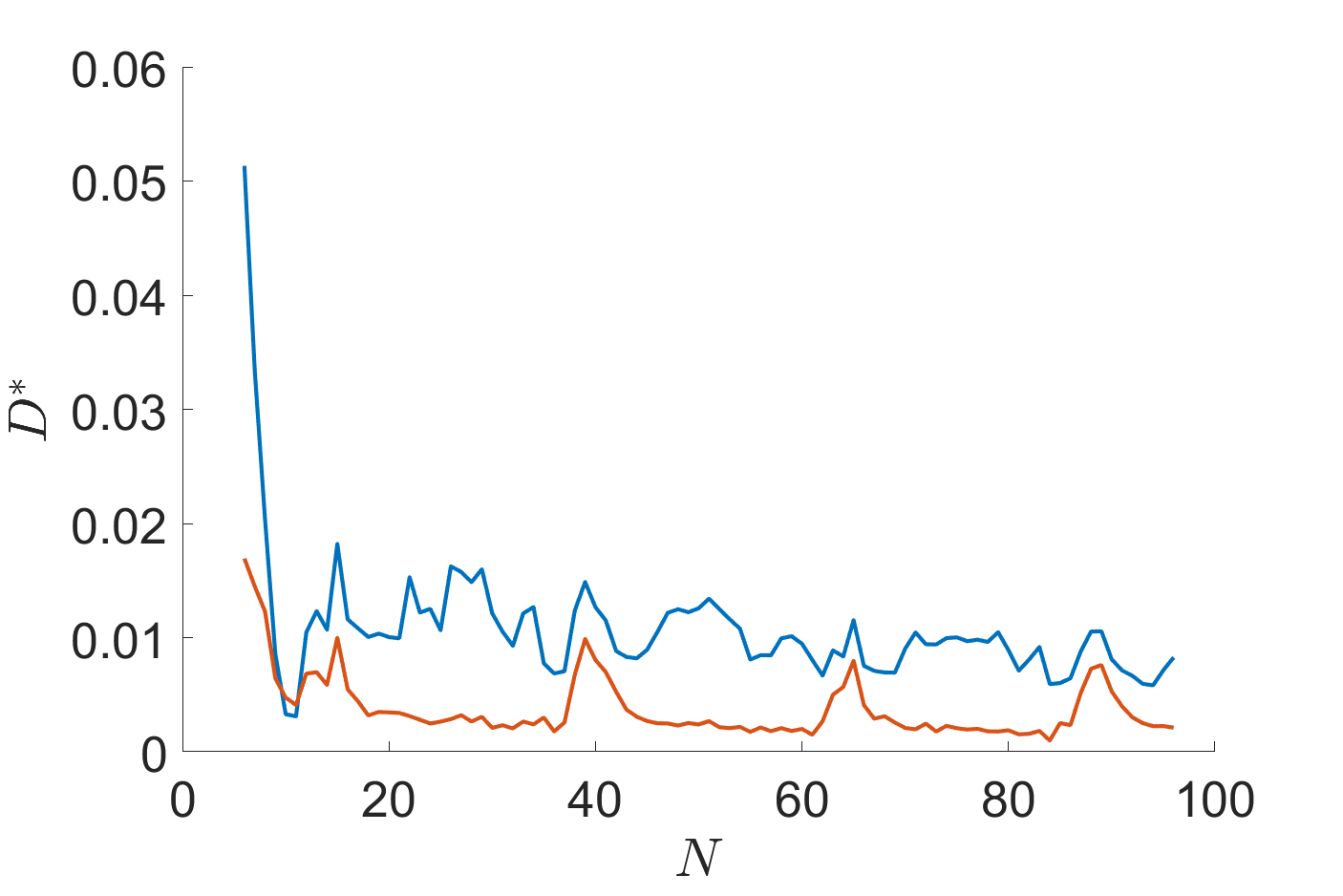}
         \caption{}
         \label{fig:zeta_marginal_error}
     \end{subfigure}
     
\caption{Logarithm of the empirical distribution of the Z-score for the real (a) and imaginary (b) part of fluctuations $\zeta$ obtained from simulating \eqref{eq:abrams_1}--\eqref{eq:abrams_2} for $N=8$, $N=16$ and $N=96$ showing convergence with increasing system size $N$ to a standard normal distribution. The vertical lines delineate deviations with $2$ standard deviations. 
(c): Kolmogorov–Smirnov statistics for the distributions of the real (blue) and imaginary (red) components of $\zeta$ quantifying the convergence to a Gaussian with increasing $N$. Equation parameters are as in Figure~\ref{fig:r}.}
        \label{fig:zetadens}
\end{figure}

Figure~\ref{fig:varN} shows how the variance of the real and imaginary components of $Z$ varies with system size $N$. It is clearly seen that the variance decreases as $1/N$. The Gaussianity of the distribution and the scaling of the variance with approximately $1/N$ suggests that finite-size fluctuations of $Z$ around its thermodynamic mean $\langle Z\rangle$ are governed by an underlying central limit theorem and 
\begin{align}\label{eq:zeta_def}
\zeta = \sqrt{N}\left( Z - \langle Z\rangle \right) + o(1/\sqrt{N})
\end{align}
is effectively a stochastic complex-valued Gaussian process. Hence, to obtain an effective stochastic dynamics capable of inducing the observed chimera switching and the approximation of $Z(t)$ by a Gaussian process, we require the number of oscillators to be sufficiently large for the central limit theorem to hold and sufficiently small so that the variance of the effective noise is non-negligible. In Figure~\ref{fig:zetadens} we show how the empirical histograms of the real and imaginary part of $\zeta$, obtained from simulations of \eqref{eq:abrams_1}--\eqref{eq:abrams_2} over $10,000$ time units, converge to a Gaussian for increasing system size $N$.  We plot in Figure~\ref{fig:zetadens}(a), (b) the empirical probability density function on a logarithmic scale after scaling the fluctuations by their empirical standard deviation and compare with a standard unit Gaussian distribution.  It is clearly seen that the fluctuations are not Gaussian for $N=8$, which is too small for the central limit theorem to work. For larger system sizes, the distribution of the components of $\zeta$ becomes approximately Gaussian, and indeed becomes more Gaussian as $N$ increases.  In particular, for $N=16$ the centre of the distribution of $\zeta$ is well approximated by a Gaussian but the tail for fluctuations deviating more than $2$ standard deviations from the mean departs from a Gaussian distribution. Note that this implies that roughly $94.5\%$ of the fluctuations are well described by a Gaussian distribution.  We remark that increasing the system size leads to a decrease in the variance of the (scaled) fluctuations $\zeta$, with saturation occurring around $N=48$ (not shown).  We further show in Figure~\ref{fig:zetadens}(c) the Kolmogorov-Smirnov statistics $D^*$ quantifying the convergence to a Gaussian distribution with increasing system size \cite{MasseyFJ51}. 
%
%

Note that the observed exponential increase of the switching time implies a mean switching time of approximately $5\times 10^{16}$ time units for $N=48$ (cf. Figure~\ref{fig:tau}(a)). This implies that for all practical purposes, switching is not observable for $N=48$ in simulations, even for large integration times of $10^6$ time units.  
We hence limit the numerical study of switching to systems of $N\le 16$.\\

\begin{figure}[htb]
     \centering
     \includegraphics[width=0.80 \linewidth]{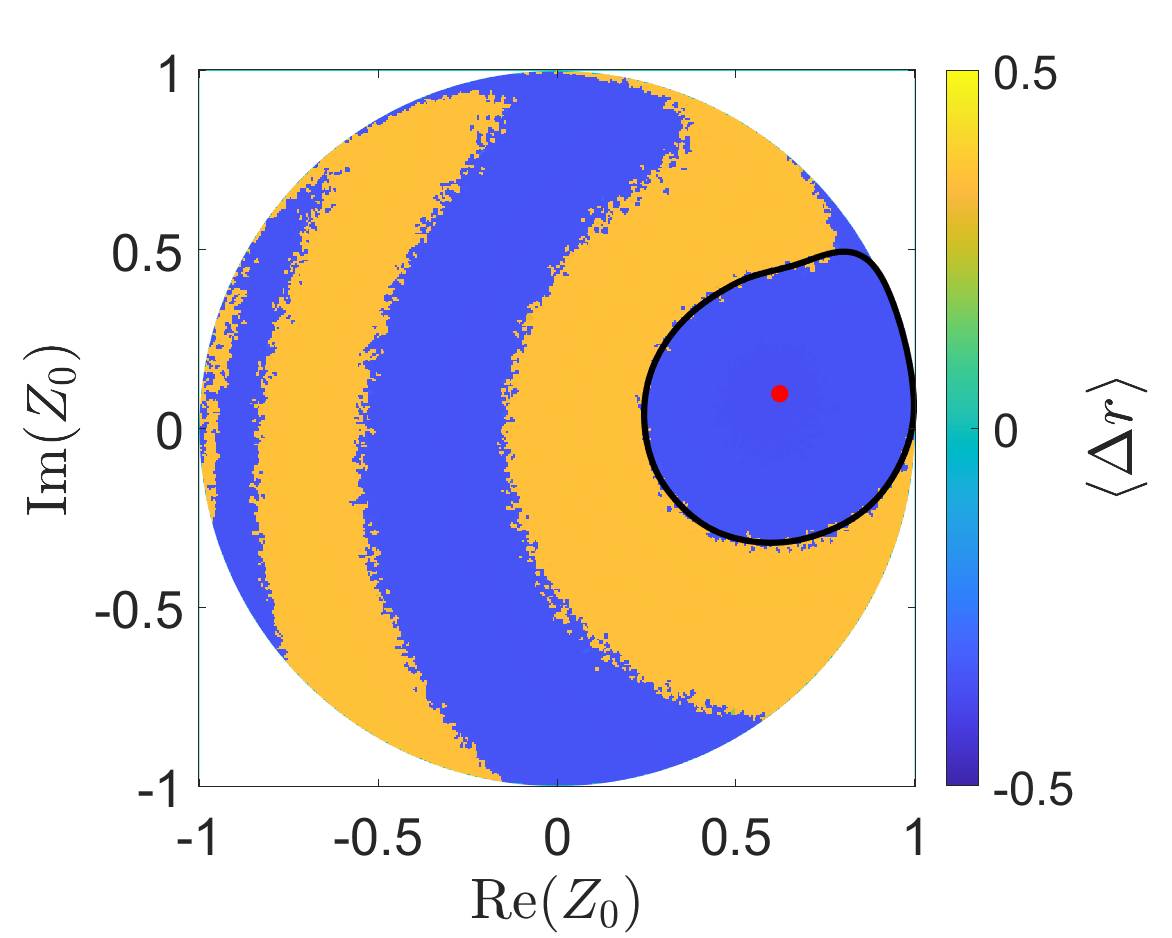}
\caption{Time-averaged difference of the order parameters, $\langle \Delta r \rangle := \langle r_2-r_1 \rangle = \langle |Z|-r_{1}\rangle$, as a function of the initial condition $Z_0=Z(0)$ of the two-population KS model \eqref{eq:abrams_1}--\eqref{eq:abrams_2}. $N=512$ oscillators were initialised with $\theta_j^{(1)}(0)=0$ for all $j\in P_1$ and $\theta_j^{(2)}(0)$ were drawn from the stationary distribution for non-entrained oscillators \eqref{eq:rhostat_nonen}.  We used $30,000$ different initial conditions.  The temporal mean was estimated during the time window $t\in[2.5,10]$. The thermodynamic equilibrium with $\langle Z\rangle =0.6230 + 0.0972i$ is shown as a red dot. The blue region denotes initial conditions that evolve into a state where $P_1$ is synchronised and $P_2$ is desynchronised, and the yellow region denotes initial conditions that evolve into states where $P_1$ is desynchronised and $P_2$ is synchronised. The black curve approximates the boundary separating these two states.  The curve was obtained by first via ray casting from $\langle Z\rangle$ at $1,000$ different angles to locate points separating the two regions. The curve is then obtained by smoothing with a Gaussian kernel with a standard deviation of $0.162$ radians.  The boundary has an average radius of $q^\star=0.386$. Equation parameters are as in Figure~\ref{fig:r}.}
    \label{fig:raycast}
\end{figure}

Recall that $Z$ encodes the effect of the desynchronised population onto the phases of the synchronised population, i.e. the second sum in the right-hand side of \eqref{eq:abrams_1} using our convention that population $P_1$ is the synchronised one. This opens up the way to employ the stochastic model reduction framework developed in \cite{YueGottwald24} to model the dynamics of the phases of the synchronised population. This will be achieved in Section~\ref{sec:stochred}. To allow for an analytical approximation of the mean switching time, we will need to determine the critical size of fluctuations that are sufficient to induce the synchronised population to switch. The asymptotic state of population $P_1$ sensitively depends on the initial condition of the desynchronised population $P_2$. Figure~\ref{fig:raycast} shows the dependency of the average value of $\Delta r := r_2-r_1= |Z|-r_{1}$ on the initial value $Z_0=Z(0)$ for a network with $N=512$ oscillators. Results were obtained by simulating the two-population KS model \eqref{eq:abrams_1}--\eqref{eq:abrams_2} for $10$ time units, during which the order parameters have equilibrated. We chose non-equilibrium initial conditions for population $P_1$ with $\theta_1^{(1)}=\theta_2^{(1)}=\dots=\theta_N^{(1)}$, and random initial conditions for population $P_2$ with $\theta^{(2)}_j$, $j=1,\dots,N$, drawn from the stationary density \eqref{eq:rhostat_nonen} consistent with the initial value $Z_0$. We observe two scenarios: either $P_1$ is synchronised and $P_2$ is desynchronised (blue region), or $P_1$ is desynchronised and $P_2$ is synchronised (yellow region). The boundary separating these two states and its distance to the fixed point $\langle Z\rangle$ (red cross) determines the size of critical fluctuations of $Z$. The near circular boundary suggests that the critical fluctuation size is approximated by the average radius of the basin boundary, which is estimated as $q^\star=0.386$. In the following section, we employ mean-field theory to determine an analytical expression for the critical size $q^\star$.

%


\section{Deterministic model reduction} 
\label{sec:detred}
Using the celebrated Ott-Antonsen ansatz valid in the thermodynamic limit, the two-population KS model \eqref{eq:abrams_1}--\eqref{eq:abrams_2} was shown to support stable chimeras, breathing chimeras, as well as saddle-node and Hopf bifurcations of chimera states \cite{AbramsEtAl08, Laing09, Laing12, MartensEtAl16, BickEtAl20}. We will follow \cite{Laing09} and derive a low-dimensional dynamical model for the complex order parameters \eqref{eq:z} in a perturbative approach. Using \eqref{eq:babyadler1}--\eqref{eq:babyadler2_2} we obtain the evolution equation for the complex order parameters 
\begin{align}
\label{eq:dzk}
\frac{d}{dt}z_{\alpha}&=\frac{1}{N}\sum_{j\in P_{\alpha}} ie^{i\theta^{(\alpha)}_j}\left(\omega^{(\alpha)}_j+\text{Im}(v_{\alpha}e^{-i\theta^{(\alpha)}_j})\right)
\nonumber\\&=\frac{1}{N}\sum_{j\in P_{\alpha}} ie^{i\theta^{(\alpha)}_j}\left(\omega^{(\alpha)}_j+\frac{v_{\alpha}e^{-i\theta^{(\alpha)}_j}-v_{\alpha}^*e^{i\theta^{(\alpha)}_j}}{2i}\right)
\nonumber\\&= i z_\alpha^{(1,1)} + \frac{1}{2}(v_\alpha-v^*_\alpha z_\alpha^{(2,0)}),
\end{align}
for the two populations with $\alpha=1,2$,  
$z_\alpha^{(m,n)}$ is defined as
\begin{align}
z_{\alpha}^{(m,n)} = \frac{1}{N}\sum_{j\in P_\alpha} (\omega^{(\alpha)}_j)^n \exp(im\theta^{(\alpha)}_j)
\label{eq:zmn}
\end{align}
for $m,n\in\mathbb{N}$, and where we again moved into a frame of reference of the synchronised cluster with mean phase $\psi_1$ and mean frequency $\Omega$, and where we have again assumed without loss of generality that population $P_1$ is synchronised with $z_1=r_1$ and $z_2=r_2 e^{i(\psi_2-\psi_1})$. Note that $z_\alpha^{(1,0)}=z_\alpha$.\\

\begin{figure}[htb]
     \centering
      \includegraphics[width=0.49\linewidth]{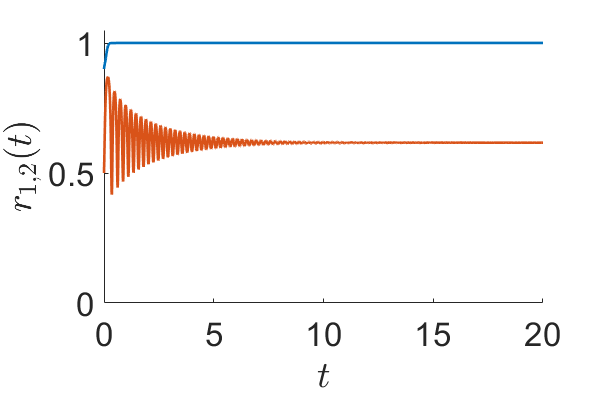}
      \includegraphics[width=0.49\linewidth]{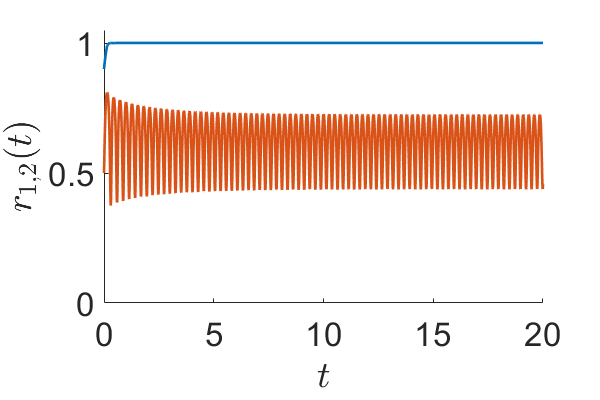}
\caption{Order parameters $r_1$ (blue) and $r_2$ (red) as a function of time for the reduced deterministic mean-field dynamics \eqref{eq:dzk_approx} with $K=100$, $\kappa=60$ and $\lambda=\pi/2-0.075$ (left) and $K=90$, $\kappa=50$ and $\lambda=\pi/2-0.1$ (right). We used $z_1(0)=0.9$ and $z_2(0)=0.5$ to generate the chimera state in the reduced model.}
\label{fig:zkapprox}
\end{figure}

Equation \eqref{eq:dzk} is not a closed equation for the complex order parameters $z_\alpha$ but instead presents a closure problem whereby higher and higher orders of the order parameter are required to close the equation with  
\begin{align}
\label{eq:dzkmn}
\frac{d}{dt}z_{\alpha}^{(m,n)} = m(i z_\alpha^{(m,n+1)} + \frac{1}{2}(v_\alpha z_\alpha^{(m-1,n)}-v^*_\alpha z_\alpha^{(m+1,n)})).
\end{align}
In order to close the equations for $z_{1,2}$, we truncate the system by employing the following two approximations
\begin{align}
\label{eq:zkapprox_cond}
z_{\alpha}^{(2,0)} = z_\alpha^2 \qquad {\rm{and}} \qquad 
z_{\alpha}^{(1,1)} = 0
\end{align}
for $\alpha=1,2$. We checked the validity of these approximations numerically (not shown) and provide a heuristic formal justification in Appendix~\ref{sec:app1}. Substituting the approximations \eqref{eq:zkapprox_cond} into \eqref{eq:dzk} we obtain the closed Ott-Antonsen equations~\cite{OttAntonsen08, Laing12, MartensEtAl16, BickEtAl20} for the dynamics of the complex order parameters with
\begin{align}
\label{eq:dzk_approx}
\frac{d}{dt}{z}_{\alpha} = \frac{1}{2}(v_\alpha-v^*_\alpha z_\alpha^2)
\end{align}
for $\alpha=1,2$. The closed equations \eqref{eq:dzk_approx} exhibit a Hopf bifurcation upon varying the intrapopulation coupling strength $K$~\cite{AbramsEtAl08}. Figure~\ref{fig:zkapprox} shows the temporal evolution of the order parameters $r_{1,2}=|z_{1,2}|$ under the Ott-Antonsen dynamics \eqref{eq:dzk_approx}. The dynamics exhibits chimera states in the reduced model characterised by a synchronised population $P_1$ with $r_1=1$ and a desynchronised population $P_2$. The desynchronised population either has a stationary limiting state (with respect to the reference frame $\Omega$) with $r_2\approx 0.62$ or supports a stable limit cycle oscillating around a similar value which are known as breathing chimera states in the full two-population KS model \eqref{eq:abrams_1}--\eqref{eq:abrams_2}. Our focus here is on the case of a stationary fixed point. The stationary fixed point of the reduced Ott-Antonsen equation \eqref{eq:dzk_approx} is $Z=0.6096 + 0.0934\,i$, matching well the results of the full model \eqref{eq:abrams_1}--\eqref{eq:abrams_2} with $\langle Z\rangle=0.6243 + 0.1007\, i$ (cf. Figure~\ref{fig:heat}).

\begin{figure}[htb]
     \centering
     \includegraphics[width=0.49\linewidth]{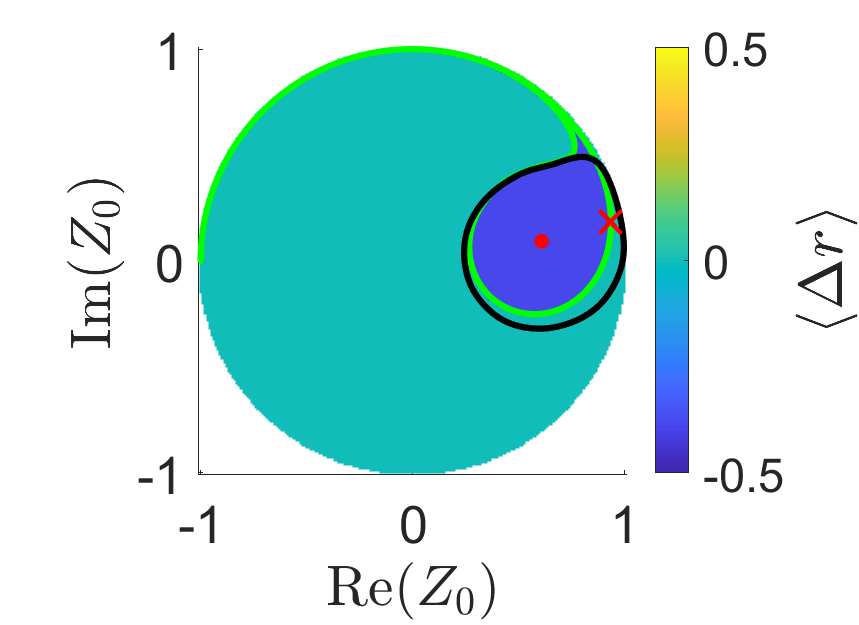}
     \includegraphics[width=0.49\linewidth]{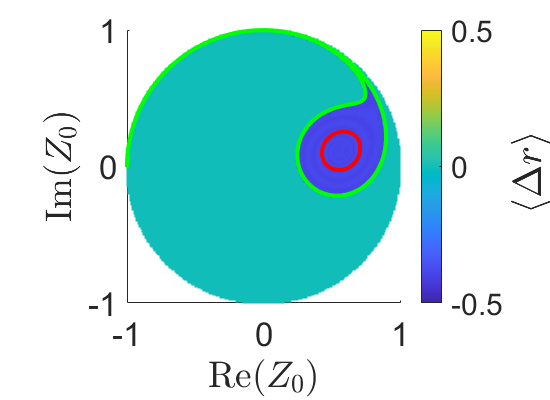}
\caption{Asymptotic value of the difference of the time-averaged order parameters, $\Delta r:= r_2-r_1= |Z|-r_{1}$, as a function of the initial condition $Z_0$ for the reduced deterministic mean-field model \eqref{eq:dZ}.
Left: Initial conditions in the blue region evolve to a fixed point with $\hat Z=0.6096+0.0934\,i$ (red dot) for $K=100$, $\kappa=60$ and $\lambda=\pi/2-0.075$. Initial conditions in the teal region evolve into a fully synchronised state with $Z=1$. The boundary between the blue and teal regions is marked in green. The red cross on the boundary marks a second unstable fixed point. We have overlaid the corresponding boundary curve (black) obtained from the full two-population KS model \eqref{eq:abrams_1}--\eqref{eq:abrams_2} for comparison (cf. Figure~\ref{fig:raycast}).
Right: Initial conditions in the blue region evolve to a limit cycle (red loop) for $K=90$, $\kappa=50$ and $\lambda=\pi/2-0.1$. Initial conditions in the teal region evolve into a fully synchronised state with $Z=1$. 
Equation parameters are as in Figure~\ref{fig:zkapprox}.}
\label{fig:Zapprox}
\end{figure}

The rapid relaxation of $r_1$ to its stable equilibrium at $r_1=1$ suggests to make the additional simplifying assumption of $r_1=1$,  which allows us to express $Z$ as $Z=z_2/z_1= r_2e^{i\Delta \psi}$ (with $\Delta \psi=\psi_1-\psi_2$), which evolves according to
\begin{align}
\label{eq:dZ}
\frac{d}{dt} Z 
&=\frac{(K Z + \kappa)e^{-i\lambda} - ( K Z |Z|^2 + \kappa Z^2 )e^{i\lambda}}{2} \nonumber \\
&
-\frac{(K Z + \kappa Z^2)e^{-i\lambda} - ( K Z + \kappa |Z|^2 )e^{i\lambda}}{2} \nonumber \\
&= \kappa \cos \lambda \,  \times \nonumber  \\
& 
\left( 1 - Z^2 - \frac{1+i\tan \lambda }{2} (1 - \frac{K}{\kappa} Z) (1 - |Z|^2)\right).
\end{align}
The real scalar factor $\kappa \cos \lambda$ controls the time scale. Figure~\ref{fig:Zapprox} shows the basin of attraction for the stationary desynchronised fixed point and the desynchronised limit cycle from Figure~\ref{fig:zkapprox} obtained from the reduced mean-field dynamics \eqref{eq:dZ}. This compares well with the corresponding plots for the full two-population KS model \eqref{eq:abrams_1}--\eqref{eq:abrams_2} (cf. Figure~\ref{fig:raycast}).\\ 

\begin{figure}[htb]
    \centering
    \includegraphics[width=0.80\linewidth]{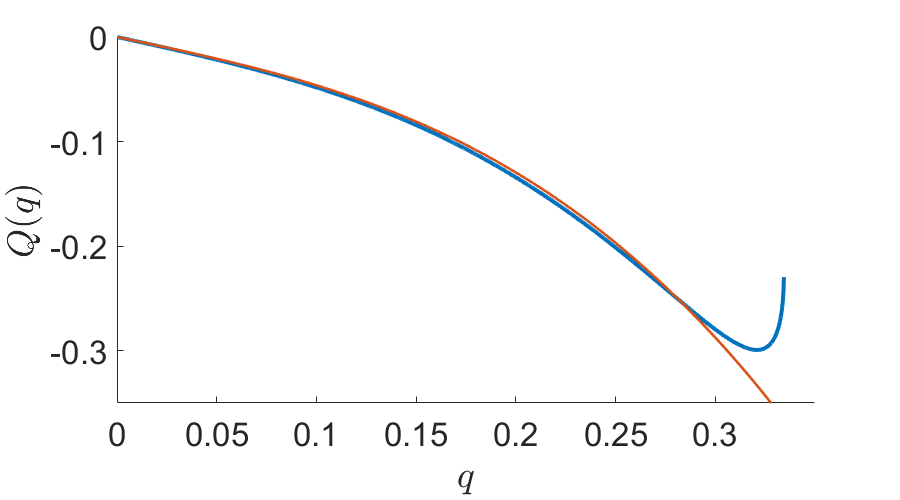}
    \caption{Approximate temporal derivative of $q=|Z-\hat Z|$, $Q(q)$, as a function of $q$ obtained from numerically solving \eqref{eq:dq} for $K=100$, $\kappa=60$, $\lambda=\frac{\pi}{2}-0.075$ (blue curve). A best-fit cubic approximation is shown as a red curve.}
    \label{fig:dqdt vs q}
\end{figure}

To approximate the critical size of fluctuations of $Z(t)$, $q^\star$,  that may lead to switching, we derive an approximation for the temporal evolution of the radial component $q:= |Z-\hat Z|$, where $\hat Z$ is the stable fixed point of \eqref{eq:dZ}. Averaging over the angular component, we obtain 
\begin{align}\label{eq:dq}
    \frac{d}{dt}q&\approx  Q(q) 
    := \frac{\left(
        \int_0^{2\pi} \frac{
            \text{Re}(F(\hat{Z}+qe^{i\vartheta})e^{-i\vartheta})
        }{
            \text{Im}(F(\hat{Z}+qe^{i\vartheta})e^{-i\vartheta})
        }d\vartheta\right)
         }{\left(\int_0^{2\pi} \frac{1}{\text{Im}(F(\hat{Z}+qe^{i\vartheta})e^{-i\vartheta})}d\vartheta\right)},
\end{align}
where  $F(Z):=\frac{d}{dt}Z$ is given by the right-hand side of \eqref{eq:dZ}. Figure~\ref{fig:dqdt vs q} shows that $Q(q)$ is well approximated by a cubic function for $q\lessapprox q^\star$ with $q^\star\approx0.339$. At $q^\star$ there is a second fixed point $Z^\star$ of \eqref{eq:dZ} with $F(Z^\star)\equiv 0$ leading to a divergence of $Q(q)$. This second fixed point $Z^\star$ is unstable and is located on the boundary of the basin of attraction of the fixed point (red cross on the green line in Figure~\ref{fig:Zapprox}). This suggests that $q^\star = |Z^\star-\langle Z\rangle|=0.339$.

\section{Stochastic model reduction}
\label{sec:stochred}
We follow here the framework of \cite{YueGottwald24} and formulate a closed stochastic equation for the synchronised population, which we assume without loss of generality to be population $P_1$. Dropping superscripts, we write the dynamics of the two-population KS model \eqref{eq:abramsz_1} for population $P_1$ as
\begin{align}
\label{eq:abramszZ}
\frac{d}{dt}\theta_i = \omega_i - \Omega + \textrm{Im}\left[ (K r_1 + \kappa Z)e^{-i(\theta_i + \lambda-\psi_1)} \right].
\end{align}

The numerical results in Section~\ref{sec:obs} suggest modelling $Z$ by a Gaussian process $\zeta_t$ as in \eqref{eq:zeta_def}. We approximate $\zeta_t$ by an Ornstein-Uhlenbeck (OU) process $\hat{\zeta}(t)$, the parameters of which are chosen to match the mean and the covariance function of the true $Z$. We hence replace the deterministic evolution equation \eqref{eq:abramszZ} by the stochastic evolution equation 
\begin{align}
\label{eq:abramszOU}
\frac{d}{dt}\theta_i &=  \omega_i - \Omega \nonumber\\ &
+ \textrm{Im}\left[ 
\left( K r_1 + \kappa \left(\langle Z \rangle +\frac{1}{\sqrt{N}}\hat{\zeta}\right) \right)
e^{-i(\theta_i + \lambda - \psi_1) } \right]
\end{align}
with a complex Gaussian mean-zero OU process 
\begin{align}
\label{eq:OU}
d\hat{\zeta}= -\gamma \hat{\zeta} dt + \sigma dW(t)
\end{align}
with complex drift and diffusion coefficients $\gamma$ and $\sigma$, respectively, and a complex-valued Wiener process $W(t)$. The mean $\langle Z\rangle$ can be either estimated numerically from a long-time simulation or by the mean-field theory in Section~\ref{sec:mft}. For the parameters we consider here with $N=12$, $K=100$, $\kappa=60$ and $\lambda=\pi/2-0.075$ and a standard Gaussian native frequency distribution with mean zero and variance $\sigma_\omega^2=1$ we obtain $\langle Z\rangle = 0.6227 + 0.1000\, i$ from a long-time simulation of the full two-population KS model \eqref{eq:abramsz_1}--\eqref{eq:abramsz_2} over $10,000$ time units which is nicely reproduced by the reduced deterministic system \eqref{eq:dzk_approx} with $\langle Z \rangle =0.6096 + 0.0934\, i$. In the following simulations, we use the thermodynamic limit obtained from \eqref{eq:dzk_approx}. 
 

\subsection{Approximation of fluctuations $\zeta$ by an Ornstein-Uhlenbeck process $\hat{\zeta}$}
\label{sec:OU}
\begin{figure}[h]
     \centering
     \includegraphics[width=0.99  \linewidth]{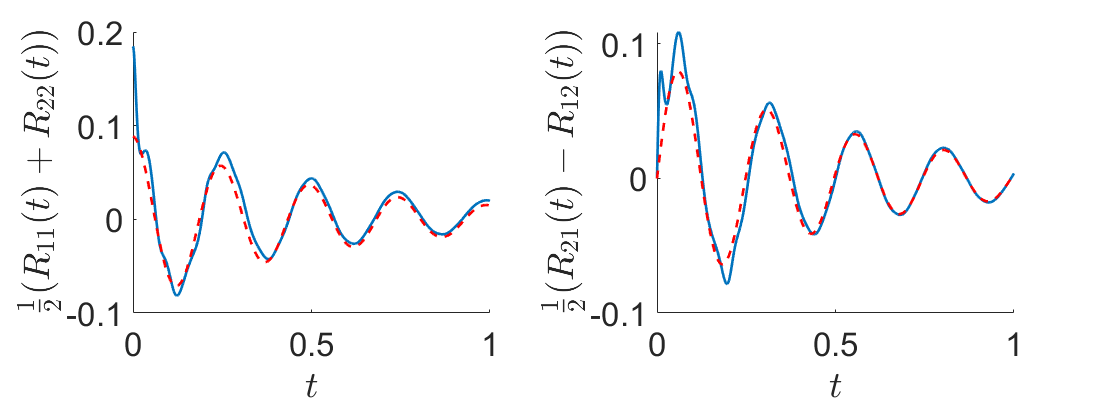}
\caption{Covariance functions $R(t)$ for the (two-dimensional) fluctuations $\zeta$ of the full two-population KS model \eqref{eq:abrams_1}--\eqref{eq:abrams_2} (blue) and $R_{\rm{OU}}(t)$ for the OU-approximation $\hat{\zeta}$ with the best-fit parameters (\ref{eq:paraOU}) (dashed red). Equation parameters are $N=12$, $K=100$, $\kappa=60$ and $\lambda=\frac{\pi}{2}-0.075$ and native frequencies $\omega$ are drawn equiprobably from a standard normal distribution $\mathcal{N}(0,1)$.}
\label{fig:covG}
\end{figure}

Any Gaussian process is entirely determined by its mean and its covariance function. It is convenient to represent the complex-valued fluctuations $\zeta$ by a $2$-dimensional real-valued Wiener process and treat $\zeta$ as a $2$-dimensional column vector with real entries. The complex drift and diffusion coefficients are then represented by $2\times 2$ real matrices $\Gamma$ and $\Sigma$, respectively. To avoid notational confusion, we denote this real-valued stochastic process of fluctuations as $\vec \zeta$. The covariance function of the full two-population KS model \eqref{eq:abrams_1}--\eqref{eq:abrams_2} is defined as
\begin{align}
\label{eq:cov}
R(t) 
&= \text{cov}(\vec\zeta(s+t),\vec\zeta(s)) \nonumber \\
&= \langle\vec \zeta(s+t)\vec\zeta^{\top}(s)\rangle_s.
\end{align}
The covariance matrix $R(t)$ can be determined numerically from a long-time simulation. The covariance function of an approximating OU process \eqref{eq:OU} is given explicitly as
\begin{align}
\label{eq:covOU}
R_{\rm{OU}}(t) &=  \exp(-\Gamma t)  \langle \vec{\hat \zeta}_0 {\vec{\hat \zeta}}_0^{\top}\rangle_{\rm{OU}},
\end{align}
where $\vec\zeta_0$ are random variables drawn from the density of the $2$-dimensional OU process \eqref{eq:OU} and the angular brackets $\langle \cdot\rangle_{\rm{OU}}$ denote the average with respect to the stationary Gaussian density of the OU process. For simplicity, we impose a symmetry assumption between the real and imaginary parts of the drift and diffusion matrices and set
\begin{align}
\Gamma &= \operatorname{Re}(\gamma) \operatorname{I} + \operatorname{Im}(\gamma) \operatorname{J},\label{eq:bigGamma} \\
\Sigma &= \sigma \operatorname{I}\label{eq:bigSigma}
\end{align}
with identity matrix $\operatorname{I}$ and skew-symmetric matrix $\operatorname{J}$ with $J_{12}=-J_{21}=1$ and $J_{11}=J_{22}=0$, rather than considering more general drift and diffusion matrices as done in \cite{YueGottwald24}. The averages in the covariance function of the OU process \eqref{eq:covOU} can be calculated explicitly, see for example \cite{YueGottwald24}. The aim is to determine the parameters of the OU process \eqref{eq:OU}, i.e. the coefficients $\gamma$ and $\sigma$, such that the covariance function of the OU process \eqref{eq:covOU} best approximates the actual covariance function \eqref{eq:cov}. The covariance function $R(t)$ of the fluctuations of the full two-population KS model exhibits small-scale variations that quickly decay, as shown in Figure~\ref{fig:covG}. This suggests only fitting the covariance functions for times larger than $t=t_{\rm{min}}$ after which the fast-scale processes have decayed. To find the approximating OU process, we therefore minimise the objective function
\begin{align}
\label{eq:cost}
E(\gamma, \sigma) &= \sum_{i,j}^2\int_{t_{\rm min}}^{t_{\rm max}} ||R_{i,j}(t) - {R_{\rm{OU}}}_{i,j}(t)||^2 dt. 
\end{align}
We use the MATLAB function \texttt{fminsearch} to perform the nonlinear least-square optimisation problem for \eqref{eq:cost}. We use $t_{\rm min}=0.5$ and $t_{\rm max}=1$ and initialise $\gamma=\gamma_0$ as the eigenvalue of the linearisation of the deterministic approximate mean-field dynamics \eqref{eq:dZ} around the stationary solution, which we numerically estimated as $\gamma_0=0.42-28.98\, i$. We found that the optimisation was less sensitive to the choice of the initial diffusion coefficient $\sigma$; we chose $\sigma$ such that $\sigma^2/(2N\operatorname{Re}(\gamma))=0.005$. The numerical minimisation of \eqref{eq:cost} yields 
\begin{align}
\label{eq:paraOU}
\gamma =1.75 -25.34i 
\quad \text{and} \quad 
\sigma=0.574.
\end{align}
Figure~\ref{fig:covG} shows how the fitted covariance function $R_{\rm{OU}}(t)$ well approximates the actual covariance function $R(t)$ for $t>t_{\rm{min}}=0.5$. This justifies our approximation of the fluctuations $\zeta$ by a Gaussian OU process.

%


\subsection{A Kramers problem for the mean switching time}
\label{sec:switching}

The reduced stochastic model \eqref{eq:abramszOU}--\eqref{eq:OU} suggests that switching is induced by sufficiently large fluctuations $\zeta$. Figure~\ref{fig:raycast} suggests that such fluctuations from the thermodynamic mean $\langle Z\rangle$ (denoted by the red dot) are those that cross the basin boundary denoted by the black line. An approximation for this basin boundary is a circle with radius $q^\star$, which we estimated numerically from the full two-population KS model as $q^\star=0.386$ and from our reduced deterministic mean-field model \eqref{eq:dZ} and its averaged approximation \eqref{eq:dq} as $q^\star=0.339$. This suggests computing the mean switching time by determining the average time it takes the OU process~\eqref{eq:OU} to generate fluctuations of magnitude $q^\star$ when started at equilibrium with $q=0$. This is the classic setting of a Kramers problem \cite{Kramers40} for a stochastic differential equation for the azimuthally averaged fluctuation 
\begin{align}
q=\frac{1}{\sqrt{N}} |\zeta|.
\end{align}
Employing It\^o's formula and the OU process \eqref{eq:OU}, we find that fluctuations $q$ evolve according to
\begin{align}
dq &= \left(\frac{\sigma^2}{2qN} -\text{Re}(\gamma)q\right)\,dt + \frac{\sigma}{\sqrt{N}} dB_t \nonumber\\
&=-\frac{d}{dq}V(q) + \frac{\sigma}{\sqrt{N}} dB_t
\label{eq:qSDE}
\end{align}
with real-valued $1$-dimensional Wiener process $B_t$ and potential 
\begin{align}
V(q) = -\frac{\sigma^2}{2N}\log(q) + \frac{1}{2}\operatorname{Re}(\gamma)\, q^2. 
\end{align}
The mean switching time $\bar \tau$ is expressed as 
\begin{align}
    \bar\tau=2\tau_e,
    \label{eq:bartau_taue}
\end{align}
where the exit time $\tau_e$ is the mean time it takes for a trajectory $q(t)$ with initial condition $q(0)=0$ to reach $q(\tau_e)=q^\star$. The exit time $\tau_e$ is the solution of Kramers problem for \eqref{eq:qSDE}
\begin{align}
\mathcal{L}\tau_e = -1
\label{eq:KP}
\end{align}
with the generator associated with the stochastic process \eqref{eq:dq}
\begin{align}
\mathcal{L} = -\frac{d}{dq}V(q)\, \partial_q +\frac{\sigma^2}{2}\partial_{xx},
\end{align}
and boundary conditions $\partial \tau_e/\partial q (q=0)=0$ and $\tau_e(q=q^\star)=0$. The solution of the Kramers problem \eqref{eq:KP} is explicitly written as
\begin{align}
\tau_e = \frac{1}{\operatorname{Re}(\gamma)}
    \int_{0}^{q^\star}\frac{1}{u} \left[ \exp(\frac{N\operatorname{Re}(\gamma)}{\sigma^2}u^2) - 1\right] du,   
    \label{eq:taue}
\end{align}
which can be evaluated numerically. Note that the exit time (and hence the mean switching time) is determined by the stochastic dynamics of the fluctuations \eqref{eq:dq}. Information about the synchronised population only enters via the value of $q^\star$.  


In the next section, we show that the reduced stochastic model \eqref{eq:abramszOU}--\eqref{eq:OU} is able to quantitatively capture the statistical behaviour of the deterministic two-population KS model \eqref{eq:abramsz_1}--\eqref{eq:abramsz_2}, including the mean switching time. 


\section{Comparison of the reduced stochastic model with the full two-population Kuramoto-Sakaguchi model}
\label{sec:results}

\begin{figure}[htb]
     \centering
      \includegraphics[width=0.4400014\linewidth]{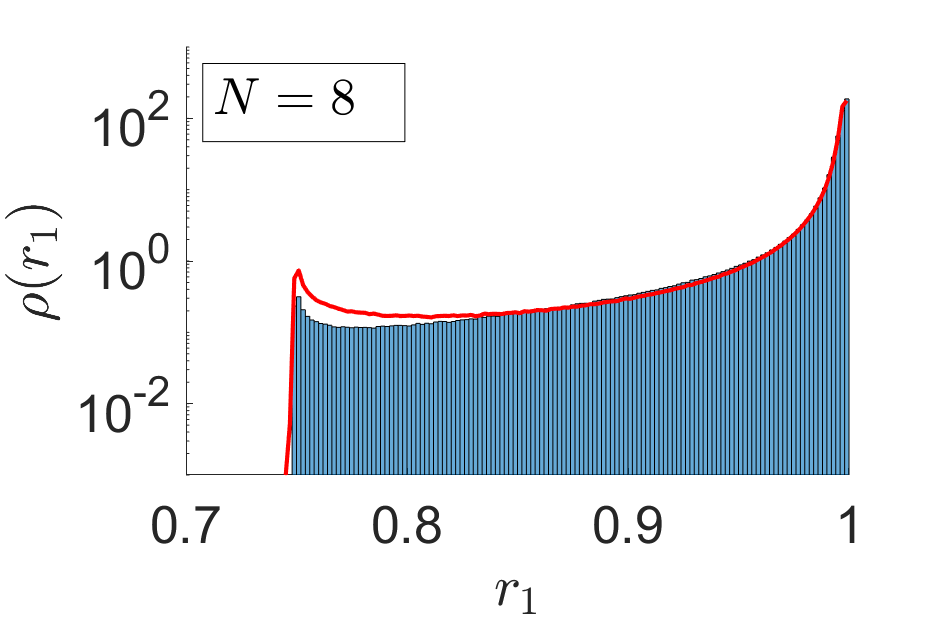}
      \includegraphics[width=0.4400014\linewidth]{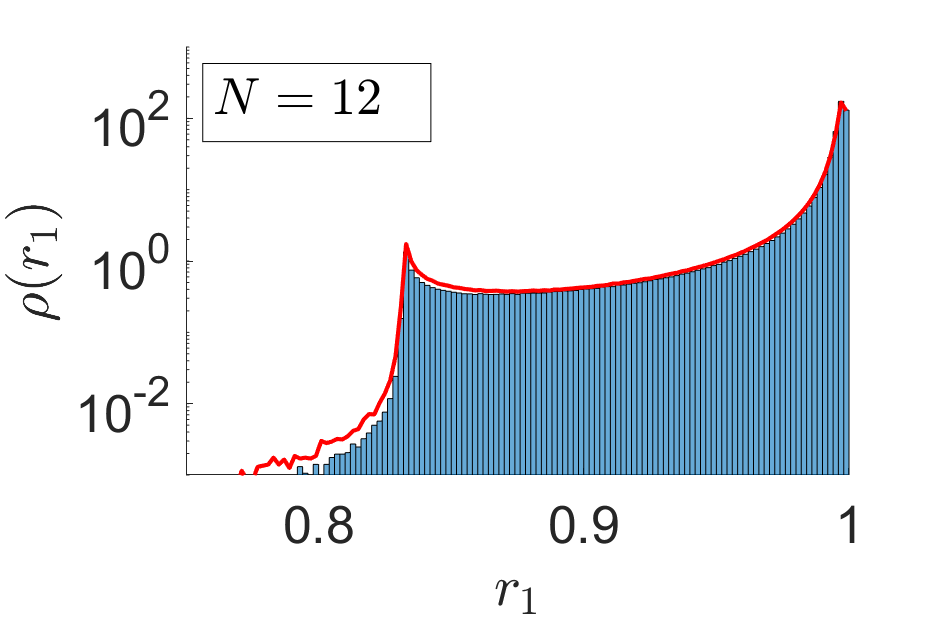}\\
      \includegraphics[width=0.4400014\linewidth]{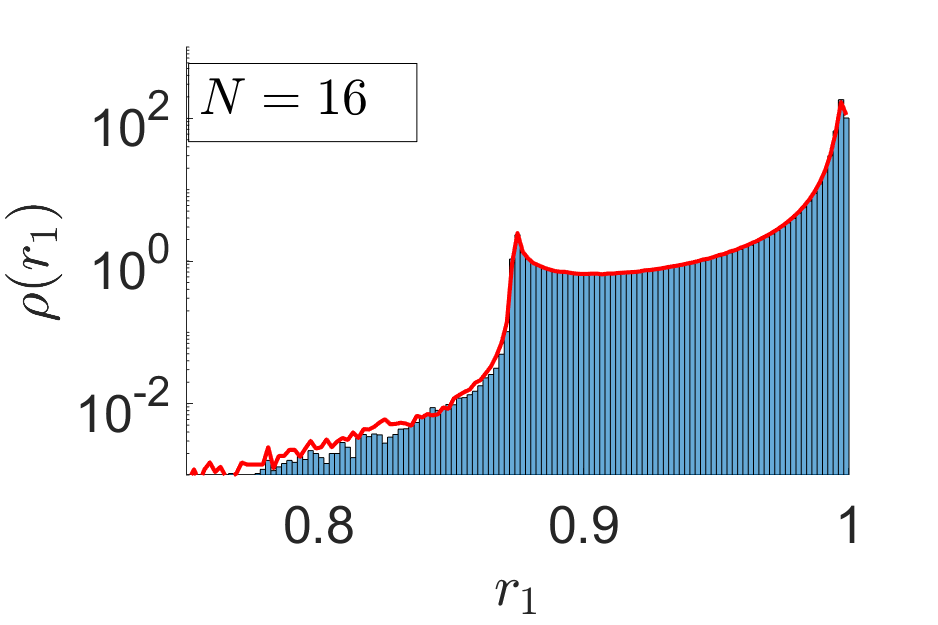}
      \includegraphics[width=0.4400014\linewidth]{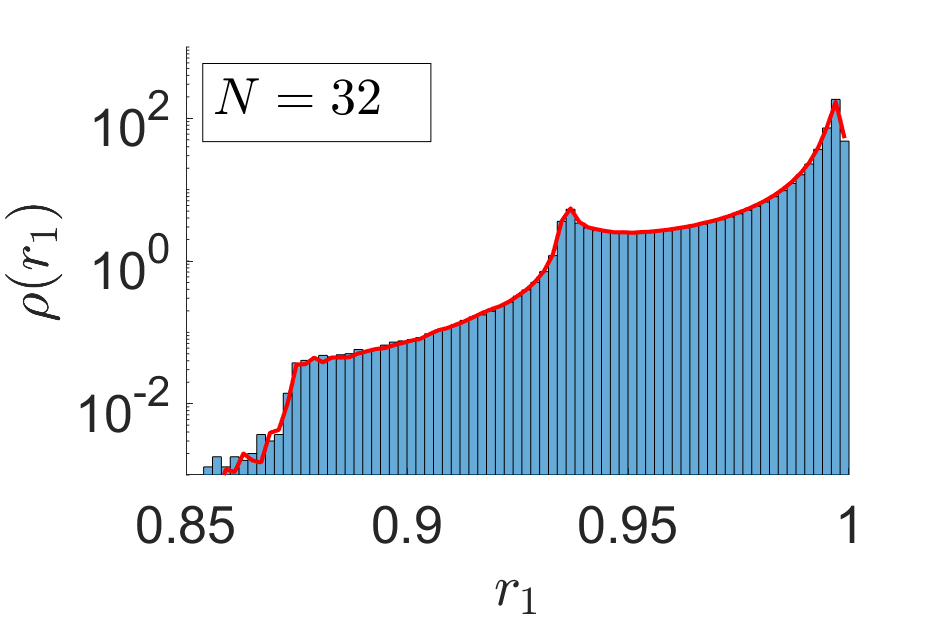}
\caption{Comparison of the empirical histograms for the order parameter $r_1$ of the synchronised population obtained from a single trajectory of the full two-population KS model \eqref{eq:abrams_1}--\eqref{eq:abrams_2} (blue histogram) and from the reduced stochastic model \eqref{eq:abramszOU}--\eqref{eq:OU} (red line) for $N=8,12,16$ and $32$. A log scale has been applied to the $y$-axis. Equation parameters are as in Figure~\ref{fig:r}. }
        \label{fig:r1histcomp}
\end{figure}
\begin{figure}[htb]
     \centering
      \includegraphics[width=0.4400014\linewidth]{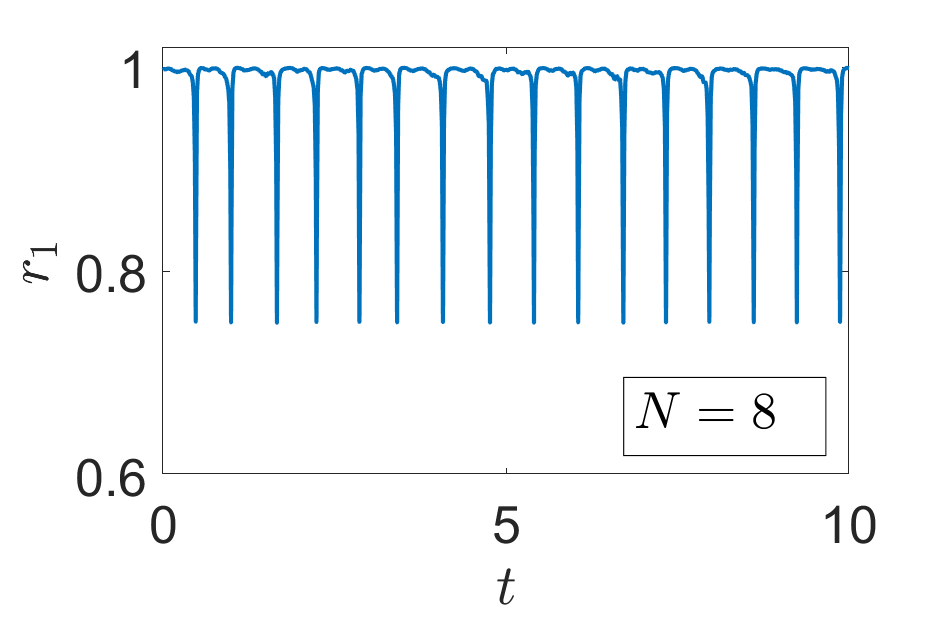}
      \includegraphics[width=0.4400014\linewidth]{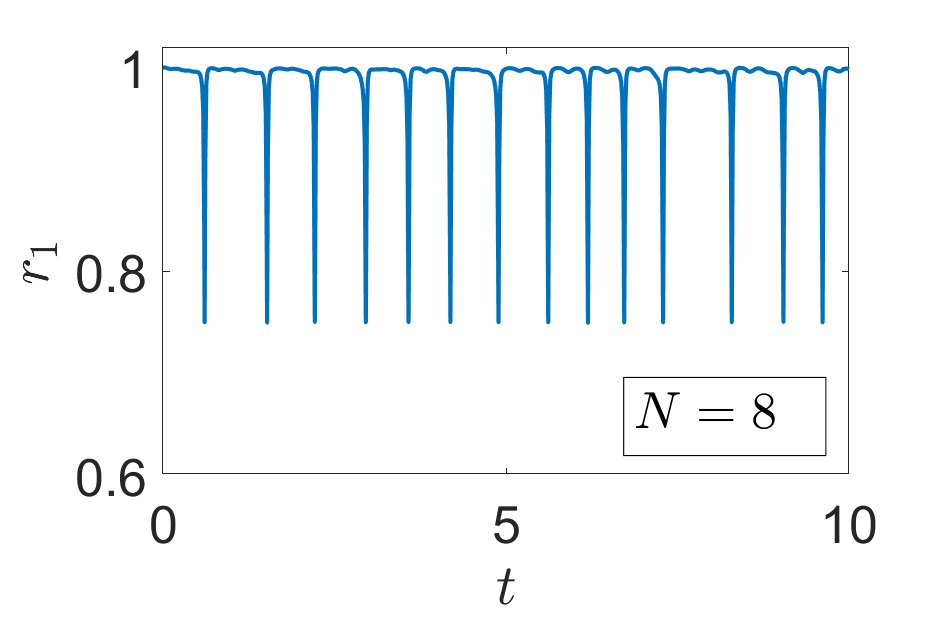}
      \includegraphics[width=0.4400014\linewidth]{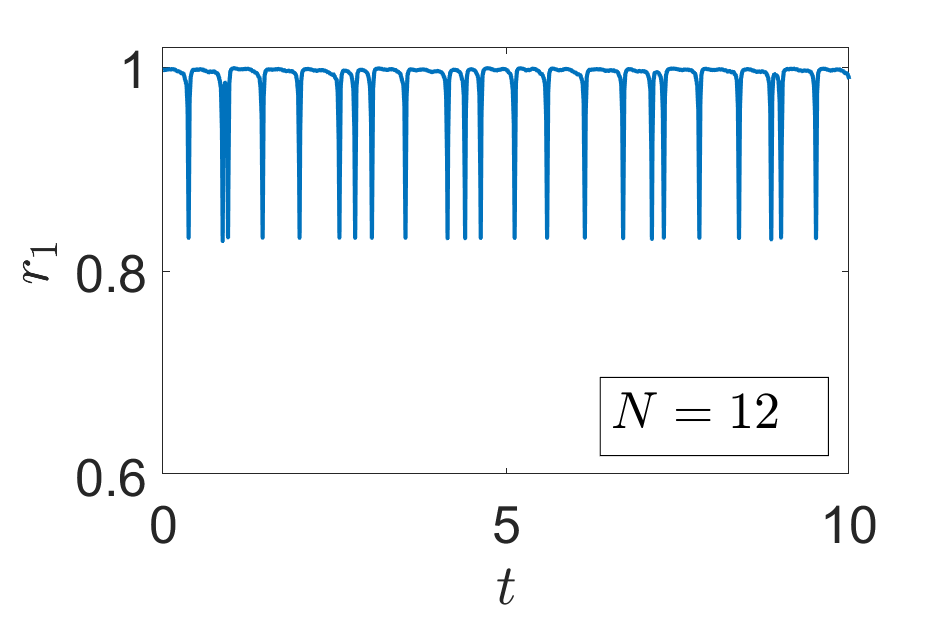}
      \includegraphics[width=0.4400014\linewidth]{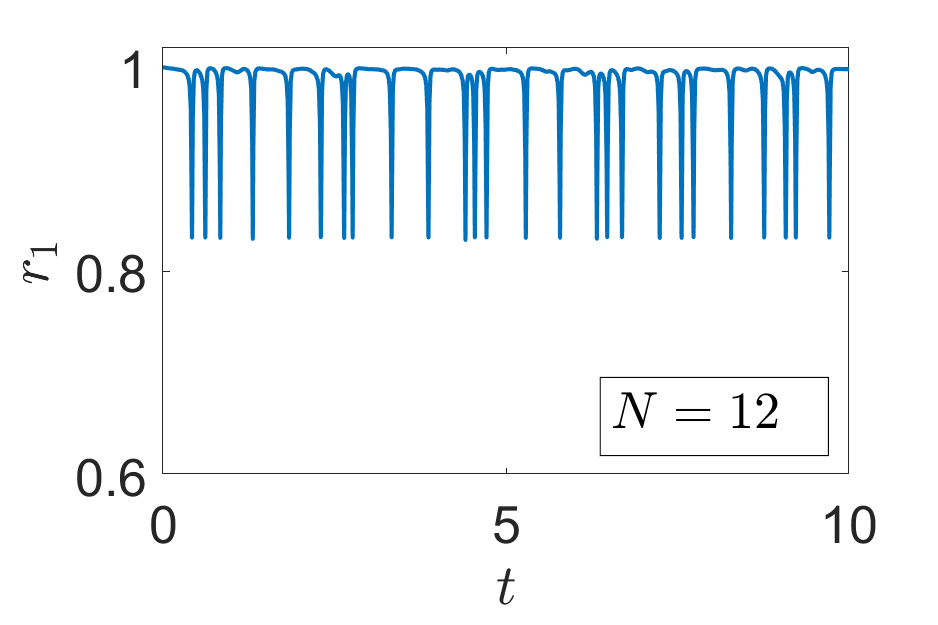}
      \includegraphics[width=0.4400014\linewidth]{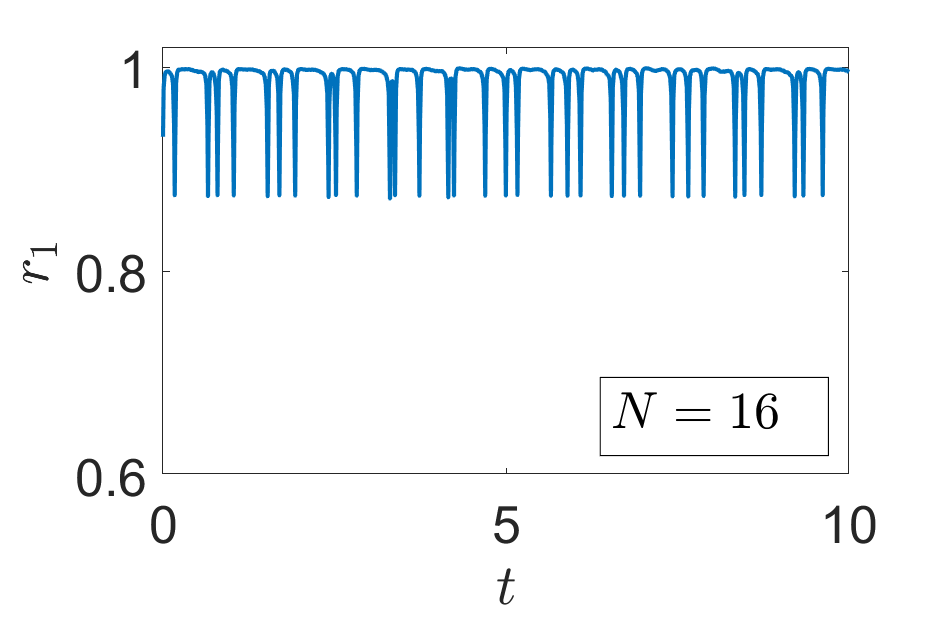}
      \includegraphics[width=0.4400014\linewidth]{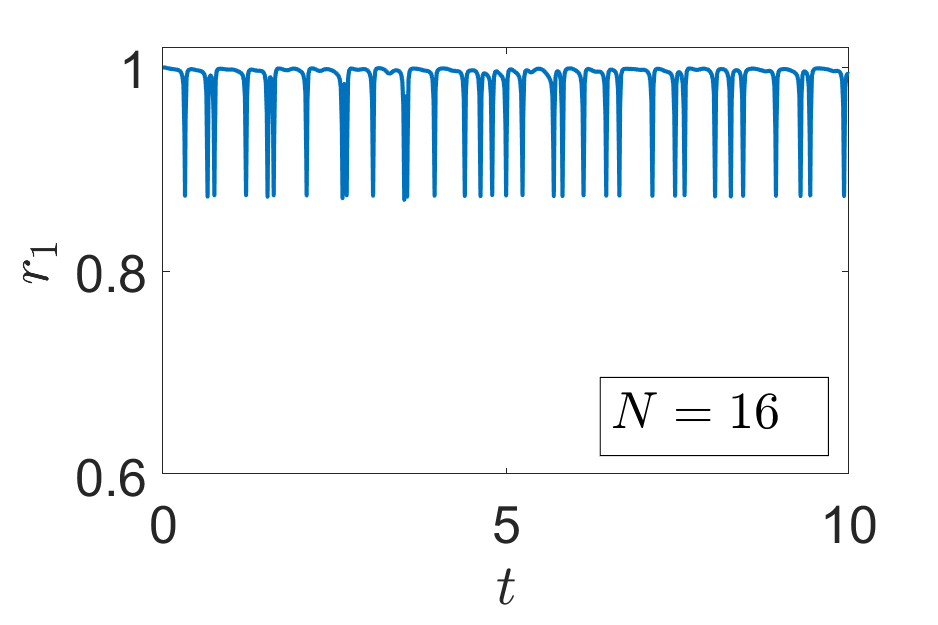}
      \includegraphics[width=0.4400014\linewidth]{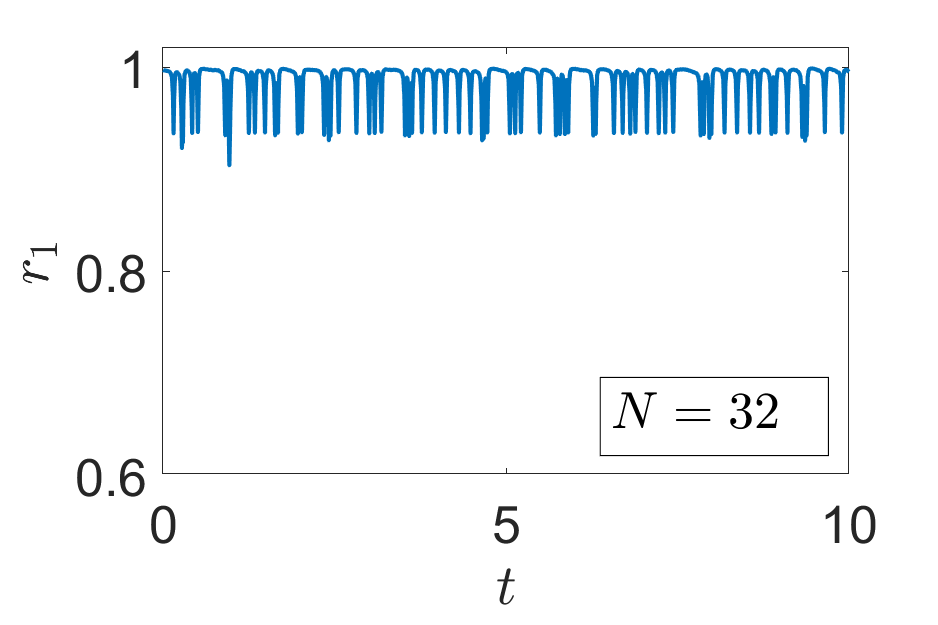}
      \includegraphics[width=0.4400014\linewidth]{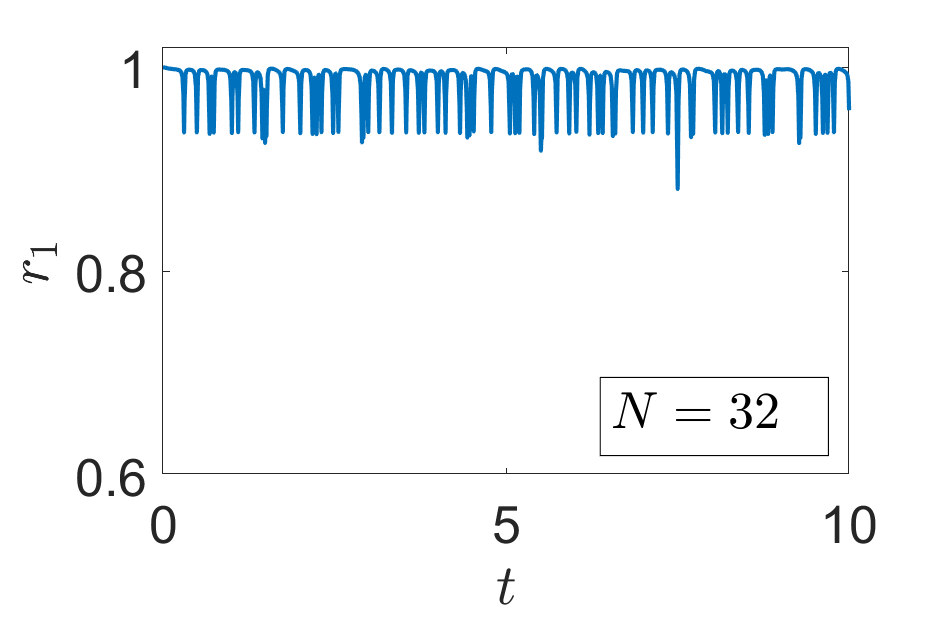}
\caption{Order parameter $r_1(t)$ as a function of time of the full two-population KS model \eqref{eq:abrams_1}--\eqref{eq:abrams_2} (left column) and of the reduced stochastic model \eqref{eq:abramszOU}--\eqref{eq:OU} (right column) for $N=8,12,16$ and $32$. Equation parameters are as in Figure~\ref{fig:r}.}
        \label{fig:r1dips}
\end{figure}
\begin{figure}[htb]
     \centering
     \includegraphics[width=0.4400014\linewidth]{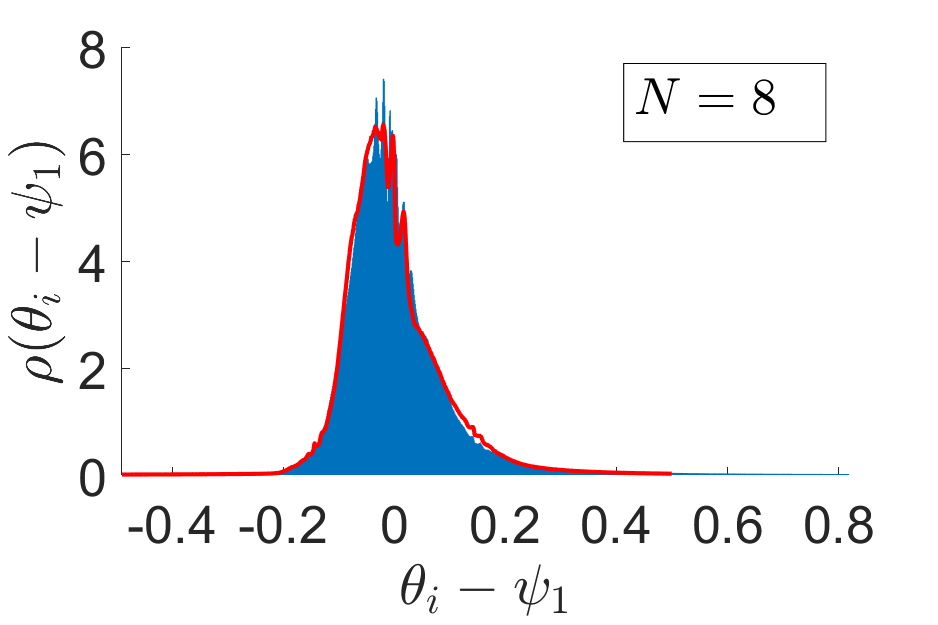}
     \includegraphics[width=0.4400014\linewidth]{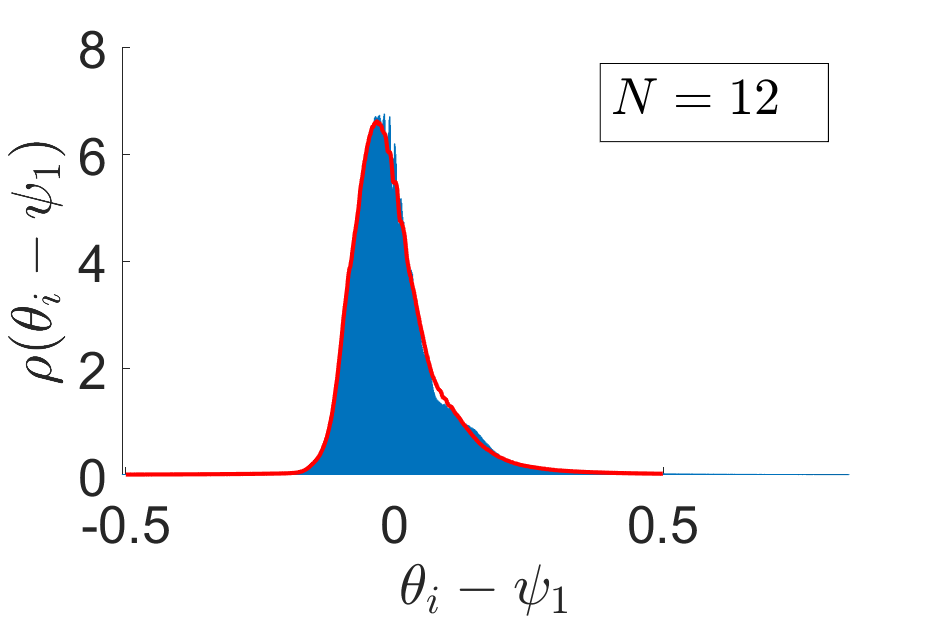}
     \includegraphics[width=0.4400014\linewidth]{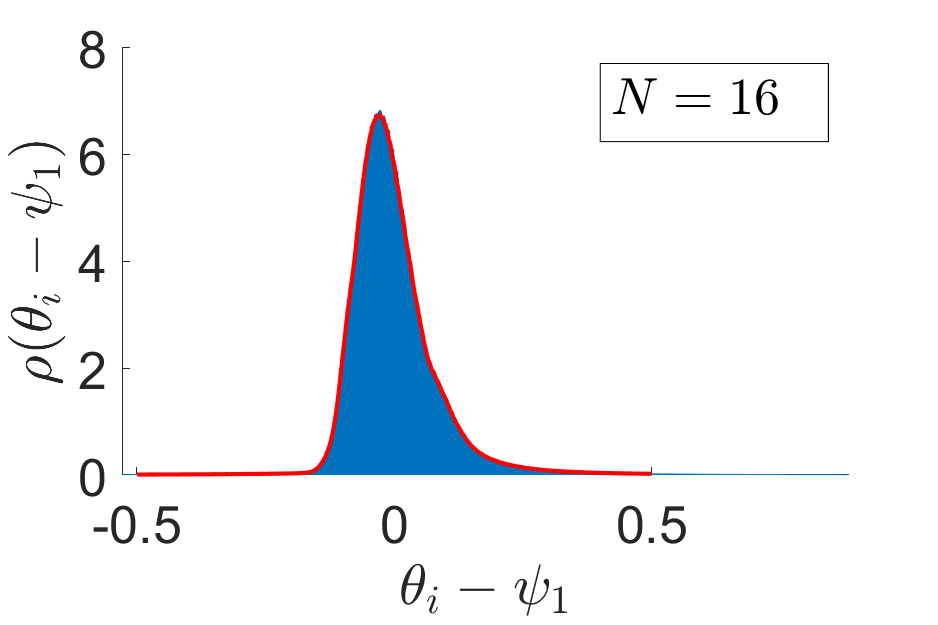}
     \includegraphics[width=0.4400014\linewidth]{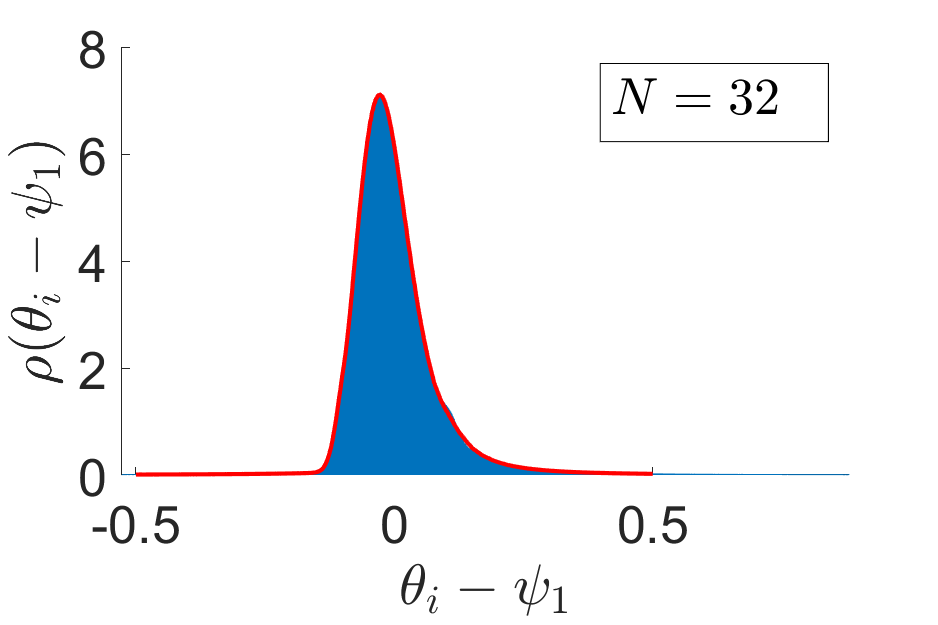}
\caption{Comparison of the empirical histograms for the fluctuations of the synchronised phases around their mean, $\theta_i-\psi_1$, obtained from a single trajectory of the full two-population KS model \eqref{eq:abrams_1}--\eqref{eq:abrams_2} (blue histogram) and of the reduced stochastic model \eqref{eq:abramszOU}--\eqref{eq:OU} (red line) for $N=8,12,16$ and $32$. Equation parameters are as in Figure~\ref{fig:r}.}
\label{fig:thetahist}
\end{figure}
\begin{figure}[htb]
    \centering
     \includegraphics[width=0.4400014\linewidth]{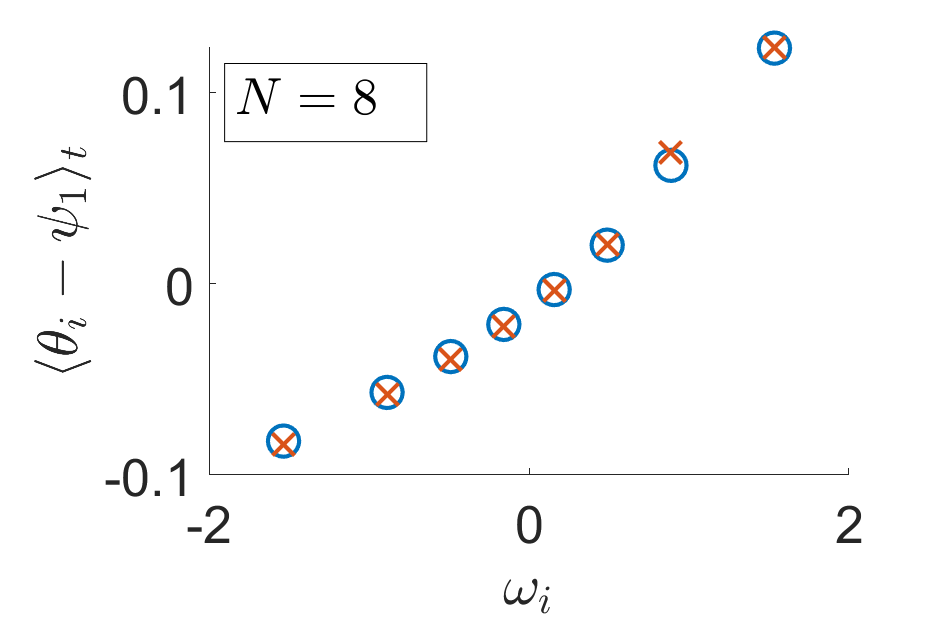}
     \includegraphics[width=0.4400014\linewidth]{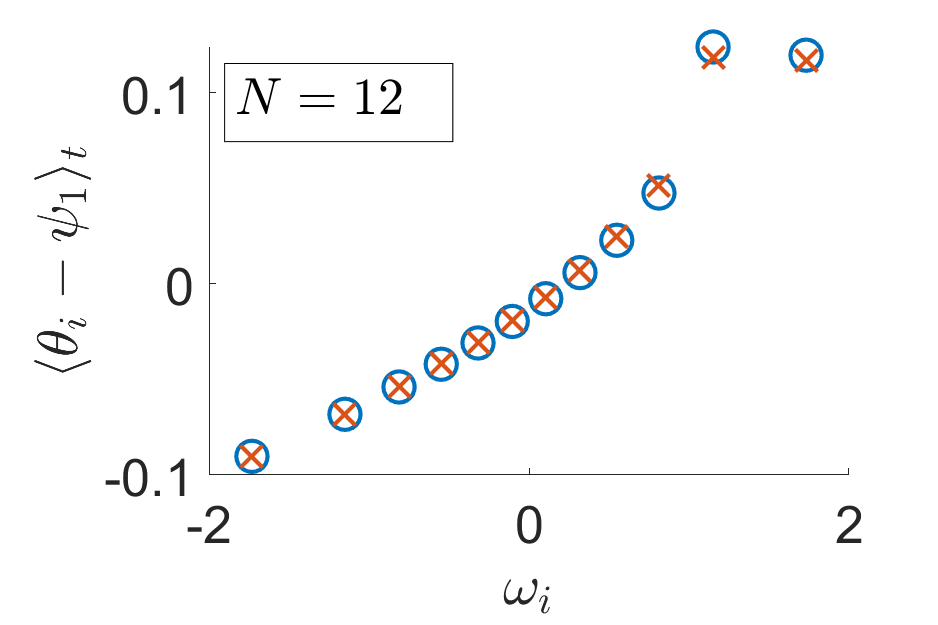}
     \includegraphics[width=0.4400014\linewidth]{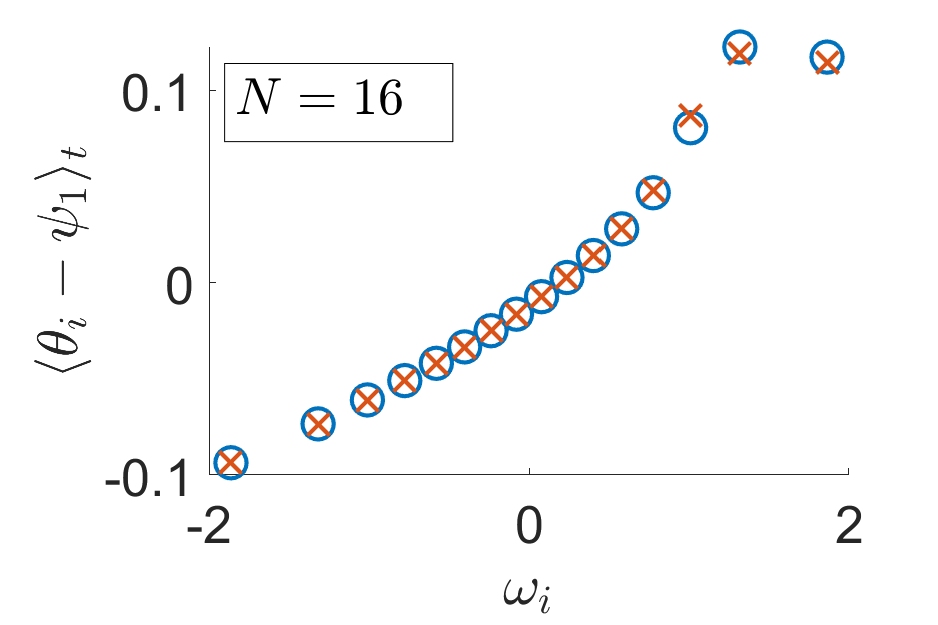}
     \includegraphics[width=0.4400014\linewidth]{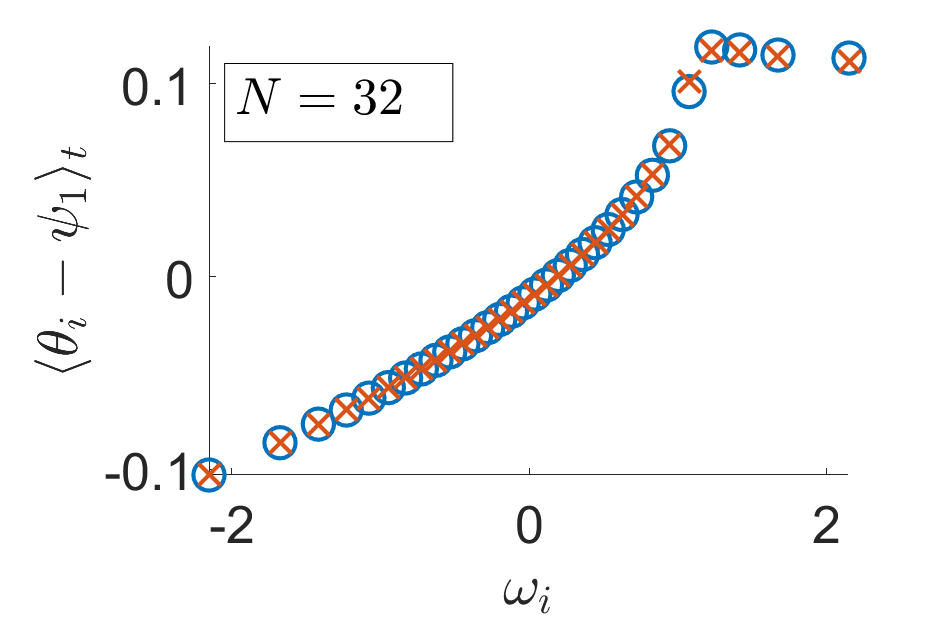}
\caption{Mean of the fluctuations of the synchronised phases around their mean, $\theta_i-\psi_1$, as a function of their native frequencies $\omega_i$ for the full two-population KS model \eqref{eq:abrams_1}--\eqref{eq:abrams_2} (blue circles) and for the reduced stochastic model \eqref{eq:abramszOU}--\eqref{eq:OU} (red crosses) for $N=8,12,16$ and $32$. Equation parameters are as in Figure~\ref{fig:r}.}
\label{fig:theta_mean}
\end{figure}
\begin{figure}[htb]
     \centering
     \includegraphics[width=0.4400014\linewidth]{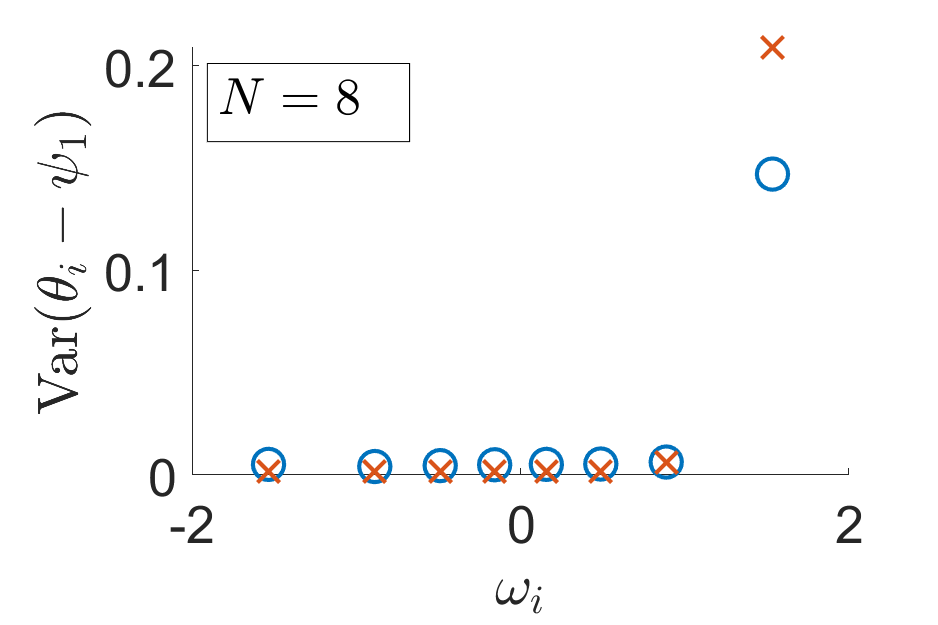}
     \includegraphics[width=0.4400014\linewidth]{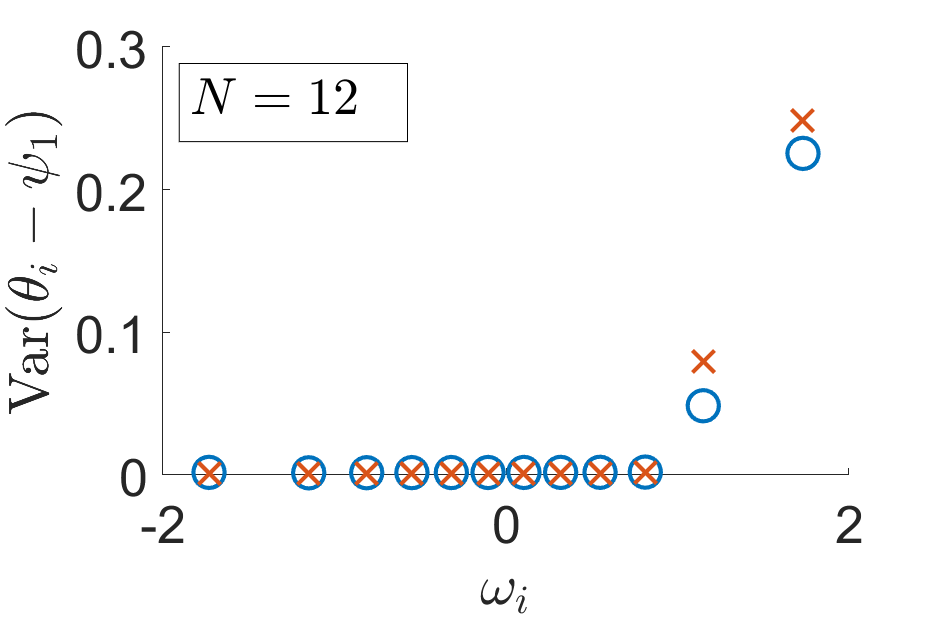}
     \includegraphics[width=0.4400014\linewidth]{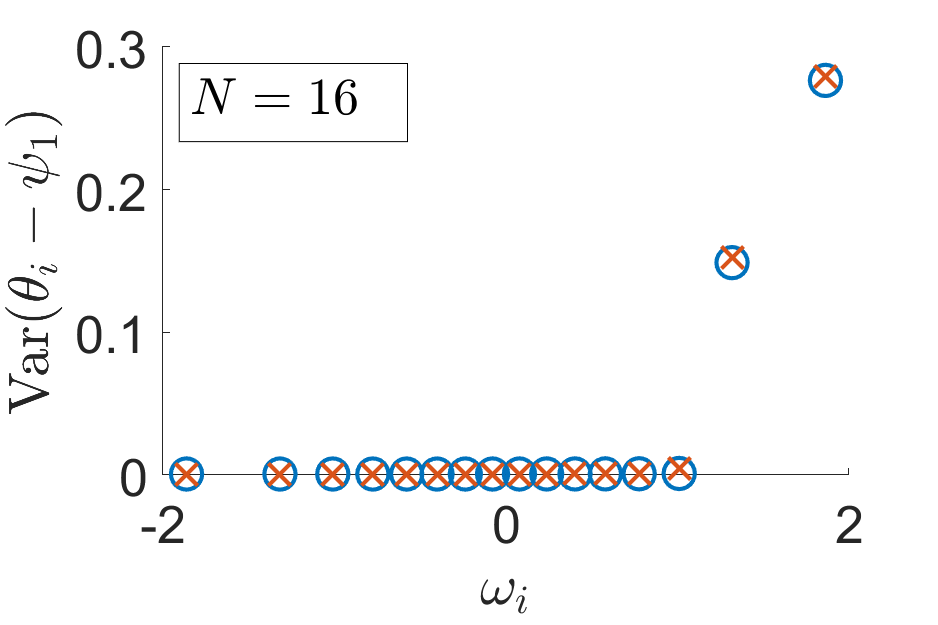}
     \includegraphics[width=0.4400014\linewidth]{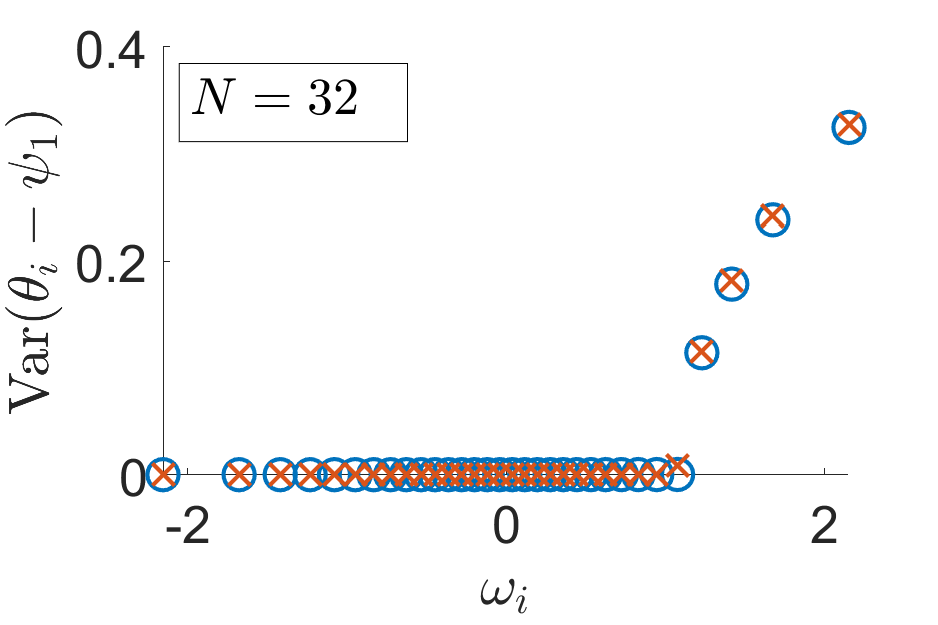}
\caption{Variance of the fluctuations of the synchronised phases around their mean, $\theta_i-\psi_1$, as a function of their native frequencies $\omega_i$ for the full two-population KS model \eqref{eq:abrams_1}--\eqref{eq:abrams_2} (blue circles) and for the reduced stochastic model \eqref{eq:abramszOU}--\eqref{eq:OU} (red crosses) for $N=8,12,16$ and $32$. Equation parameters are as in Figure~\ref{fig:r}.}
\label{fig:theta_var}
\end{figure}
\begin{figure}[htb]
     \centering
      \includegraphics[width=0.4400014\linewidth]{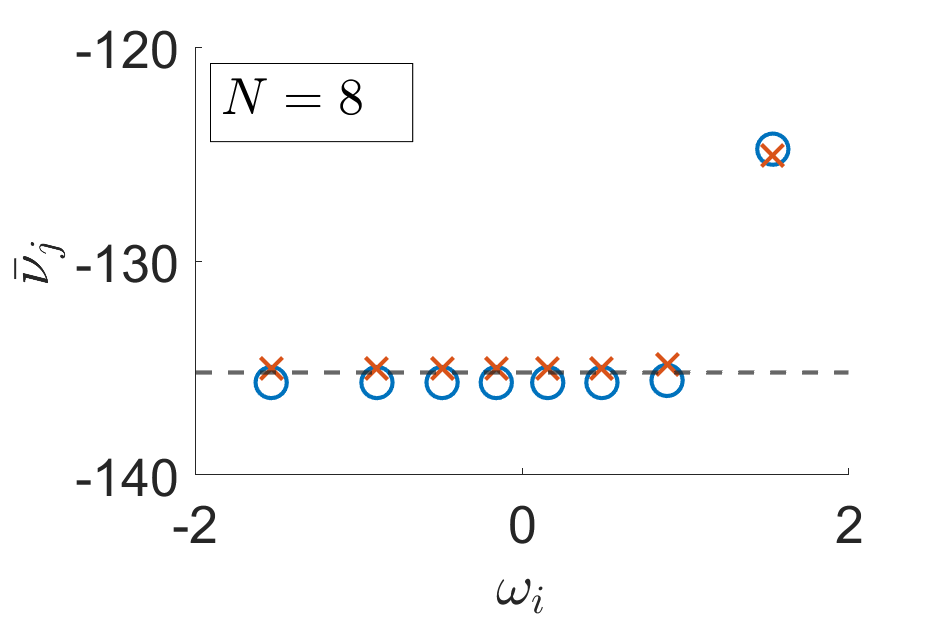}
      \includegraphics[width=0.4400014\linewidth]{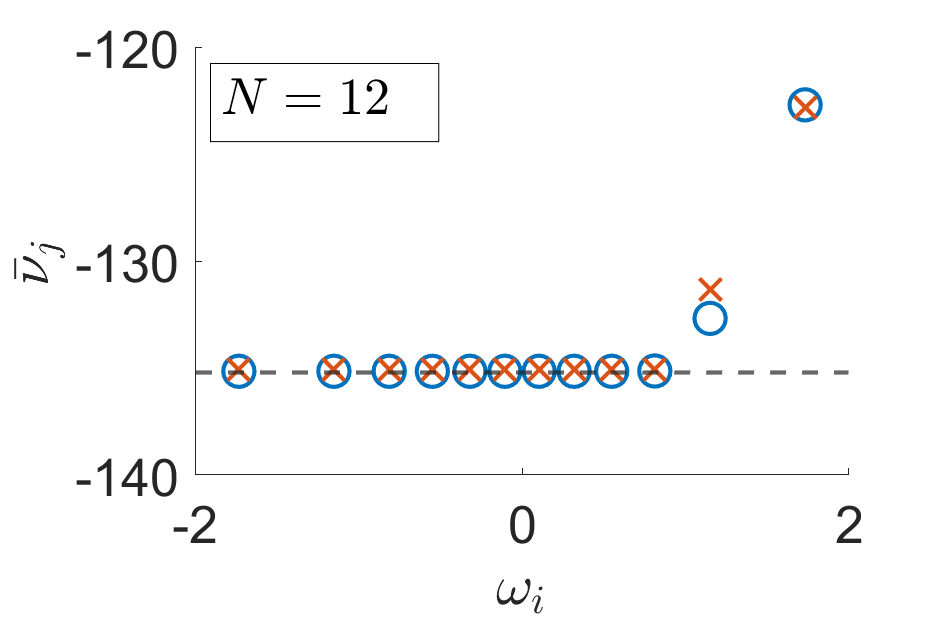}
      \includegraphics[width=0.4400014\linewidth]{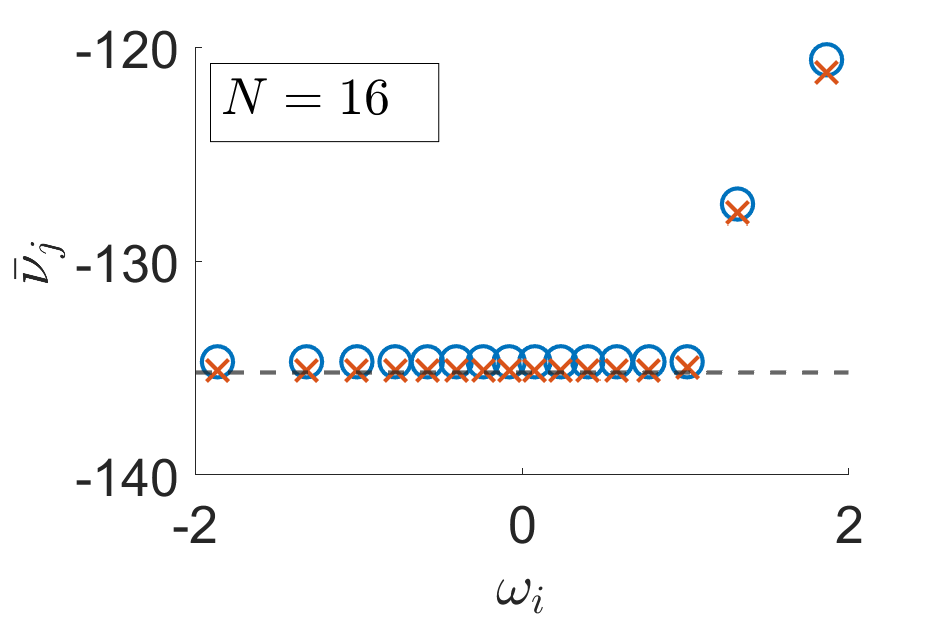}
      \includegraphics[width=0.4400014\linewidth]{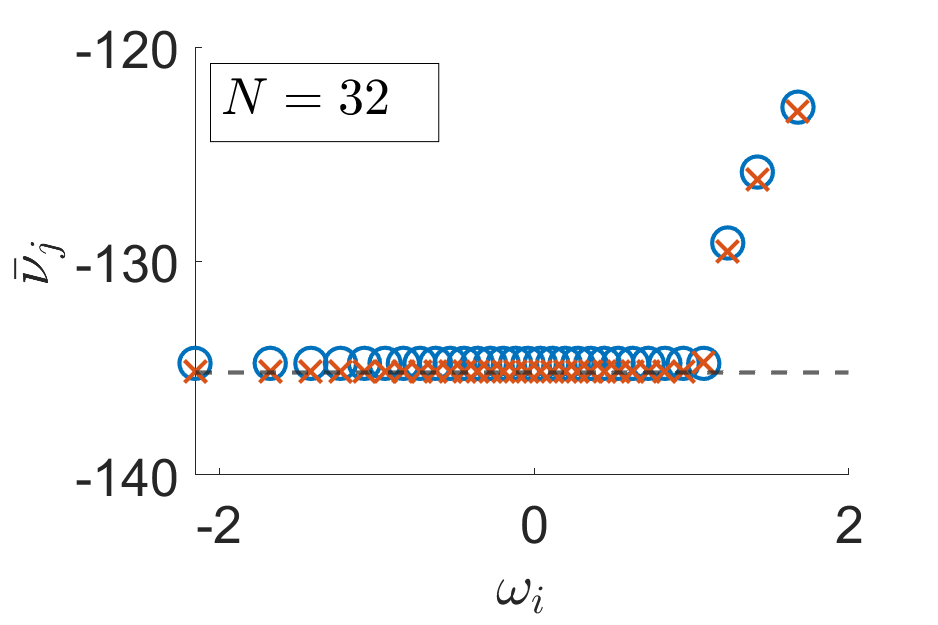}
\caption{Mean rotation frequency ${\bar \nu}_i $ of the synchronised oscillators as a function of their native frequencies $\omega_i$ for the full two-population KS model \eqref{eq:abrams_1}--\eqref{eq:abrams_2} (blue circles) and for the reduced stochastic model \eqref{eq:abramszZ}--\eqref{eq:abramszOU} (red crosses) for $N=8,12,16$ and $32$. The dashed lines denote the mean frequency $\Omega$ as calculated from the self-consistency relations \eqref{eq:selconst_r}--\eqref{eq:selconst_Z} of the mean-field theory. Equation parameters are as in Figure~\ref{fig:r}.}
        \label{fig:theta_dot}
\end{figure}

We now assess the ability of the reduced stochastic model of the synchronised population \eqref{eq:abramszOU}--\eqref{eq:OU} to describe the dynamics of the full two-population KS model \eqref{eq:abrams_1}--\eqref{eq:abrams_2}. We begin by first assessing how the model is able to describe the invariant density of the order parameter $r_1$ and of the relative phases $\theta_i-\psi_1$ of individual synchronised oscillators. 
The reduced stochastic model \eqref{eq:abramszOU}--\eqref{eq:OU} is integrated with a time step of $\Delta t=0.001$ using a 4th order Runge-Kutta scheme with pregenerated noise $\hat{\zeta}$.  We remark that this scheme is not fourth-order in time due to the presence of the non-smooth noise.  To generate the OU process $\hat{\zeta}$ we employ a numerical scheme that exactly reproduces the transition probabilities independent of the employed time step (see Appendix~\ref{sec:app2} for details). 

We first show in Figure~\ref{fig:r1histcomp} a comparison of the empirical histogram of the order parameter $r_1$ for the synchronised population for $N=8, 12, 16$ and $N=32$. Remarkably, even $N=8$ oscillators are sufficient to allow for the reduced stochastic model to accurately approximate the density of the order parameter. The empirical histograms shown in Figure~\ref{fig:r1histcomp} have pronounced peaks. These peaks are caused by one or more oscillators of the synchronised population to phase-slip, which causes isolated drops of the order parameter as illustrated in Figure~\ref{fig:r1dips}, which shows the order parameter $r_1$ as a function of time, during periods over which no switching occurred. The frequency of the dips and their size are well recovered by the reduced stochastic model for $N\ge 12$. The amplitude of the peaks can be explicitly estimated via a simple argument: Under the assumption that a single oscillator, which we label here without loss of generality as $\theta_N$, desynchronises with $\theta_N=\pi$ from the synchronised population, which we approximate by $\theta_1=\theta_2=\dots \theta_{N-1}=0$, the order parameter can be approximated as $r_1\approx 1-\frac{2}{N}$, which fits the observed peaks in the histograms in Figure~\ref{fig:r1histcomp} and in the time series in Figure~\ref{fig:r1dips} very well, in particular for $N>12$.  

The reduced stochastic model \eqref{eq:abramszOU}--\eqref{eq:OU} further reproduces the statistics of fluctuations of the synchronised phases around their mean,  $\theta_i-\psi_1$. Figure~\ref{fig:thetahist} shows the empirical histograms of $\theta_i-\psi_1$ for $N=8,12,16$ and $N=32$, over all oscillators of the synchronised population. It is seen again that even for a small system size with $N=8$ the reduced stochastic model is able to describe the statistical properties of the full two-population KS model \eqref{eq:abrams_1}--\eqref{eq:abrams_2}. The accuracy of the reduced stochastic model increases, as expected, with increasing system size $N$. The histograms of the full two-population KS model exhibit isolated sharp peaks for $N<16$ at the mean of the phases of individual oscillators. These undulations get smeared out for a sufficiently large number of oscillators. 

To see in how far the reduced stochastic model captures the statistics of individual oscillators we show in Figures~\ref{fig:theta_mean} and \ref{fig:theta_var} the mean and variance for each synchronised oscillator, and in Figure~\ref{fig:theta_dot} the average effective frequency ${\bar \nu}_i\coloneq  \langle \theta_i(t)\rangle$ for each synchronised oscillator. Remarkably, the mean of the fluctuations is well approximated by the reduced stochastic model even for $N=8$. These figures show that the synchronised population involves rogue oscillators at the edge of the population, which oscillate with markedly different mean rotation frequencies and mean phases. The reduced stochastic model is able to capture the mean phase as well as the variance of all oscillators very well, even for the rogue oscillators. 

\begin{figure}[htb]
     \centering
      \includegraphics[width=0.80\linewidth]{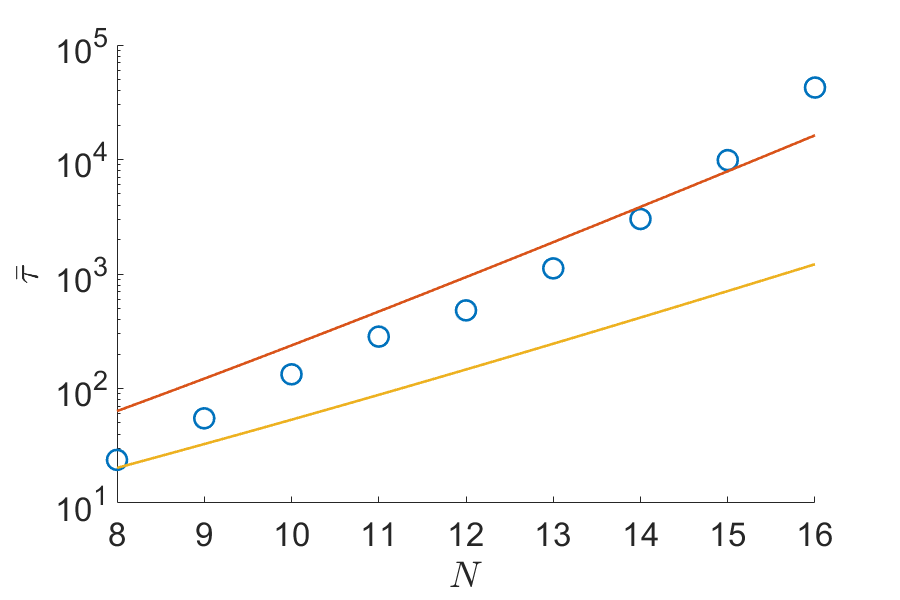}
\caption{Comparison of the mean switching time $\bar{\tau}$ for different system sizes $N$, obtained from long-time simulations of the full two-population KS model \eqref{eq:abrams_1}--\eqref{eq:abrams_2} (blue circles) and from the analytical expressions \eqref{eq:bartau_taue} and \eqref{eq:taue} based on solving Kramers problem using our reduced stochastic model. We show results of the Kramers problem for $q^\star=0.386$ (red line), which was estimated from the full two-population KS model \eqref{eq:abrams_1}--\eqref{eq:abrams_2} (cf. Figure~\ref{fig:raycast}), and for $q^\star=0.339$ (yellow line), which was estimated from the reduced deterministic mean-field model \eqref{eq:dZ} (cf. Figures~\ref{fig:Zapprox} and \ref{fig:dqdt vs q}). Equation parameters are as in Figure~\ref{fig:r}.}
        \label{fig:tau_bar}
\end{figure}

 The empirical histograms of the order parameter and the statistics of the fluctuations of individual synchronised oscillators are relatively robust to the numerically estimated covariance function. In contrast, rare switching events are induced by large excursions of the fluctuations and hence depend crucially on correctly estimating the tail behaviour of their distribution.   Figure~\ref{fig:tau_bar} shows the mean switching time $\bar{\tau}$ as a function of the system size $N$. We compare the mean switching time of the full two-population KS model \eqref{eq:abrams_1}--\eqref{eq:abrams_2} with the expected switching times $\bar{\tau}=2\tau_e$ of the reduced stochastic model \eqref{eq:abramszOU}--\eqref{eq:OU} which we estimate from the solution of the Kramers problem \eqref{eq:taue} for both $q^\star=0.386$ as estimated from the full two-population KS model \eqref{eq:abrams_1}--\eqref{eq:abrams_2} (cf. Figure~\ref{fig:raycast}) and for $q^\star=0.339$ as estimated from the reduced deterministic mean-field model (cf. Figures~\ref{fig:Zapprox} and \ref{fig:dqdt vs q}). For each value of $N$, the mean switching time is estimated over $1,000$ switching events in the full model. It is remarkable how our simple approximation captures the mean switching time reasonably well. Setting a uniform critical fluctuation size $q^\star$ by azimuthally averaging is a significant simplification (which clearly fails at the top right of the blue teardrop depicted in Figure~\ref{fig:raycast}). We have checked that for parameter values that exhibit a less circular basin boundary, the approximation of the mean switching times becomes worse.\\


%

\section{Summary and Outlook}
\label{sec:discussion}

We have shown that a deterministic two-population Kuramoto-Sakaguchi model exhibits switching of chimera states whereby the two populations switch between a synchronised and desynchronised state in a seemingly random way. The switching events were shown to be uncorrelated events and to follow a Poisson distribution. The average time between switches was shown to increase exponentially with the system size $N$ and hence switching constitutes a finite-size effect. As such, the effective stochastic behaviour cannot be described by the standard techniques valid in the thermodynamic limit. We evoked the central limit theorem to describe the fluctuations around the mean behaviour and employed the stochastic modelling framework from \cite{YueGottwald24}. The effective reduced stochastic model captures the statistics of both the synchronised cluster and the switching of this cluster into a desynchronised state. Using mean-field theory, we obtained an approximation of the critical size of finite-size fluctuations required to induce chimera switching. Our model produces relatively accurate predictions for the rate at which switching events become less frequent with increasing system size.

The chimera switching in the two-population KS model is a non-transient persistent process with a stationary probability density. in contrast, in a single population KS model with identical oscillators, chimera states were shown to be neutrally stable in the thermodynamic limit and were in fact only chaotic transients in finite-size systems, collapsing to a stable synchronised state \cite{WolfrumOmelchenko11, WolfrumEtAl11}. In our two-population system, the state where both populations are synchronised is not stable. This, we conjecture, leads to the emergence of persistent chimera switching. To fully analyse this mechanism is planned for further research.

 For illustrative purposes, we have presented results for a single set of parameters and with equiprobably drawn native frequencies. Our results are robust with respect to changes in the parameters (see Appendix~\ref{sec:appNR}). For random realisations of native frequency draws, the probability of observing switching in a finite time window sensitively depends on the details of the realisation. In particular, we find that switching is suppressed if at least one of the populations has a smaller than expected variance of their native frequencies; this can be understood by realising that tighter native frequency distributions promote stability of the synchronised state. Conversely, switching is promoted by an overall larger spread of the native frequencies. This, we argue, in turn increases the tail in the distribution of the fluctuations $\zeta$, responsible for inducing switching (see Appendix~\ref{sec:appNR_R}, for details).

Our approach here is entirely heuristic and relies on numerically fitting an Ornstein-Uhlenbeck process to the observed statistics of the fluctuations. It would be interesting to see if the approximation of the OU process can at least be formally justified. It was argued that the approximation of $Z$ is related to a trigonometric approximation of Gaussian processes, where instead of trigonometric functions, the near-periodic solutions of the desynchronised oscillators are used \cite{YueGottwald24}. This suggests a theoretical avenue to put the heuristic method employed here on a more rigorous footing, following the recent work of \cite{OmelchenkoGottwald25}.

The $N+2$-dimensional reduced stochastic model is still relatively high-dimensional, and it would be interesting to see if it can be further reduced. Since the system-size $N$ is not sufficiently large to allow for mean-field theory to capture the finite-size induced switching we need to resort to methods which are valid for finite-size systems such as the method of collective coordinates \cite{Gottwald15, HancockGottwald18, SmithGottwald19, SmithGottwald20, YueEtAl20} which has already been shown to work for stochastically driven Kuramoto models \cite{Gottwald17}. This would allow for a straightforward calculation of the Kramers problem to calculate the transition probabilities describing switching chimeras.

In Section~\ref{sec:detred} we saw that for certain parameters, i.e., $K=85$, $\kappa=55$ and $\lambda=\pi/2-0.09$, the chimera state forms around a limit cycle. In this case, the interaction term $Z$ will not be a centred Gaussian, but its density will be supported on a circular stripe. It is planned for further research to extend our stochastic modelling framework to this case of a breathing chimera. 


\begin{acknowledgments}
GAG would like to acknowledge Wenqi Yue, from whom we learned about this interesting phenomenon.
\end{acknowledgments}

%


\appendix
\section{Justification for the approximations \eqref{eq:zkapprox_cond}}
\label{sec:app1}

We present here a more formal justification for the approximations  \eqref{eq:zkapprox_cond} used to close the mean-field dynamics \eqref{eq:dzk}. We begin by expressing 
\begin{align}
    z_{\alpha}^{(2,0)}-z_{\alpha}^2&=\var(\text{Re}(e^{i\theta^{(\alpha)}})) - \var(\text{Im}(e^{i\theta^{(\alpha)}})) \nonumber \\
    &
    + 2i\operatorname{cov}(\text{Re}(e^{i\theta^{(\alpha)}}),\text{Im}(e^{i\theta^{(\alpha)}})),
    \label{eq:error term epsilon}
\end{align}
where $\var$ and $\operatorname{cov}$ denote the empirical variance and covariance across $P_\alpha$, respectively, which is readily bounded by 
\begin{align}
    |z_{\alpha}^{(2,0)}-z_{\alpha}^2|&\leq\var(\text{Re}(e^{i\theta_j^{(\alpha)}})) + \var(\text{Im}(e^{i\theta_j^{(\alpha)}}))\nonumber\\
    &
    =1-r_\alpha^2.
    \label{eq:zkbound1}
\end{align}
For the synchronised population with $\alpha=1$, we have $r_1\approx 1$, which immediately yields the first approximation in \eqref{eq:zkapprox_cond}. For $\alpha=2$, the bound \eqref{eq:zkbound1} is less sharp.

To approximate $z^{(1,1)}_{\alpha}$, we expand 
\begin{align}
z_{\alpha}^{(1,1)}&=\frac{1}{N}\sum_{j\in P_\alpha} \omega^{(\alpha)}_j e^{i\theta^{(\alpha)}_j} \nonumber \\
&=\langle\omega^{(\alpha)}_j\rangle_{j\in P_{\alpha}}\left(\frac{1}{N}\sum_{j\in P_\alpha} e^{i\theta^{(\alpha)}_j}\right) \nonumber\\
&
+ \operatorname{cov}(\omega^{(\alpha)}_j,\text{Re}(e^{i\theta^{(\alpha)}_j})) \nonumber\\
&
+ i\operatorname{cov}(\omega^{(\alpha)}_j,\text{Im}(e^{i\theta^{(\alpha)}_j})).
\label{eq:z11 approx explain}
\end{align}
The native frequencies $\omega^{(\alpha)}_j$ are drawn from a Gaussian distribution $\mathcal{N}(0,\sigma_\omega^2)$ and hence \eqref{eq:z11 approx explain} can be bounded by 
\begin{align}
|z_{\alpha}^{(1,1)}-0|&\lesssim\sigma_\omega\sqrt{\var(\text{Re}(e^{i\theta_j^{(\alpha)}})) + \var(\text{Im}(e^{i\theta_j^{(\alpha)}}))}\nonumber\\
    &=\sigma_\omega\sqrt{1-r_\alpha^2}.
    \label{eq:z11 approx bound}
\end{align}
For the synchronised population with $\alpha=1$ and $r_1\approx 1$ the second approximation in \eqref{eq:zkapprox_cond} then follows immediately.

A necessary condition of the validity of the approximations \eqref{eq:zkapprox_cond} is that they need to be consistent with the dynamics \eqref{eq:dzkmn} in the sense that if they are satisfied at any time, they remain satisfied for all times. Unfortunately, this is not the case as can be readily seen from \eqref{eq:dzkmn}, which involves higher moments. However, if initially we require the stronger conditions
\begin{align}
\label{eq:zkmnapprox_cond}
z_\alpha^{(m,n) }& = z_\alpha^m, \quad \text{for}\quad n=0, \nonumber\\ 
z_\alpha^{(m,n)} &=0,                   \quad \text{for}\quad n\neq0,
\end{align} 
for $\alpha=1,2$ and $m,n\in \mathbb{N}$, we can show that they remain satisfied if so at any time. We provide the motivation behind the assumptions \eqref{eq:zkmnapprox_cond} in Appendix~\ref{sec:app1_s1} below. Note that for $(m=0, n=0)$ and $(m=1, n=0)$ the conditions are trivially satisfied, and $(m=2,n=0)$ and $(m=1,n=1)$ correspond to \eqref{eq:zkapprox_cond}. 

Using the unperturbed dynamics \eqref{eq:dzk}, conditions \eqref{eq:zkmnapprox_cond} imply that
\begin{align}
\label{eq:dzkmnapprox_app}
\frac{d}{dt} z_\alpha^{(m,0)} &= \frac{d}{dt} z_\alpha^m \nonumber\\
&= m \frac{1}{2} (v_\alpha z_\alpha^{m-1} - v_\alpha^*z_\alpha^{(2,0)} z_\alpha^{m-1})
\end{align}
for $\alpha=1,2,$ and $v_\alpha$ defined in \eqref{eq:babyadler2_1}--\eqref{eq:babyadler2_2}. Since \eqref{eq:zkmnapprox_cond} implies $z_\alpha^{(m,0)}  z_\alpha^{m^{\prime}} =z_\alpha^{(m+{m^{\prime}},0)}$ for all $m,m^{\prime}\in\mathbb{N}$, \eqref{eq:dzkmnapprox_app} simplifies to
\begin{align}
\label{eq:dzkmnapprox_app2}
\frac{d}{dt} z_\alpha^{(m,n)}=m ( i z_\alpha^{(m,n+1)} + \frac{1}{2}(v_\alpha z_\alpha^{(m-1,n)} -v_\alpha^* z_\alpha^{(m+1,n)}))
\end{align}
for $n=0$ and $\alpha=1,2$. Note that \eqref{eq:dzkmnapprox_app2} is trivially valid for $n\neq 0$, and corresponds to the unperturbed dynamics \eqref{eq:dzkmn}. This implies that if conditions \eqref{eq:zkmnapprox_cond} hold for any point in time, then they must be true for all times. We remark that there is an infinite family of conditions with this self-consistency property, which we describe in Appendix~\ref{sec:app1_s2} below.


\subsection{Motivation for conditions \eqref{eq:zkmnapprox_cond}}
\label{sec:app1_s1}
For small variance of the native frequencies, $\sigma_\omega^2$, we approximate $\omega^{(\alpha)}_j\approx \bar \omega^{(\alpha)}$, where $\bar \omega^{(\alpha)}$ is the mean of the native frequencies of population $P_\alpha$. Assuming further that the dynamics is near a stable equilibrium at the thermodynamic limit, we approximate \eqref{eq:zmn} as
\begin{align}
    z_\alpha^{(m,n)} \approx (\bar{\omega}^{(\alpha)})^n \langle e^{im\theta^{(\alpha)}}\rangle.
\end{align}
For entrained oscillators, it is readily shown that 
\begin{align}
    \langle e^{im\theta^{(\alpha)}}\rangle&=(\langle e^{i\theta^{(\alpha)}}\rangle)^m.
\end{align}
Non-entrained oscillators are distributed according to the stationary distribution \eqref{eq:rhostat_nonen}, and we can write 
\begin{align}
    \langle e^{im\theta^{(\alpha)}}\rangle &= \int_{-\pi}^{\pi}e^{im\theta}\rho^{(\alpha)}(\theta)d\theta
    \nonumber\\
    &= \frac{1}{\int_{-\pi}^{\pi}\frac{1}{\nu^{(\alpha)}}d\theta}\, \int_{-\pi}^{\pi}\frac{e^{im\theta}}{\nu^{\alpha}}d\theta
    \nonumber\\
    &= \frac{1}{2\pi} \int_{-\pi}^{\pi} e^{im\theta}
    \frac{\sqrt{(\bar{\omega}^{(\alpha)} - \Omega)^2-|v_\alpha|^2}}{\bar{\omega}^{(\alpha)} - \Omega + \text{Im}(v_{\alpha}e^{-i\theta})}d\theta.
    \label{eq:appendix approximation for zmn}
\end{align}
Introducing for ease of exposition
\begin{align}
    p&=\frac{\bar{\omega}^{(\alpha)}-\Omega}{\left|v_\alpha\right|}-\sqrt{\frac{\bar{\omega}^{(\alpha)}-\Omega}{\left|v_\alpha\right|}-1}\sqrt{\frac{\bar{\omega}^{(\alpha)}-\Omega}{\left|v_\alpha\right|}+1},\\
    {\rm{v}}&=i\frac{v_{\alpha}}{|v_\alpha|},
\end{align}
we write
\begin{align}
    \langle e^{im\theta^{(\alpha)}}\rangle &=\frac{1}{2\pi} \int_{-\pi}^{\pi} e^{im\theta}\frac{1-p^{2}}{1+p^{2}-2p\operatorname{Re}\left({\rm{v}}e^{-i\theta}\right)}d\theta
    \nonumber\\
    &=\frac{1}{2\pi}\int_{-\pi}^{\pi} e^{im\theta}\frac{1-p^{2}}{1+p^{2}-p\left(\frac{{\rm{v}}}{e^{i\theta}}+\frac{e^{i\theta}}{{\rm{v}}}\right)}d\theta
    \nonumber\\
    &=\frac{1}{2\pi}\int_{-\pi}^{\pi} e^{im\theta} \left(1+\frac{p\frac{{\rm{v}}}{e^{i\theta}}}{1-p\frac{{\rm{v}}}{e^{i\theta}}}+\frac{p\frac{e^{i\theta}}{{\rm{v}}}}{1-p\frac{e^{i\theta}}{{\rm{v}}}}\right)d\theta
    \nonumber\\
    &=\frac{1}{2\pi}\int_{-\pi}^{\pi} e^{im\theta} \sum_{n=-\infty}^{\infty}p^{\left|n\right|}\left(\frac{{\rm{v}}}{e^{i\theta}}\right)^{n}d\theta
    \nonumber\\
    &=\sum_{n=-\infty}^{\infty}p^{\left|n\right|}{\rm{v}}^{n}\int_{-\pi}^{\pi}\frac{e^{i\left(m-n\right)\theta}}{2\pi}d\theta
    \nonumber\\
    &=p^{\left|m\right|}{\rm{v}}^{m}.
\end{align}
Therefore, we are left with the approximation of 
\begin{align}
    z_\alpha^{(m,n)} \approx (\bar{\omega}^{(\alpha)})^nz_{\alpha}^m.
\end{align}
With $\bar\omega=0$ conditions  \eqref{eq:zkmnapprox_cond} immediately follow. 


\subsection{Self-consistent approximations for $z^{(m,n)}_{\alpha}$}
\label{sec:app1_s2}

Here we derive a more general set of approximations for $z^{(m,n)}_\alpha$, which satisfies the set of self-consistency conditions \eqref{eq:zkmnapprox_cond}. We restrict the approximations to be of the form 
\begin{align}
    z^{(m,n)}_\alpha=c^ne^{mi\psi_\alpha}f_m(r_\alpha),\label{eq:zmn form}
\end{align}
where $f_m$ is an unknown function to be determined. Special cases are $f_0(r_\alpha)=1$ and $f_1(r_\alpha)=r_\alpha$. The temporal evolution of $z_\alpha^{(m,n)}$ is given by
\begin{align}
\frac{d}{dt}z_{\alpha}^{(m,n)} &= c^ne^{mi\psi_\alpha}(f'_m(r_\alpha)\frac{dr_{\alpha}}{dt}+i\, mf_m(r_\alpha)\frac{d\psi_{\alpha}}{dt})\nonumber\\
&= c^ne^{mi\psi_\alpha}\left(f'_m(r_\alpha)\frac{(1-f_2(r_\alpha))\operatorname{Re}(v_\alpha e^{-i\psi_\alpha})}{2}\right.\nonumber\\
&\left.+i\,mf_m(r_\alpha)(c+\frac{(1+f_2(r_\alpha))\operatorname{Im}(v_\alpha e^{-\psi_\alpha})}{2r_\alpha})\right).
\label{eq:appendix dzmndt}
\end{align}
Equating \eqref{eq:appendix dzmndt} with \eqref{eq:dzkmn} determines a recursive relationship for $f_m$ with
\begin{align}
f_{m+1}(r_\alpha)&=f_m(r_\alpha)\frac{1+f_2(r_\alpha)}{r_\alpha}-f_{m-1}(r_\alpha),\label{eq:fm+1}\\
f'_m(r_\alpha)&=m\frac{f_{m-1}(r_\alpha)-f_{m+1}(r_\alpha)}{1-f_2(r_\alpha)}.\label{eq:dfm}
\end{align}
Note that if \eqref{eq:dfm} holds for $m=2$ and \eqref{eq:fm+1} holds for all $m$, then \eqref{eq:dfm} holds for all $m$.
This means that to find a valid set of functions $f_m$, we must only find some $f_2$ that satisfies 
\begin{align}
f'_2(r_\alpha)
&=2\frac{2r_\alpha-f_2(r_\alpha)\frac{1+f_2(r_\alpha)}{r_\alpha}}{1-f_2(r_\alpha)}-1.
\label{eq:df2}
\end{align}
For $m>2$, $f_m$ is determined recursively from $f_{0,1,2}(r_\alpha)$. We find that \eqref{eq:df2} is satisfied by the family of solutions to the equations
\begin{align}
    f_2(r_\alpha)=2r_\alpha\frac{r_\alpha(4-g^2)+g\sqrt{4-r_\alpha^2(4-g^2)}}{4}-1 \label{eq:f2}
\end{align}
for any $g\in \mathbb{R}$. For $g\to \infty$, for example, we obtain $f_2(r)=r^2$. Substituting $f_m$ into \eqref{eq:zmn form} we find
\begin{align}
\label{eq:selfconsist_set}
 z_\alpha^{(m,n)} &= \lim_{g_0\rightarrow g}c^n e^{mi\psi_\alpha} 
 \sum_{s=-1,1} \left( \frac{1}{2} + s\frac{r_{\alpha}-C_\alpha(g_0))}{2\sqrt{C_\alpha(g_0)^{2}-1}} \right)  \nonumber \\
 &
 \qquad  \qquad 
 \left( C_\alpha(g_0) + s\sqrt{C_\alpha(g_0)^{2}-1} \right)^{m}
\end{align} 
for any constants $c\in\mathbb{C}$, $g\in \mathbb{R}\cup\{\infty\}$ and with
\begin{align}
 C_\alpha(g_0) = \frac{r_{\alpha}\left(4-g_0^{2}\right)+g_0\sqrt{4-r_{\alpha}^{2}\left(4-g_0^{2}\right)\ }}{4}.
\end{align}
Note that for $c=0$ and $g=\infty$ \eqref{eq:selfconsist_set} yields conditions \eqref{eq:zkmnapprox_cond} with $z_\alpha^{(m,n)} = 0^n z_\alpha^m$.


\section{Numerical integrator for the OU process}
\label{sec:app2}
Consider a general $d$-dimensional stochastic OU process
\begin{align}
    d\zeta=-\Gamma\zeta dt+\Sigma dW_t,\label{eq:general_OU}
\end{align}
where $\Gamma\in\mathbb{R}^{d\times d}$ and $\Sigma\in\mathbb{R}^{d\times d}$ are the drift and diffusion matrices, respectively, and $W_t$ is a $d$-dimensional Wiener process.

To numerically integrate the multivariate OU process \eqref{eq:general_OU} with a fixed time step $\Delta t$, we employ a modified Euler-Maruyama scheme
\begin{align}
\label{eq:EMHenry}
\zeta_{n+1} = \zeta_n - \hat\Gamma\, \zeta_n \Delta t + \hat\Sigma\, \Delta W_n,
\end{align}
with $\Delta W \sim \mathcal{N}(0,\Delta t \Id_d)$ and 
\begin{align}
\hat{\Gamma} &= \frac{\Id-e^{-\Gamma \Delta t}}{\Delta t}
\label{eq:EMGam}
\end{align}
and
\begin{align}
\hat{\Sigma} &= \frac{1}{\sqrt{\Delta t}} \sqrt{\int_0^{\Delta t} \exp(-\Gamma s)\Sigma \Sigma^{\top} \exp(-\Gamma^{\top}s)\, ds}.
\label{eq:EMSig}
\end{align}
Note that $\lim_{\Delta t\to 0}\hat \Gamma = \Gamma$ and $\lim_{\Delta t\to 0}\hat \Sigma = \Sigma$. The modified Euler-Maruyama scheme \eqref{eq:EMHenry} has the desirable property that it preserves the true transition density $\rho(\vec\zeta_{n+1}|\vec\zeta_n)$ of the underlying process \eqref{eq:general_OU}; see, for example, \cite{VatiwutipongPhewchean19}. 

The integral over the matrix exponential in \eqref{eq:EMSig} 
can be explicitly calculated for diagonalizable drift matrices $\Gamma=VDV^{-1}$ with diagonal $D$ as
\begin{align}
        &\int_0^t \exp(-\Gamma s)\Sigma\Sigma^\top \exp(-\Gamma^{\top}s)\, ds \nonumber \\
        &= \int_0^t \exp(-VDV^{-1} s)\Sigma\Sigma^\top \exp(-V^{-\top}DV^{\top}s)\, ds \nonumber \\
        &= V \left(\int_0^t \exp(-D s)V^{-1}\Sigma\Sigma^\top V^{-\top}\exp(-Ds)\, ds\right) V^{\top} \nonumber \\
        &= V \left( V^{-1}\Sigma\Sigma^\top V^{-\top} \odot \int_0^t \exp_{\circ}(-(D\1+\1D) s)\, ds\right) V^{\top} \nonumber \\
        &= V \left( V^{-1}\Sigma\Sigma^\top V^{-\top}\oslash(D\1+\1D) \right.\nonumber\\
        &\left. 
        \odot (\1- \exp_{\circ}(-(D\1+\1D)t)) \right) V^{\top},
        \label{eq:matrix_transpose_integral}
\end{align}
where $\odot$, $\oslash$ and $\exp_{\circ}$ denote Hadamard multiplication, division and exponentiation respectively, and $\1\in\R^{d\times d}$ denotes a matrix consisting of $1$s.\\
For the special case of drift and diffusion matrices that are of the form \eqref{eq:bigGamma}--\eqref{eq:bigSigma}, \eqref{eq:matrix_transpose_integral} simplifies to
\begin{align}
    \int_0^t \exp(-\Gamma s)\Sigma\Sigma^\top \exp(-\Gamma^{\top}s)\, ds \nonumber\\=\frac{\sigma^2}{2\operatorname{Re}(\gamma)} \exp(-2\operatorname{Re}(\gamma))\,\operatorname{I}.
\end{align}
We remark that for more general cases, the analytical formula for the variance can be costly to compute in high dimensions, and one may be well advised to employ a naive approach of repeated Euler-Maruyama steps~\cite{SachsEtAl17}. 



\section{Additional numerical results}
\label{sec:appNR}
To show that our stochastic modelling framework is generally valid, we present results for a different choice of parameters with $K=85$, $\kappa=55$ and $\lambda=\pi/2-0.09$. For these parameters we obtain $\langle Z\rangle= 0.689 + 0.131\, i$ from the mean-field theory. Fitting the covariance function by minimising the cost function \eqref{eq:cost} in the range from $t_{\rm{min}}=0.5$ to $t_{\rm{max}}=1$ yields $\gamma =1.918 -18.05i$ and $\sigma=0.335$. 

\begin{figure}[htb]
     \centering
      \includegraphics[width=0.49\linewidth]{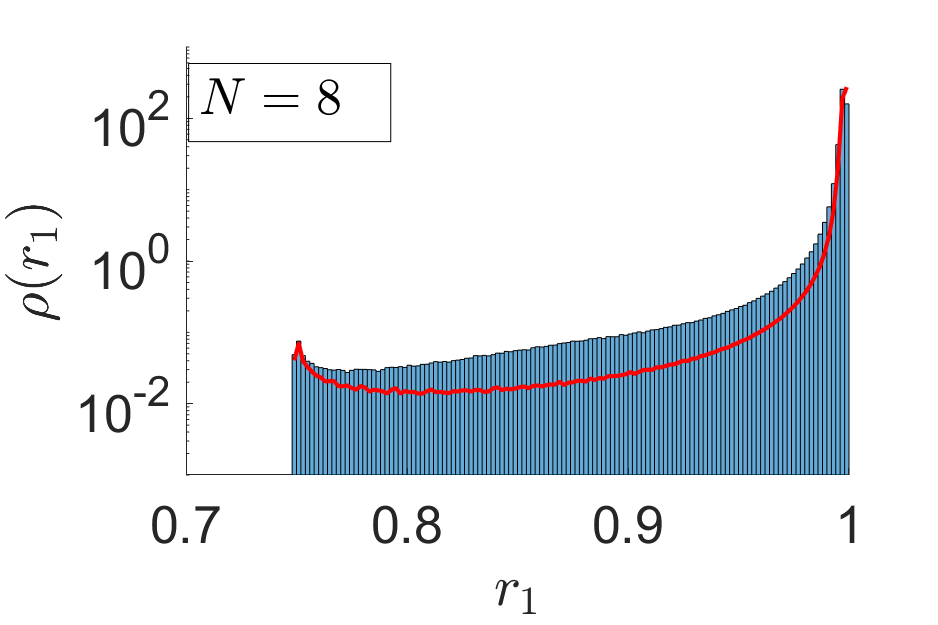}
      \includegraphics[width=0.49\linewidth]{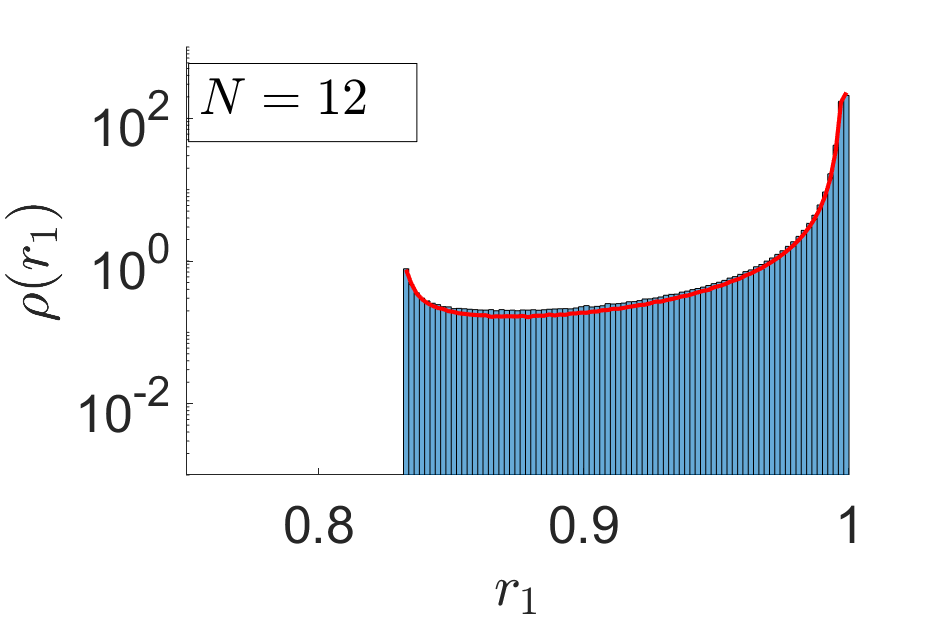}
      \includegraphics[width=0.49\linewidth]{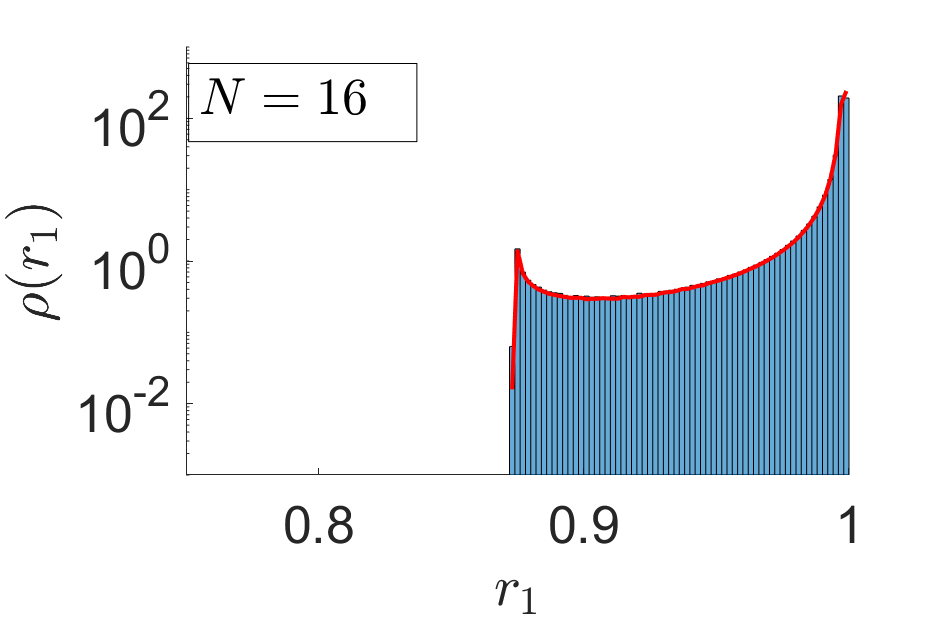}
      \includegraphics[width=0.49\linewidth]{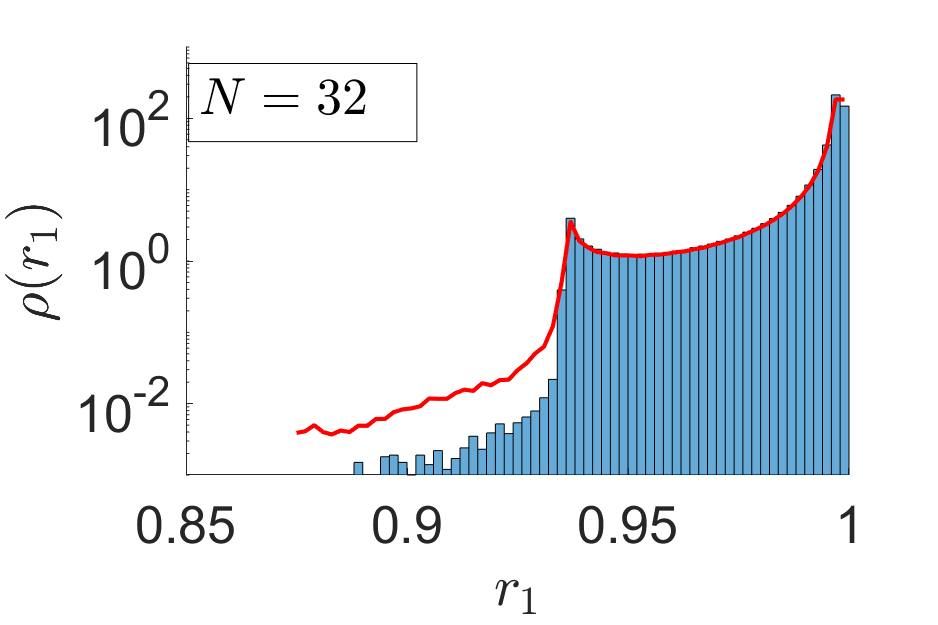}
\caption{Comparison of the empirical histograms for the order parameter $r_1$ of the synchronised population obtained from a single trajectory of the full two-population KS model \eqref{eq:abrams_1}--\eqref{eq:abrams_2} (blue histogram) and from the reduced stochastic model \eqref{eq:abramszOU}--\eqref{eq:OU} (red line) for $N=8,12,16$ and $32$. A log scale has been applied to the $y$-axis. Equation parameters are as $K=85, \kappa=55$ and $\lambda = \pi/2-0.09$.}
        \label{fig:ALT1_r1histcomp}
\end{figure}

\begin{figure}[htb]
     \centering
     \includegraphics[width=0.49\linewidth]{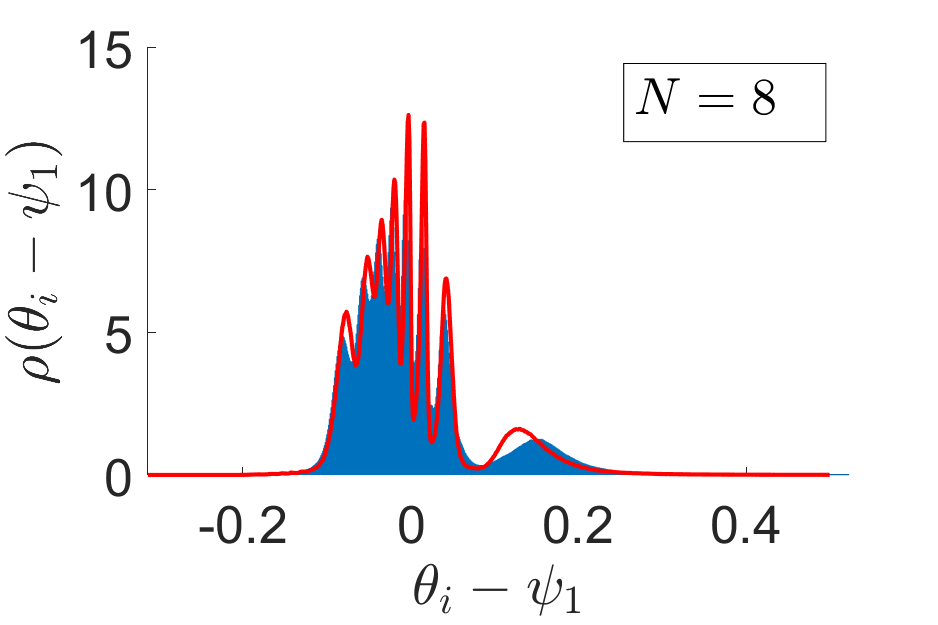}
     \includegraphics[width=0.49\linewidth]{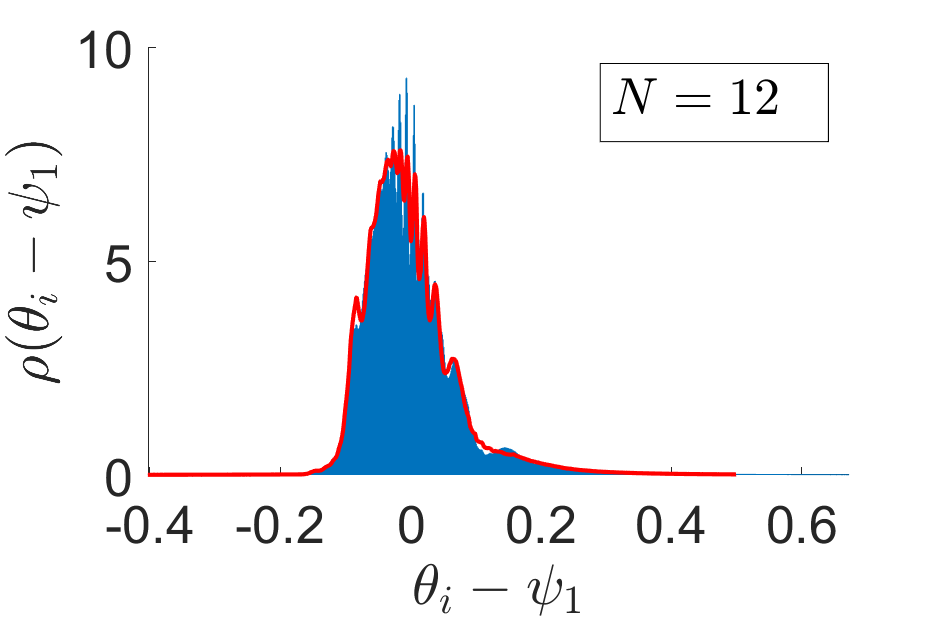}
     \includegraphics[width=0.49\linewidth]{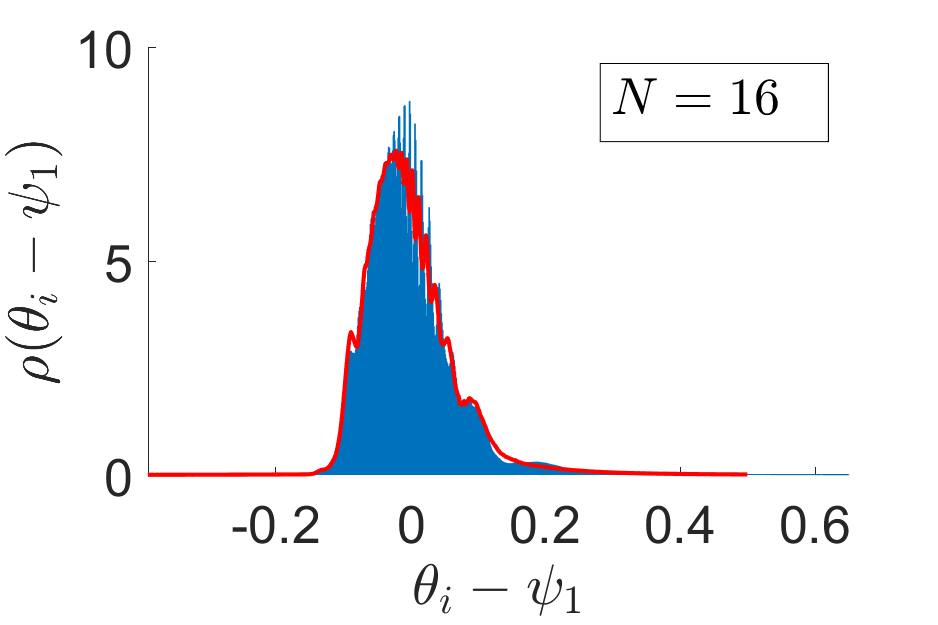}
     \includegraphics[width=0.49\linewidth]{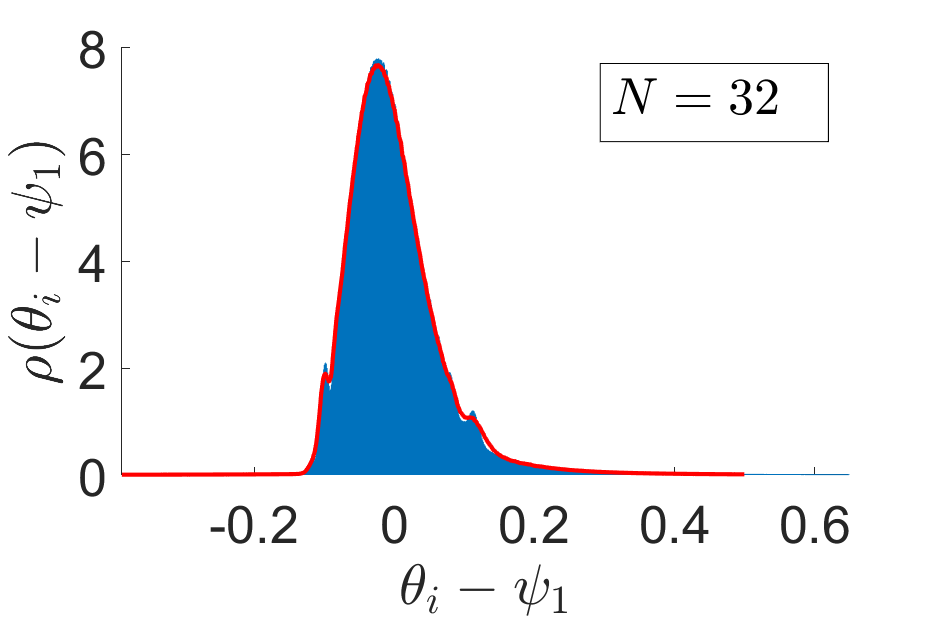}
\caption{Comparison of the empirical histograms for the fluctuations of the synchronised phases around their mean, $\theta_i-\psi_1$, obtained from a single trajectory of the full two-population KS model \eqref{eq:abrams_1}--\eqref{eq:abrams_2} (blue histogram) and of the reduced stochastic model \eqref{eq:abramszOU}--\eqref{eq:OU} (red line) for $N=8,12,16$ and $32$. Equation parameters are as in Figure~\ref{fig:ALT1_r1histcomp}.}
\label{fig:ALT1_thetahist}
\end{figure}

\begin{figure}[htb]
    \centering
     \includegraphics[width=0.49\linewidth]{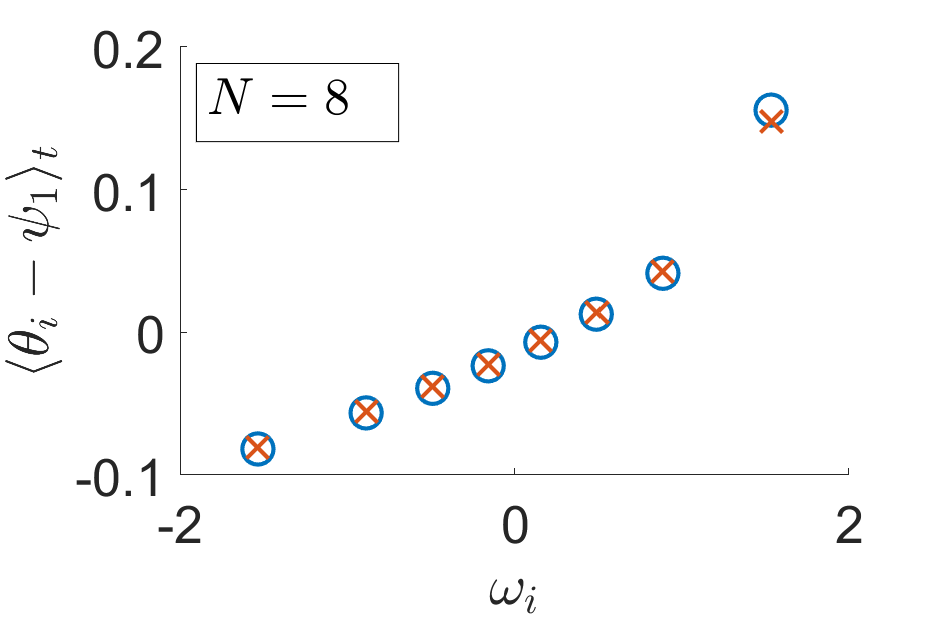}
     \includegraphics[width=0.49\linewidth]{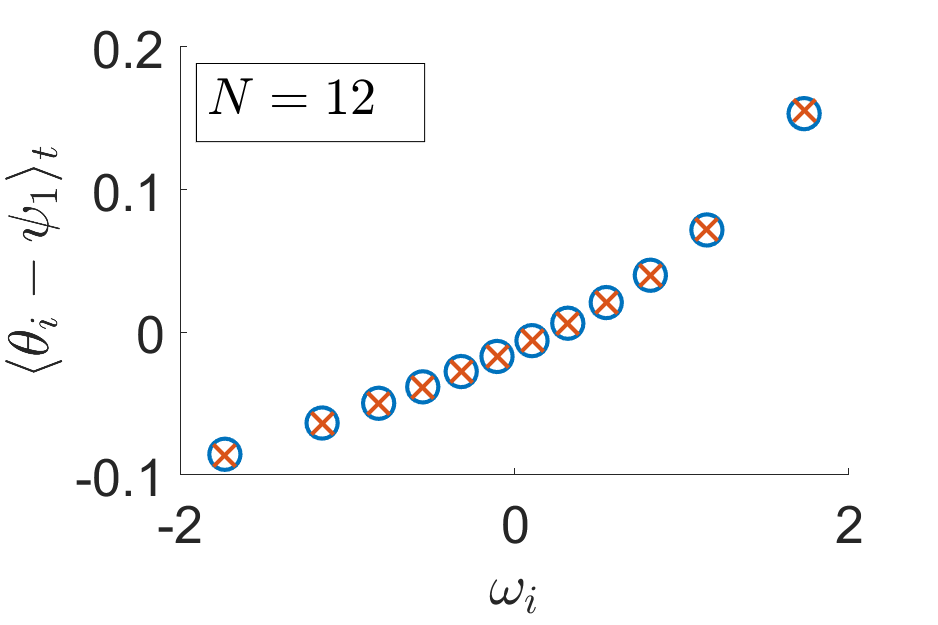}
     \includegraphics[width=0.49\linewidth]{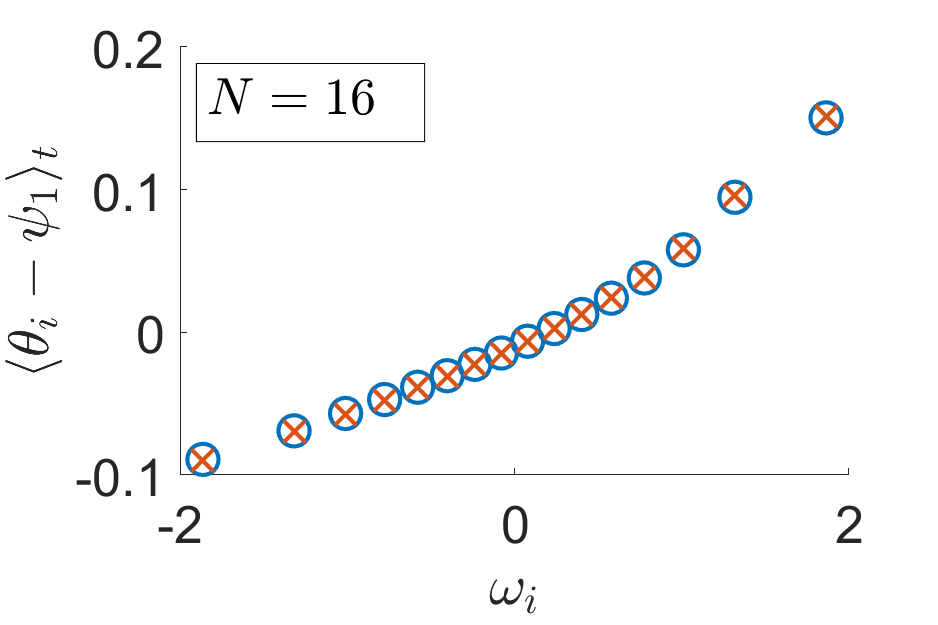}
     \includegraphics[width=0.49\linewidth]{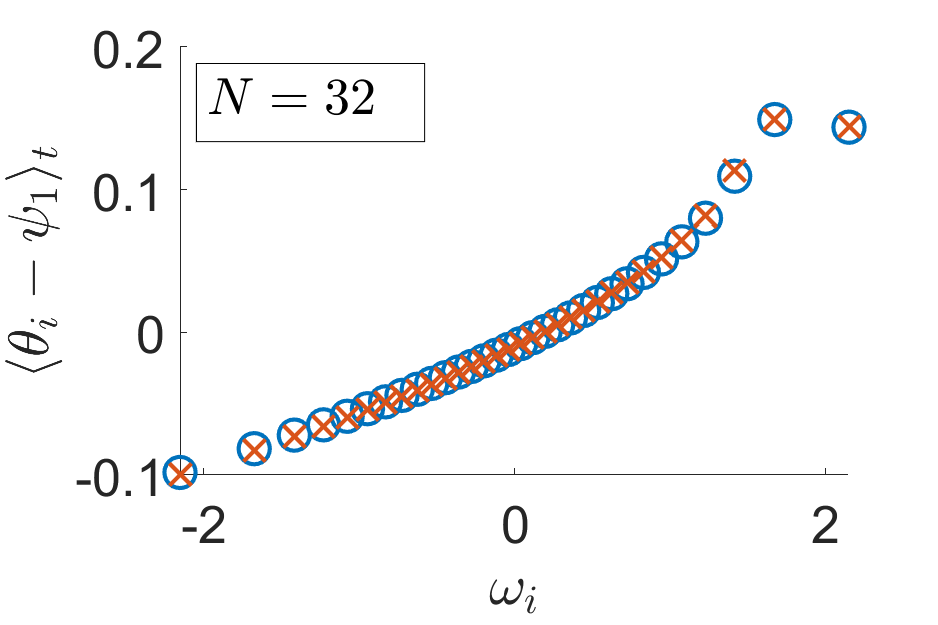}
\caption{Mean of the fluctuations of the synchronised phases around their mean, $\theta_i-\psi_1$, as a function of their native frequencies $\omega_i$ for the full two-population KS model \eqref{eq:abrams_1}--\eqref{eq:abrams_2} (blue circles) and for the reduced stochastic model \eqref{eq:abramszOU}--\eqref{eq:OU} (red crosses) for $N=8,12,16$ and $32$. Equation parameters are as in Figure~\ref{fig:ALT1_r1histcomp}.}
\label{fig:ALT1_theta_mean}
\end{figure}

\begin{figure}[htb]
     \centering
     \includegraphics[width=0.49\linewidth]{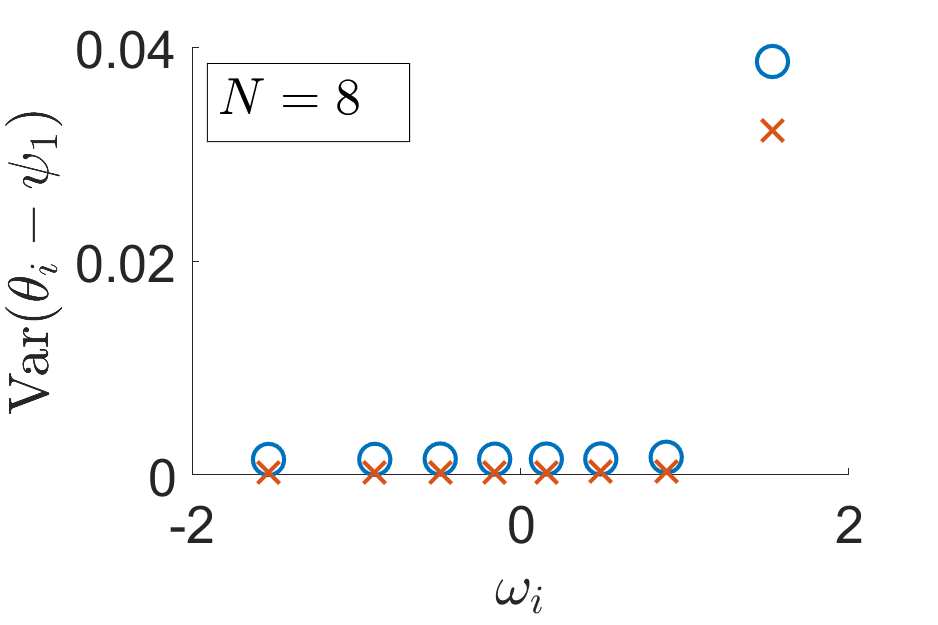}
     \includegraphics[width=0.49\linewidth]{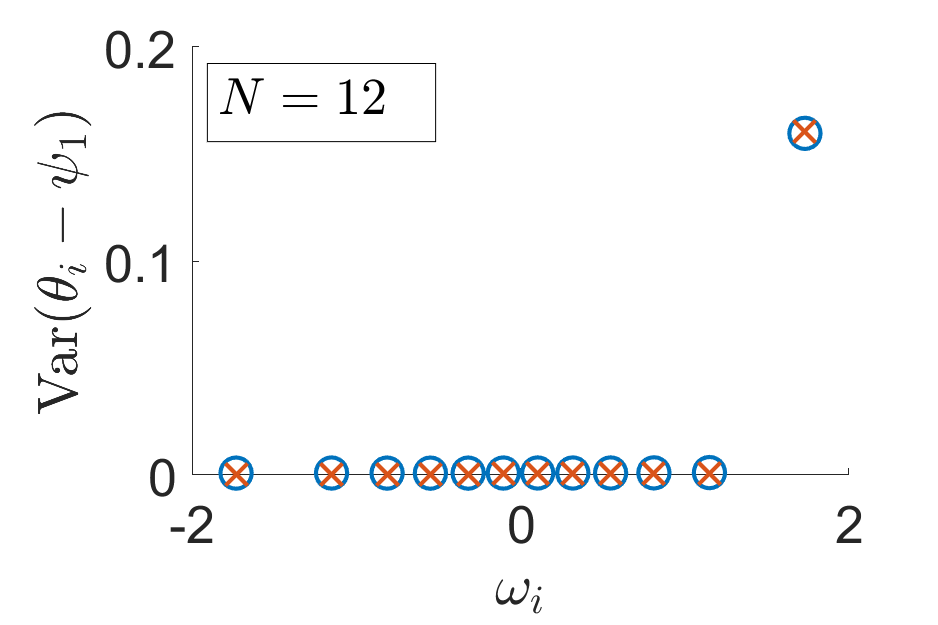}
     \includegraphics[width=0.49\linewidth]{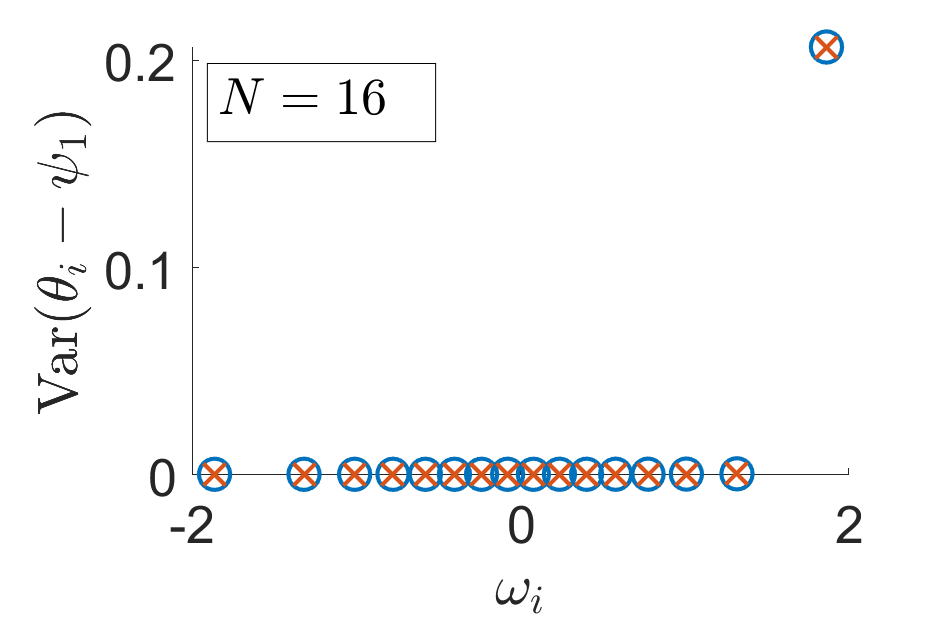}
     \includegraphics[width=0.49\linewidth]{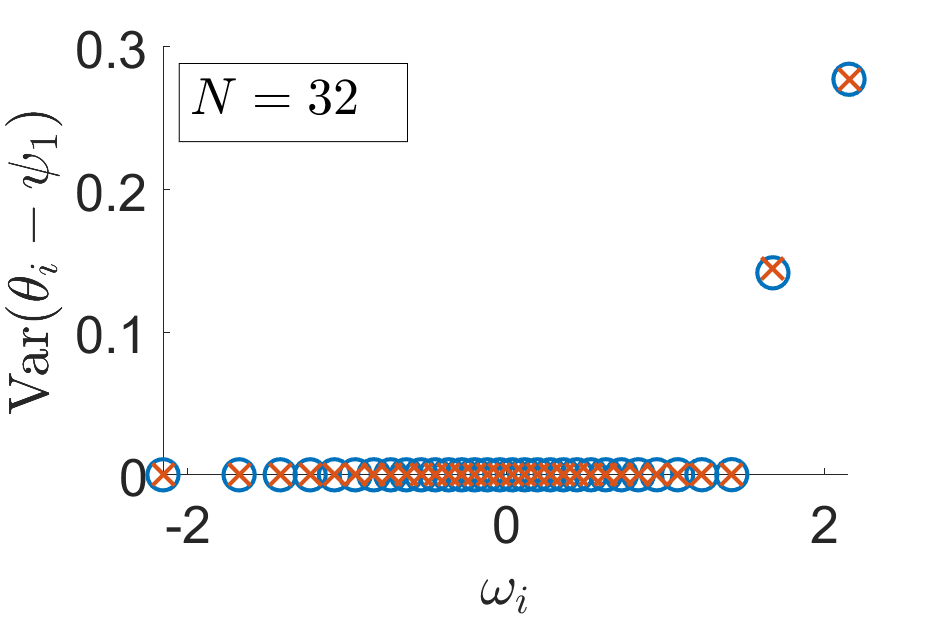}
\caption{Variance of the fluctuations of the synchronised phases around their mean, $\theta_i-\psi_1$, as a function of their native frequencies $\omega_i$ for the full two-population KS model \eqref{eq:abrams_1}--\eqref{eq:abrams_2} (blue circles) and for the reduced stochastic model \eqref{eq:abramszOU}--\eqref{eq:OU} (red crosses) for $N=8,12,16$ and $32$. Equation parameters are as in Figure~\ref{fig:ALT1_r1histcomp}.}
\label{fig:ALT1_theta_var}
\end{figure}

\begin{figure}[htb]
     \centering
      \includegraphics[width=0.49\linewidth]{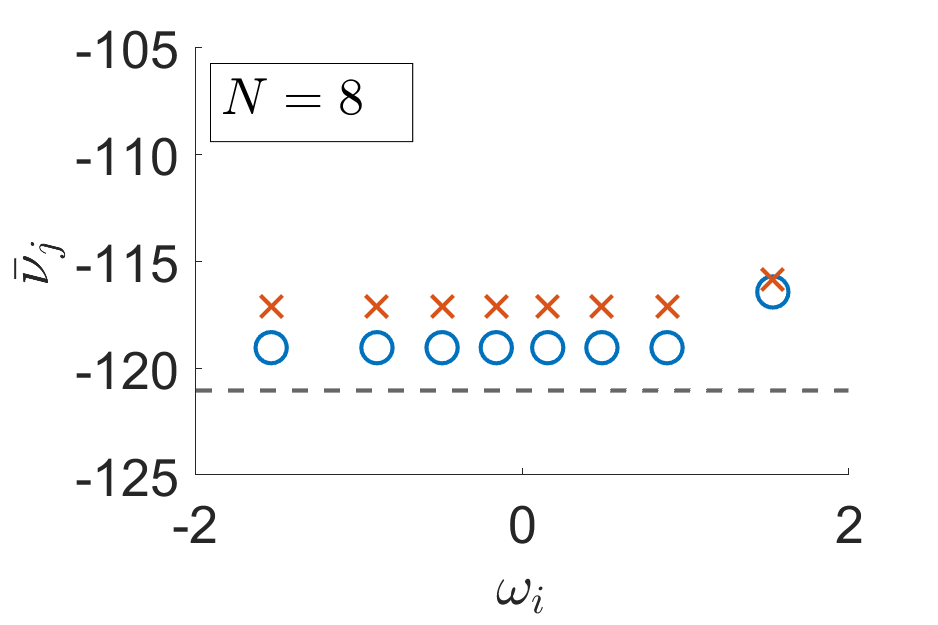}
      \includegraphics[width=0.49\linewidth]{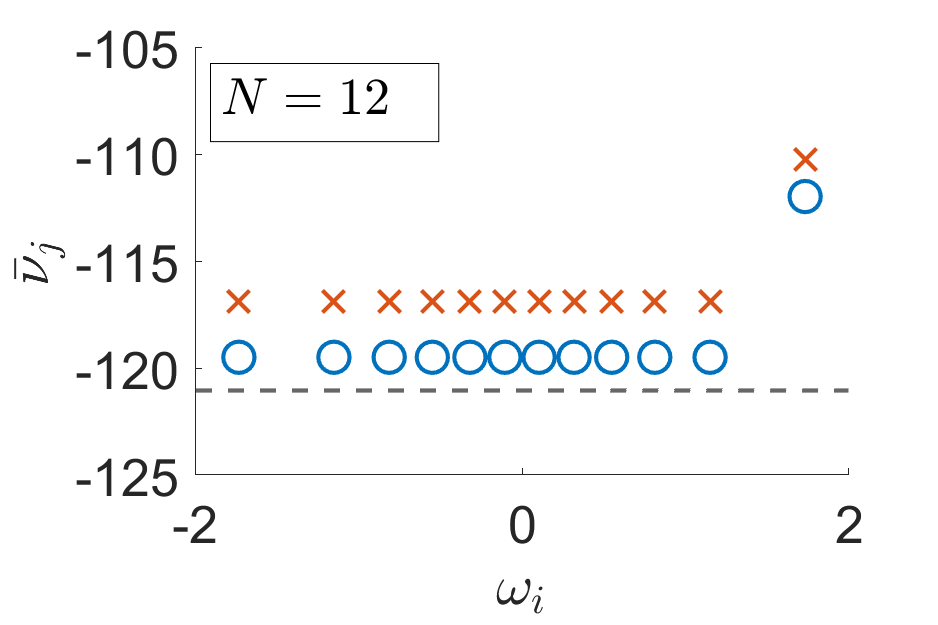}
      \includegraphics[width=0.49\linewidth]{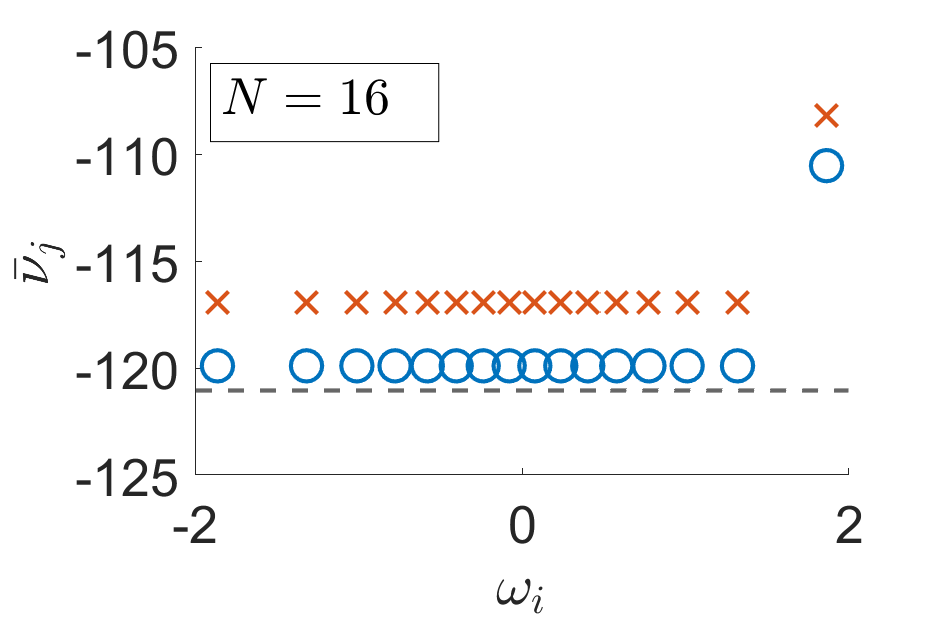}
      \includegraphics[width=0.49\linewidth]{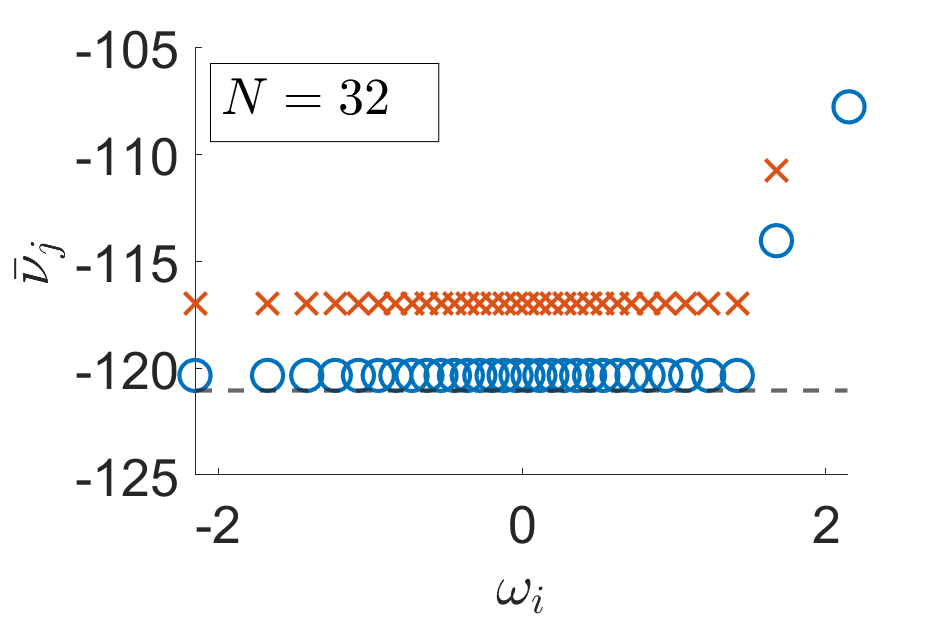}
\caption{Mean rotation frequency ${\bar \nu}_i $ of the synchronised oscillators as a function of their native frequencies $\omega_i$ for the full two-population KS model \eqref{eq:abrams_1}--\eqref{eq:abrams_2} (blue circles) and for the reduced stochastic model \eqref{eq:abramszOU}--\eqref{eq:OU} (red crosses) for $N=8,12,16$ and $32$. The dashed lines denote the mean frequency $\Omega$ as calculated from the self-consistency relations \eqref{eq:selconst_r}--\eqref{eq:selconst_Z} of the mean-field theory. Equation parameters are as in Figure~\ref{fig:ALT1_r1histcomp}.}
        \label{fig:ALT1_theta_dot}
\end{figure}

\begin{figure}[htb]
     \centering
      \includegraphics[width=0.7\linewidth]{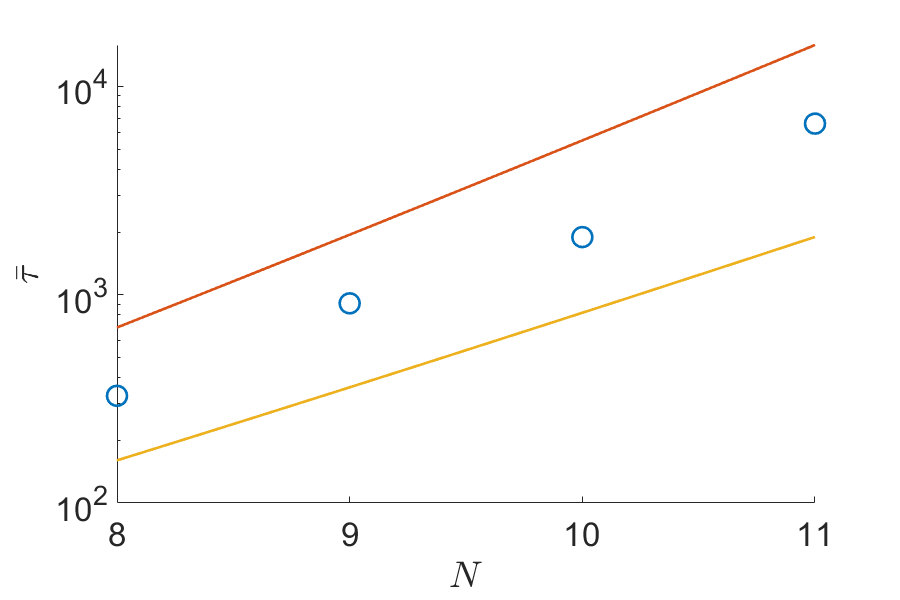}
\caption{Comparison of the mean switching time $\bar{\tau}$ for different system sizes $N$, obtained from a long-time simulations of the full two-population KS model \eqref{eq:abrams_1}--\eqref{eq:abrams_2} (blue circles) and from the analytical expressions \eqref{eq:bartau_taue} and \eqref{eq:taue}\ based on solving Kramers problem using our reduced stochastic model. We show the result of the Kramers problem for $q^\star=0.261$ (red line) which was estimated from the full two-population KS model \eqref{eq:abrams_1}--\eqref{eq:abrams_2}, and for $q^\star=0.236$ (yellow line) which was estimated from the reduced deterministic mean-field model \eqref{eq:dZ}. Equation parameters are as in Figure~\ref{fig:ALT1_r1histcomp}.}
        \label{fig:ALT1_tau_bar}
\end{figure}

We show in Figures~\ref{fig:ALT1_r1histcomp}--\ref{fig:ALT1_theta_dot} that our stochastically driven model \eqref{eq:abramszOU}--\eqref{eq:OU} reproduces the statistics of the order parameter and the phases very well. The case of a small network with $N=8$ is less well approximated. Figure~\ref{fig:ALT1_theta_dot} shows the mean rotation frequency ${\bar \nu}_i$ for various values of $N$. We used the mean-field result for $\langle Z\rangle$ in the reduced stochastic model \eqref{eq:abramszOU}--\eqref{eq:OU}. We remark that if we used the empirical mean estimated from simulations of the two-population KS model \eqref{eq:abrams_1}--\eqref{eq:abrams_2} for each value of $N$, the mean rotation frequencies are as accurate as those shown in Figure~16 in the main manuscript. Figure~\ref{fig:ALT1_tau_bar} shows that the azimuthally averaged equation for the fluctuations is capable to provide good estimates of the mean switching time using the Kramers problem solution~\eqref{eq:taue}.

\section{Random draws of the native frequencies}
\label{sec:appNR_R}
We present results for the case where the native frequencies are drawn randomly rather than equiprobably, i.e. where the two populations are not exactly the same. We show results for the same set of parameters as used in the main manuscript.

\begin{figure}[htb]
     \centering
      \includegraphics[width=0.49\linewidth]{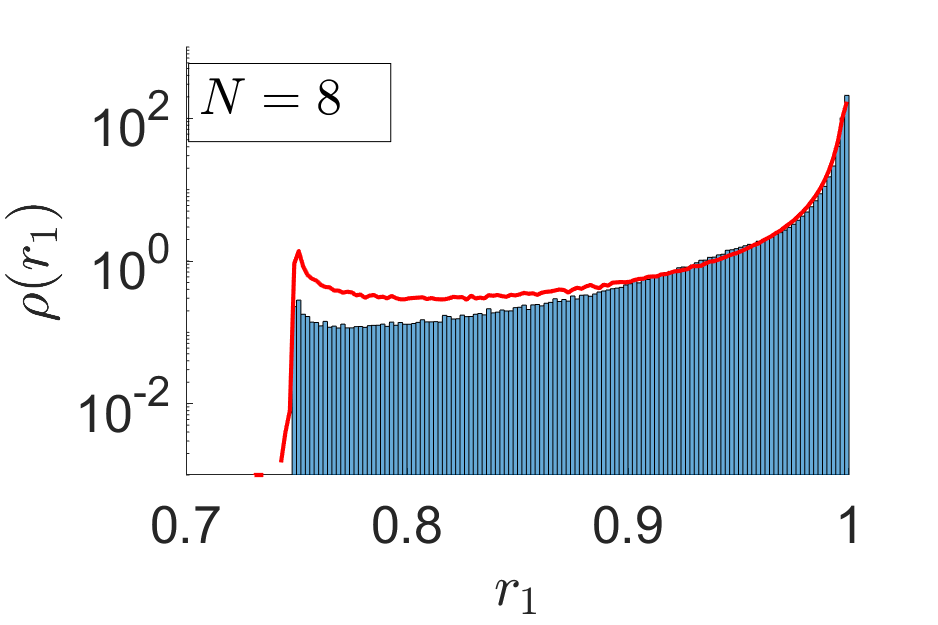}
      \includegraphics[width=0.49\linewidth]{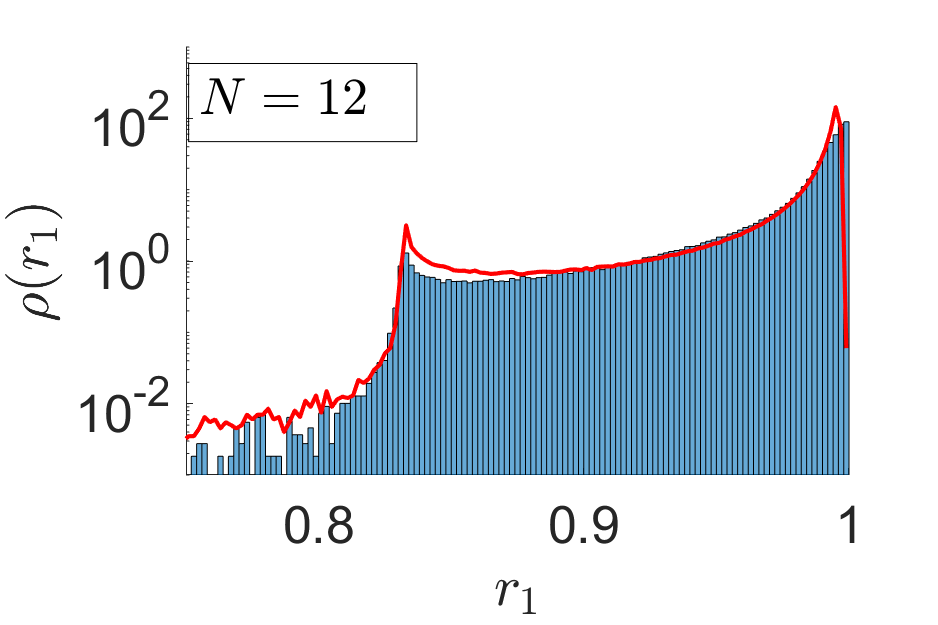}
      \includegraphics[width=0.49\linewidth]{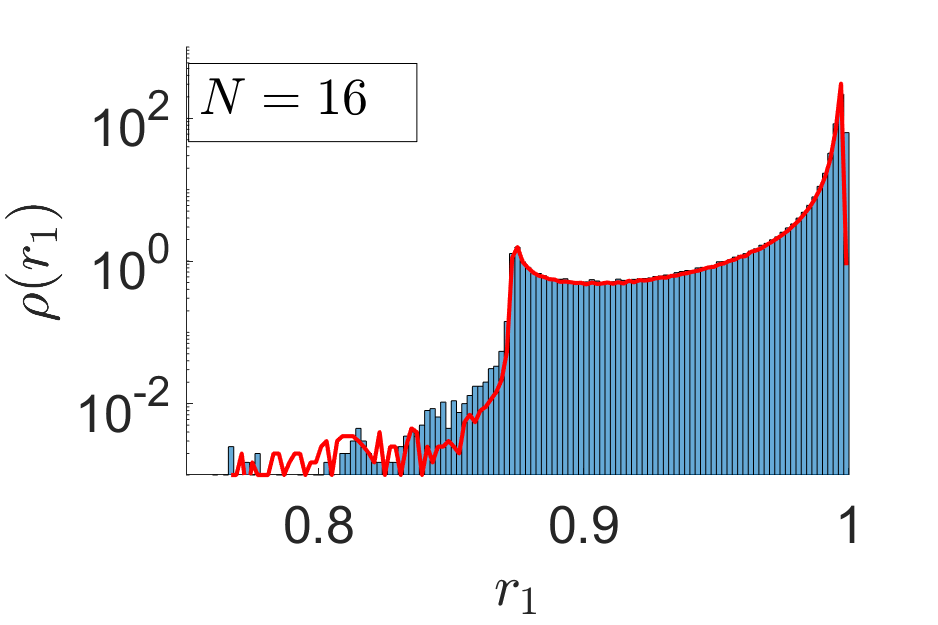}
      \includegraphics[width=0.49\linewidth]{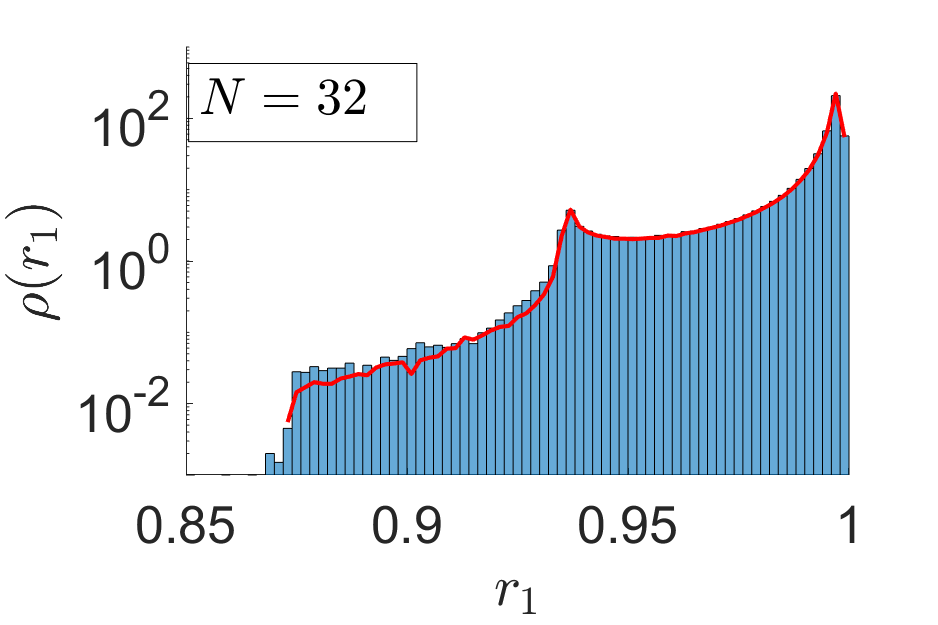}
\caption{Comparison of the empirical histograms for the order parameter $r_1$ of the synchronised population obtained from a single trajectory of the full two-population KS model \eqref{eq:abrams_1}--\eqref{eq:abrams_2} (blue histogram) and from the reduced stochastic model \eqref{eq:abramszOU}--\eqref{eq:OU} (red line) for $N=8,12,16$ and $32$, where the native frequencies $\omega_i$ are drawn randomly and independently from a standard normal distribution $\mathcal{N}(0,1)$. Equation parameters are $K=100$, $\kappa=60$ and $\lambda=\frac{\pi}{2}-0.075$.
}
        \label{fig:ALTSAMP1_r1histcomp}
\end{figure}

The empirical histograms of the order parameters are not very sensitive to the particular realisation of the native frequencies. We show in Figure~\ref{fig:ALTSAMP1_r1histcomp} that our stochastically driven model \eqref{eq:abramszOU}--\eqref{eq:OU} (using the same fitted values of $\gamma=1.75 -25.34\,i$ and $\sigma=0.574$ as used in the main manuscript) reproduces the statistics of the order parameter and the phases very well even in the case of random samplings of native frequencies $\omega_i$. As expected, the case of a small network with $N=8$ is less well approximated. 

\begin{figure}[htb]
    \centering
     \includegraphics[width=0.49\linewidth]{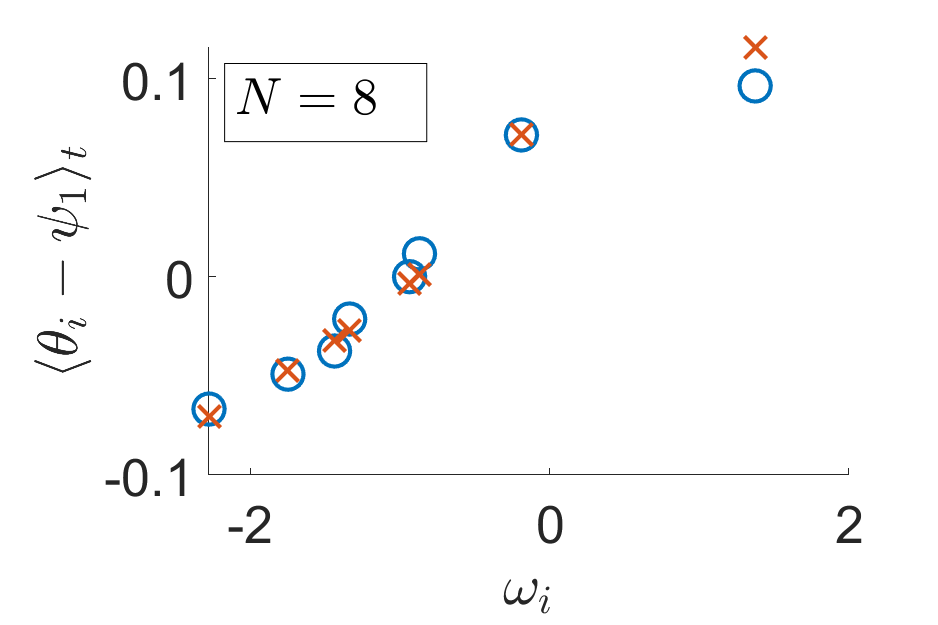}
     \includegraphics[width=0.49\linewidth]{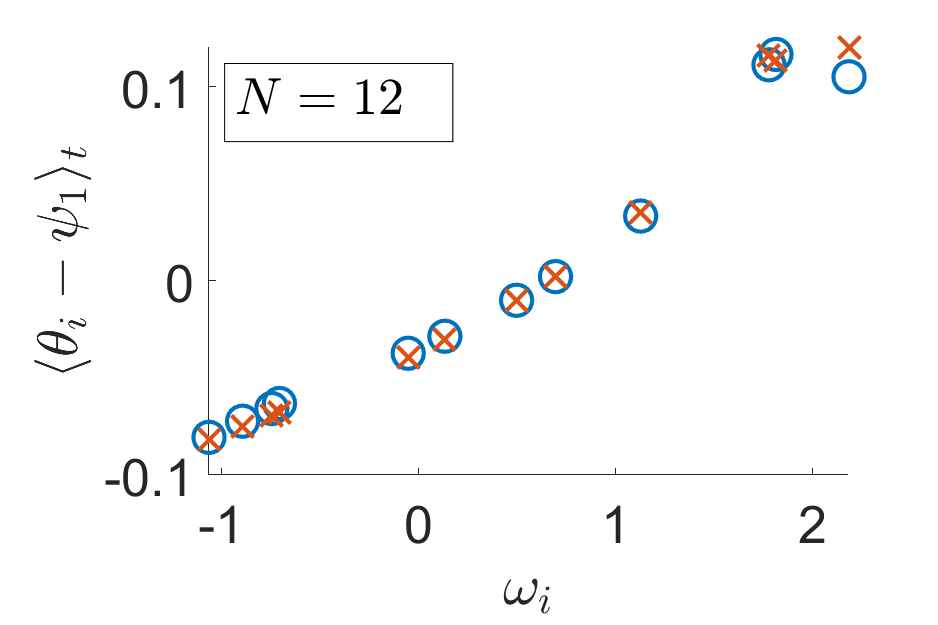}
     \includegraphics[width=0.49\linewidth]{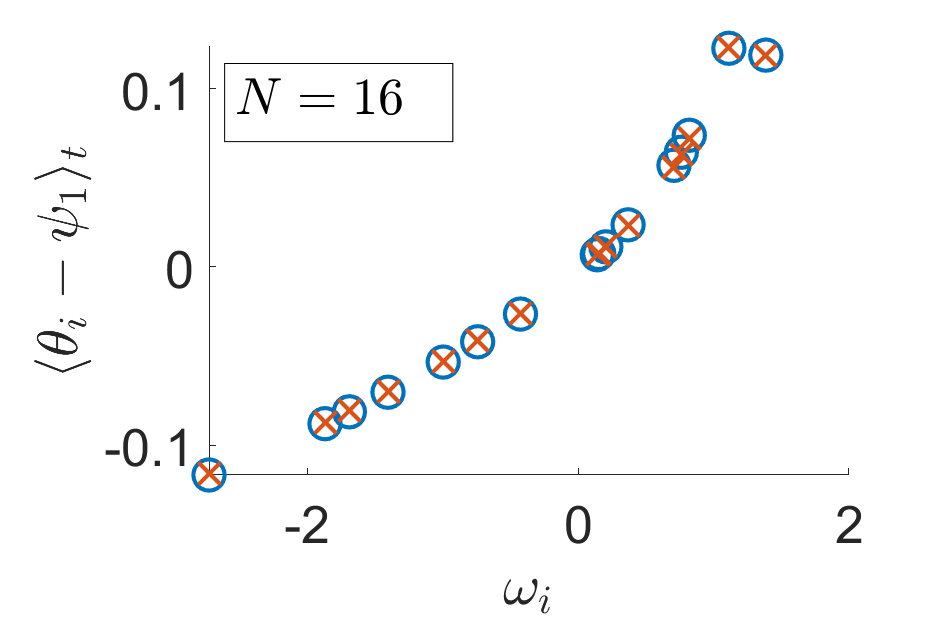}
     \includegraphics[width=0.49\linewidth]{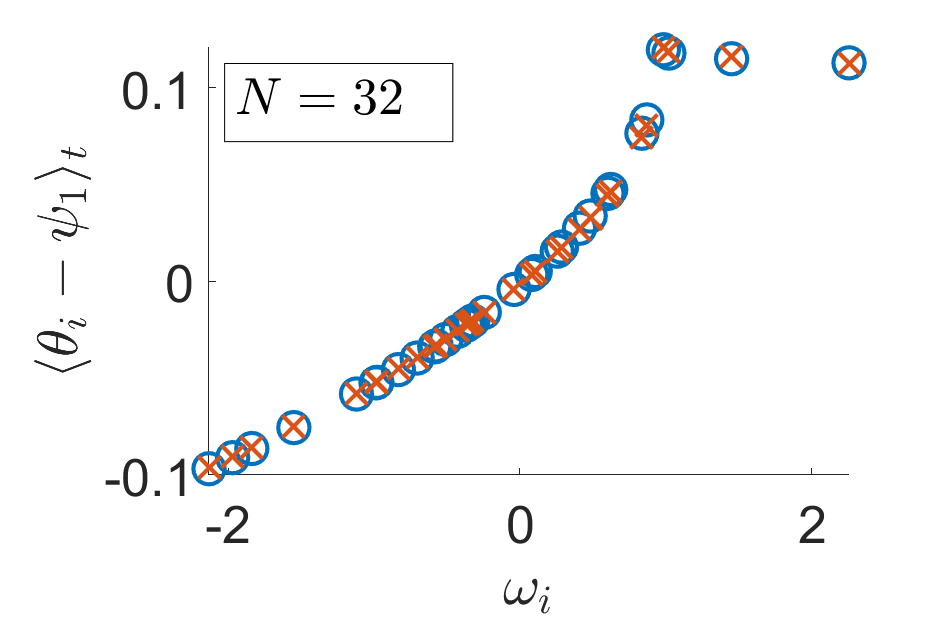}
\caption{Mean of the fluctuations of the synchronised phases around their mean, $\theta_i-\psi_1$, as a function of their native frequencies $\omega_i$ for the full two-population KS model \eqref{eq:abrams_1}--\eqref{eq:abrams_2} (blue circles) and for the reduced stochastic model \eqref{eq:abramszOU}--\eqref{eq:OU} (red crosses) for $N=8,12,16$ and $32$, where the native frequencies $\omega_i$ are drawn randomly and independently from a standard normal distribution $\mathcal{N}(0,1)$. Equation parameters are as in Figure~1 of the main manuscript.}
\label{fig:ALTSAMP1_theta_mean}
\end{figure}

\begin{figure}[htb]
     \centering
     \includegraphics[width=0.49\linewidth]{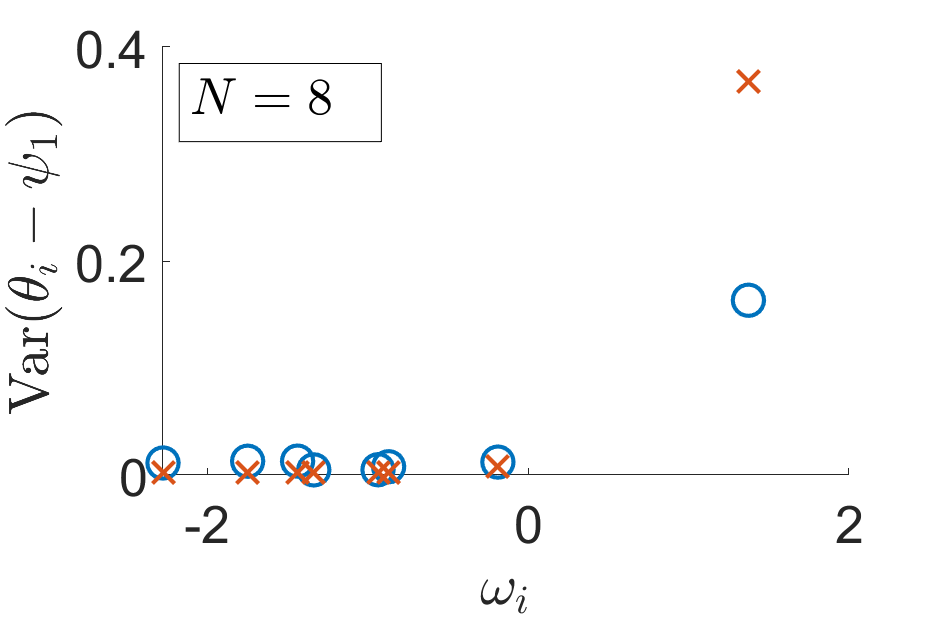}
     \includegraphics[width=0.49\linewidth]{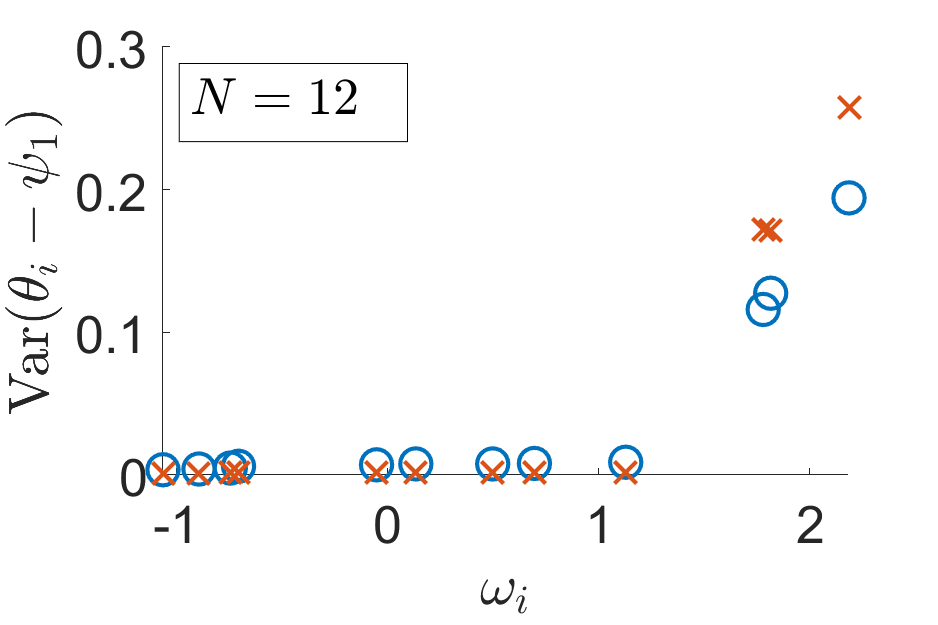}
     \includegraphics[width=0.49\linewidth]{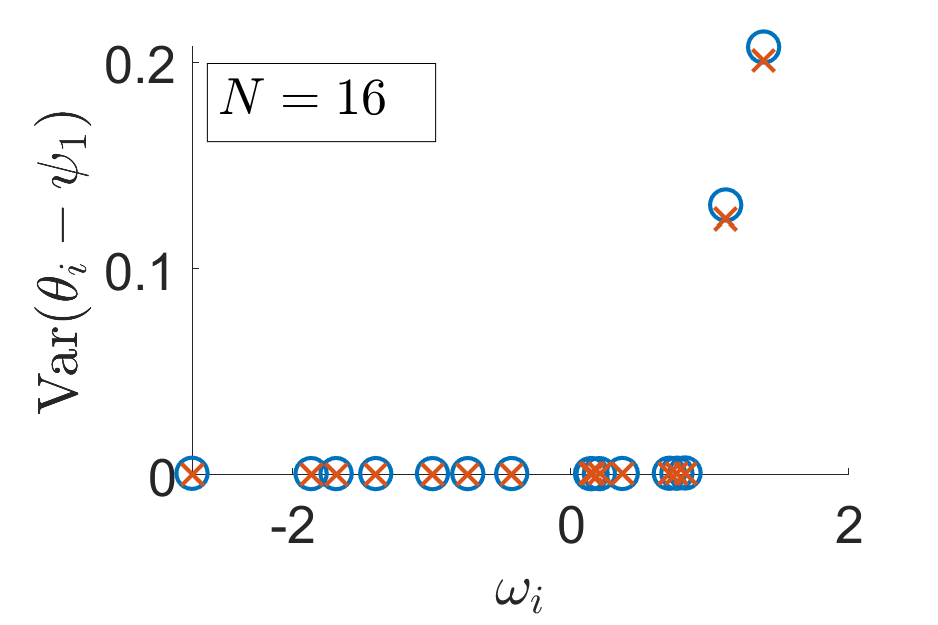}
     \includegraphics[width=0.49\linewidth]{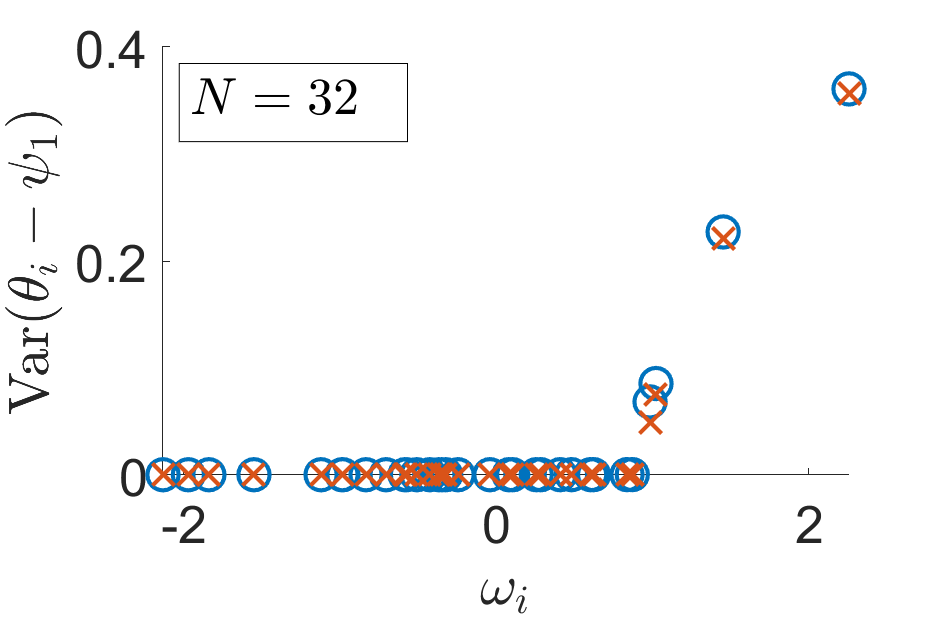}
\caption{Variance of the fluctuations of the synchronised phases around their mean, $\theta_i-\psi_1$, as a function of their native frequencies $\omega_i$ for the full two-population KS model \eqref{eq:abrams_1}--\eqref{eq:abrams_2} (blue circles) and for the reduced stochastic model \eqref{eq:abramszOU}--\eqref{eq:OU} (red crosses) for $N=8,12,16$ and $32$, where the native frequencies $\omega_i$ are drawn randomly and independently from a standard normal distribution $\mathcal{N}(0,1)$. Equation parameters are as in Figure~1 of the main manuscript.}
\label{fig:ALTSAMP1_theta_var}
\end{figure}

\begin{figure}[htb]
     \centering
      \includegraphics[width=0.49\linewidth]{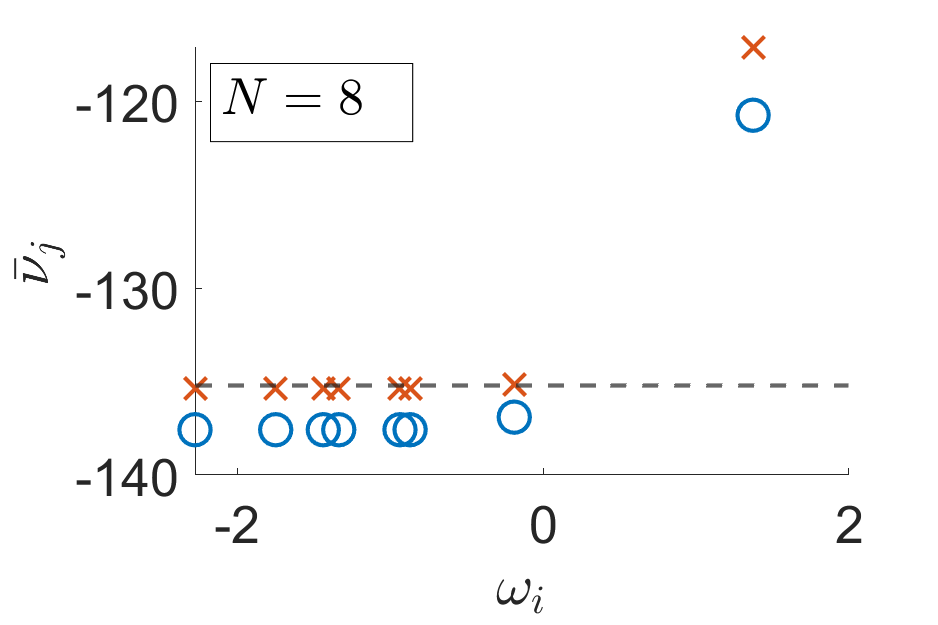}
      \includegraphics[width=0.49\linewidth]{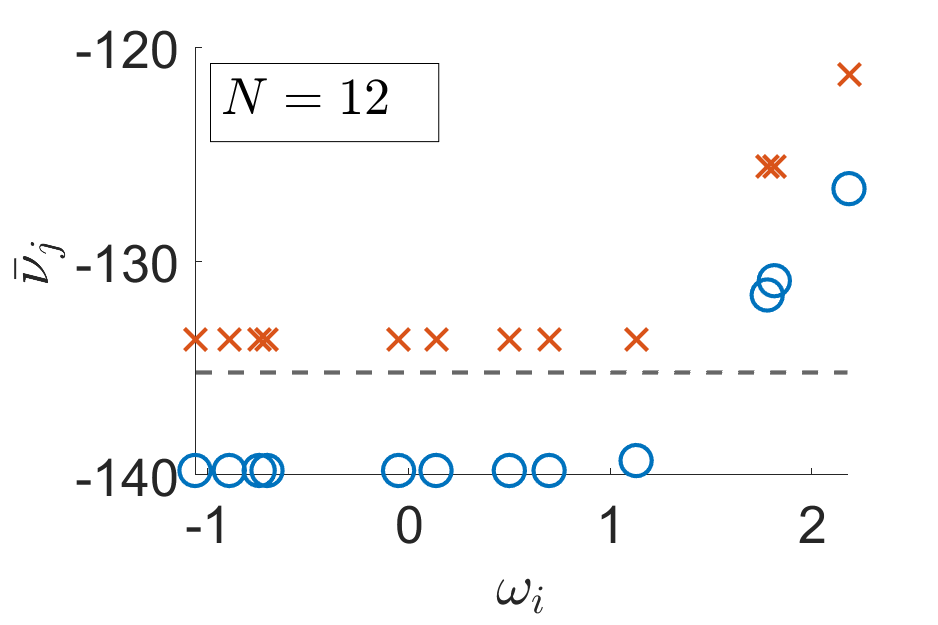}
      \includegraphics[width=0.49\linewidth]{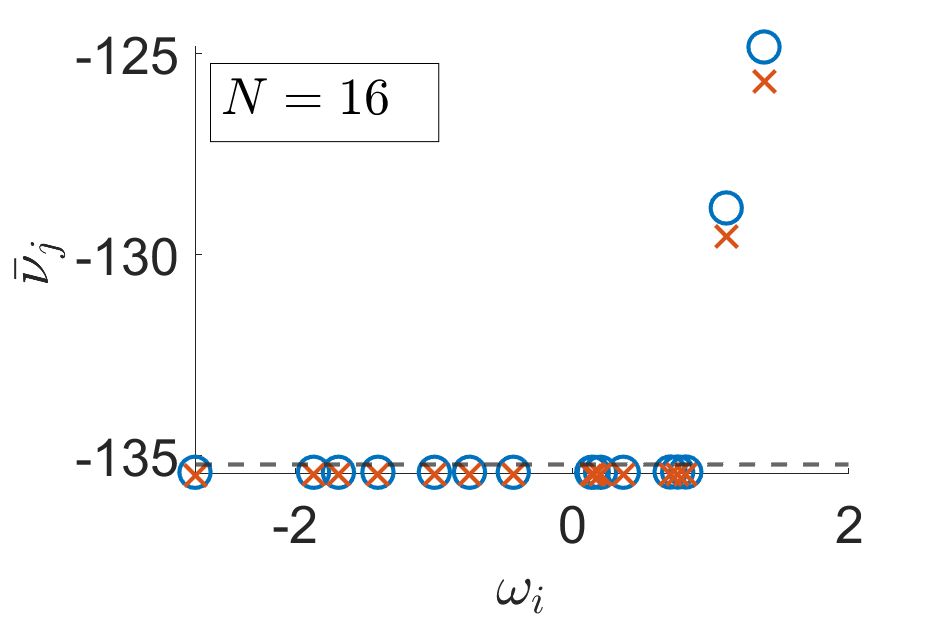}
      \includegraphics[width=0.49\linewidth]{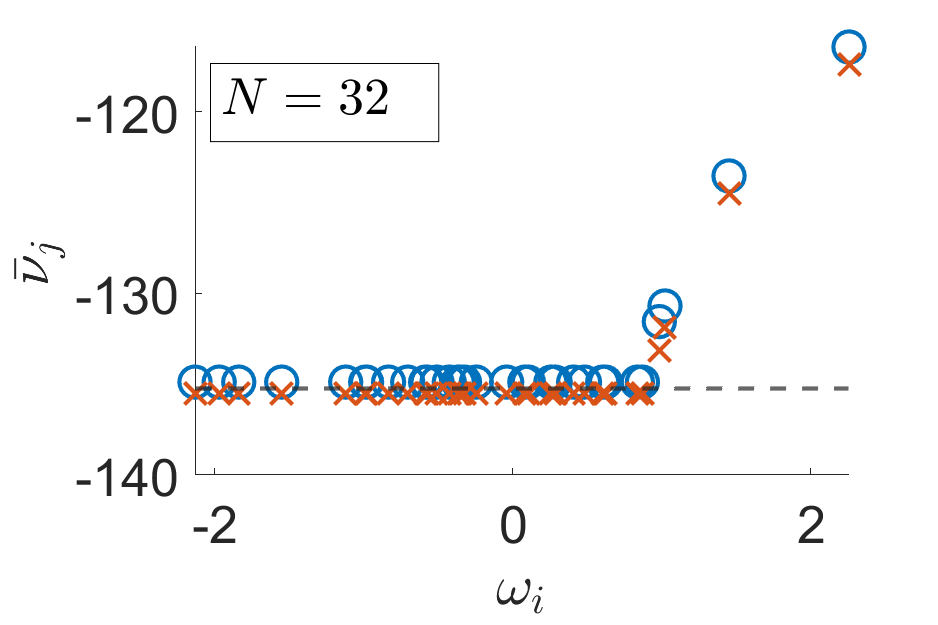}
\caption{Mean rotation frequency ${\bar \nu}_i $ of the synchronised oscillators as a function of their native frequencies $\omega_i$ for the full two-population KS model \eqref{eq:abrams_1}--\eqref{eq:abrams_2} (blue circles) and for the reduced stochastic model \eqref{eq:abramszOU}--\eqref{eq:OU} (red crosses) for $N=8,12,16$ and $32$, where the native frequencies $\omega_i$ are drawn randomly and independently from a standard normal distribution $\mathcal{N}(0,1)$. The dashed lines denote the mean frequency $\Omega$ as calculated from the self-consistency relations \eqref{eq:selconst_r}--\eqref{eq:selconst_Z} of the mean-field theory. Equation parameters are as in Figure~1 of the main manuscript.}
        \label{fig:ALTSAMP1_theta_dot}
\end{figure}

Similarly, on the level of reproducing the statistics of individual oscillators, Figures~\ref{fig:ALTSAMP1_theta_mean}--\ref{fig:ALTSAMP1_theta_dot} show that the stochastically driven model \eqref{eq:abramszOU}--\eqref{eq:OU} reproduces the mean and the variance of the fluctuations of individual synchronised oscillators around their mean values as well as their mean rotation frequencies.  

\begin{figure}[htb]
     \centering
     \begin{subfigure}{0.49\linewidth}
         \centering
         \includegraphics[width=\linewidth]{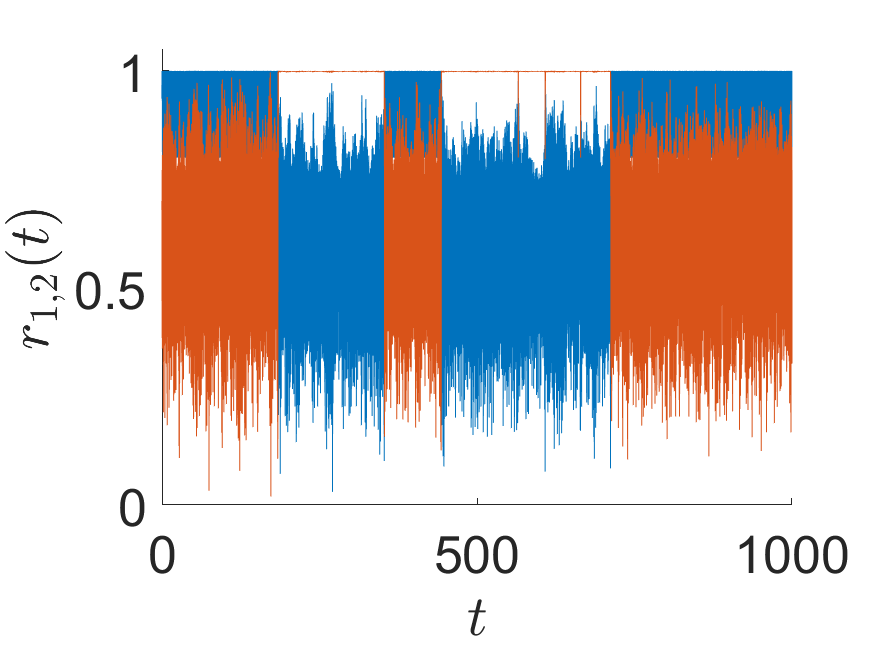}
         \label{fig:ALTSAMP1}
     \end{subfigure}
     \hfill
     \hfill
     \begin{subfigure}{0.49\linewidth}
         \centering
         \includegraphics[width=\linewidth]{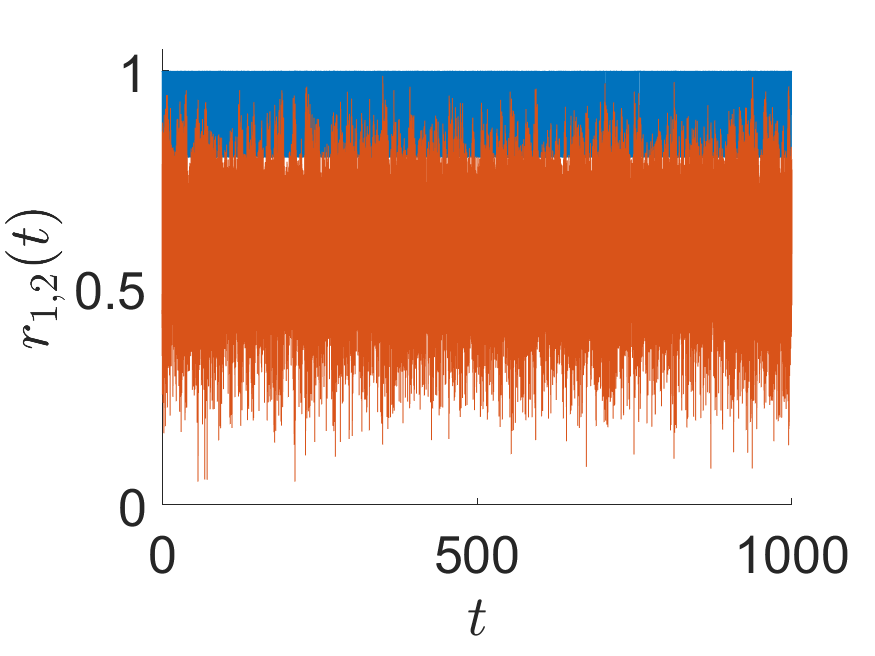}
         \label{fig:ALTSAMP7}
     \end{subfigure}
        \caption{Order parameters $r_1$ (blue) and $r_2$ (red) for the two-population KS model \eqref{eq:abrams_1}--\eqref{eq:abrams_2} as a function of time for $N=10$ oscillators. We show results for two different realisations of native frequencies drawn randomly and independently from a normal distribution with unit variance, $g(\omega)\sim{\mathcal{N}}(0,1)$. Equation parameters are as in Figure~1 of the main manuscript. }
        \label{fig:ALTSAMPr}
\end{figure}
\begin{figure}[htb]
     \centering
     \includegraphics[width=0.99\linewidth]{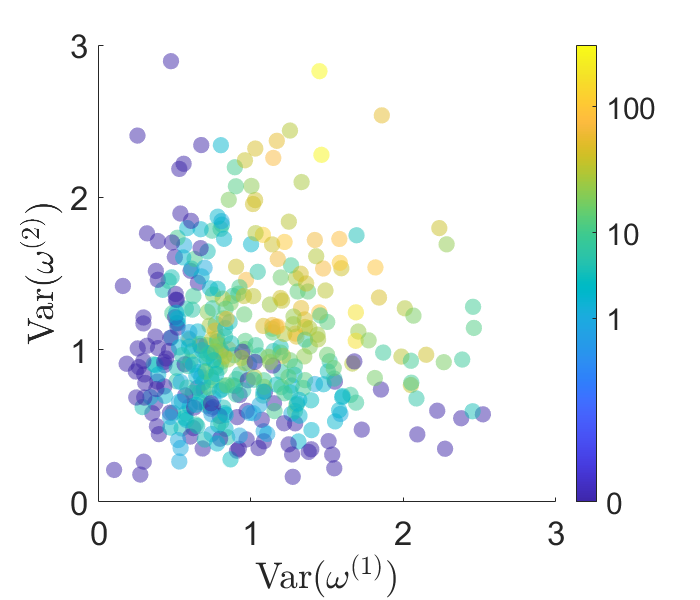} 
\caption{ Dependency of the 
probability of chimera switching 
on the realisation of the native frequency draw. Each of $400$ independent realisations was integrated for $2,000$ time units 
and the number of switching events in that time were recorded. 
Realisations, characterised by the variances of their respective populations, ${\rm{Var}}(\omega^{(1)})$ and ${\rm{Var}}(\omega^{(2)})$, are coloured according to the number of observed switching events; 
purple denotes realisations for which no switching has occurred. 
Equation parameters are as in Figure~1 of the main manuscript.
%
}
        \label{fig:ALTSAMPtaubar_vs_varomega}
\end{figure}

Whereas the statistics of the order parameter and the fluctuations are not very sensitive to the particular realisation of the native frequencies, the probability of switching events sensitively depends on the particular realisation of the draw. In Figure~\ref{fig:ALTSAMPr} we show time series of the order parameters $r_{1,2}$ for $N=10$ oscillators for two different realisations of the native frequencies drawn from $\omega \sim{\mathcal{N}}(0,1)$. Whereas for one realisation we observe frequent random switching during the $1,000$ time units, another realisation exhibits no switching events during the same period. This illustrates that the probability of switching to occur in a given time window crucially depends on the particular realisation of the native frequencies. 
To explain this, we hypothesise that the variance of a particular native frequency draw is a major factor determining this probability. More concretely, we hypothesise that a larger spread of the native frequencies increases the tail in the distribution of the fluctuations $\zeta$, responsible for inducing switching. Conversely, we hypothesise that realisations with significantly smaller variance in one population increases the stability of its synchronised state, and hence decreases the probability of random switches. Figure~\ref{fig:ALTSAMPtaubar_vs_varomega} shows the observed number of switches for $400$ random native frequency draws, according to their variances ${\rm{Var}}(\omega^{(1)})$ and ${\rm{Var}}(\omega^{(2)})$. It is clearly seen that realisations that do not exhibit any switching in the given time period are characterised by native frequency draws where at least one of the populations has a significantly smaller than expected variance (denoted by purple points). Moreover, the figure supports our hypothesis that increased overall variance promotes switching. 
The average switching frequency across all $400$ realisations, $\langle \bar{\tau}^{-1} \rangle_{\omega \sim g(\omega)} = 1/137.6$, is approximately equal to the empirical switching frequency $\bar{\tau}^{-1} = 133.0^{-1}$ for equiprobably drawn natural frequencies as reported in the main manuscript. This supports the conclusion that equiprobable samples provide a representative realisation of the distribution~$g(\omega)$. Note that we report here on the frequency of switching rather than on the switching times, because of the occurrence of realisations which do not exhibit any switching in the given time window. 


%

%

%
%

%

\clearpage


\begin{thebibliography}{61}%
\makeatletter
\providecommand \@ifxundefined [1]{%
 \@ifx{#1\undefined}
}%
\providecommand \@ifnum [1]{%
 \ifnum #1\expandafter \@firstoftwo
 \else \expandafter \@secondoftwo
 \fi
}%
\providecommand \@ifx [1]{%
 \ifx #1\expandafter \@firstoftwo
 \else \expandafter \@secondoftwo
 \fi
}%
\providecommand \natexlab [1]{#1}%
\providecommand \enquote  [1]{``#1''}%
\providecommand \bibnamefont  [1]{#1}%
\providecommand \bibfnamefont [1]{#1}%
\providecommand \citenamefont [1]{#1}%
\providecommand \href@noop [0]{\@secondoftwo}%
\providecommand \href [0]{\begingroup \@sanitize@url \@href}%
\providecommand \@href[1]{\@@startlink{#1}\@@href}%
\providecommand \@@href[1]{\endgroup#1\@@endlink}%
\providecommand \@sanitize@url [0]{\catcode `\\12\catcode `\$12\catcode
  `\&12\catcode `\#12\catcode `\^12\catcode `\_12\catcode `\%12\relax}%
\providecommand \@@startlink[1]{}%
\providecommand \@@endlink[0]{}%
\providecommand \url  [0]{\begingroup\@sanitize@url \@url }%
\providecommand \@url [1]{\endgroup\@href {#1}{\urlprefix }}%
\providecommand \urlprefix  [0]{URL }%
\providecommand \Eprint [0]{\href }%
\providecommand \doibase [0]{http://dx.doi.org/}%
\providecommand \selectlanguage [0]{\@gobble}%
\providecommand \bibinfo  [0]{\@secondoftwo}%
\providecommand \bibfield  [0]{\@secondoftwo}%
\providecommand \translation [1]{[#1]}%
\providecommand \BibitemOpen [0]{}%
\providecommand \bibitemStop [0]{}%
\providecommand \bibitemNoStop [0]{.\EOS\space}%
\providecommand \EOS [0]{\spacefactor3000\relax}%
\providecommand \BibitemShut  [1]{\csname bibitem#1\endcsname}%
\let\auto@bib@innerbib\@empty
\bibitem [{\citenamefont {Crook}\ \emph {et~al.}(1997)\citenamefont {Crook},
  \citenamefont {Ermentrout}, \citenamefont {Vanier},\ and\ \citenamefont
  {Bower}}]{CrookEtAl97}%
  \BibitemOpen
  \bibfield  {author} {\bibinfo {author} {\bibfnamefont {S.~M.}\ \bibnamefont
  {Crook}}, \bibinfo {author} {\bibfnamefont {G.~B.}\ \bibnamefont
  {Ermentrout}}, \bibinfo {author} {\bibfnamefont {M.~C.}\ \bibnamefont
  {Vanier}}, \ and\ \bibinfo {author} {\bibfnamefont {J.~M.}\ \bibnamefont
  {Bower}},\ }\bibfield  {title} {\enquote {\bibinfo {title} {The role of
  axonal delay in the synchronization of networks of coupled cortical
  oscillators},}\ }\href@noop {} {\bibfield  {journal} {\bibinfo  {journal}
  {Journal of Computational Neuroscience}\ }\textbf {\bibinfo {volume} {4}},\
  \bibinfo {pages} {161--172} (\bibinfo {year} {1997})}\BibitemShut {NoStop}%
\bibitem [{\citenamefont {Sheeba}, \citenamefont {Stefanovska},\ and\
  \citenamefont {McClintock}(2008)}]{SheebaEtAl08}%
  \BibitemOpen
  \bibfield  {author} {\bibinfo {author} {\bibfnamefont {J.~H.}\ \bibnamefont
  {Sheeba}}, \bibinfo {author} {\bibfnamefont {A.}~\bibnamefont {Stefanovska}},
  \ and\ \bibinfo {author} {\bibfnamefont {P.~V.~E.}\ \bibnamefont
  {McClintock}},\ }\bibfield  {title} {\enquote {\bibinfo {title} {Neuronal
  synchrony during anesthesia: {A} thalamocortical model},}\ }\href@noop {}
  {\bibfield  {journal} {\bibinfo  {journal} {Biophysical Journal}\ }\textbf
  {\bibinfo {volume} {95}},\ \bibinfo {pages} {2722--2727} (\bibinfo {year}
  {2008})}\BibitemShut {NoStop}%
\bibitem [{\citenamefont {Bhowmik}\ and\ \citenamefont
  {Shanahan}(2012)}]{BhowmikShanahan12}%
  \BibitemOpen
  \bibfield  {author} {\bibinfo {author} {\bibfnamefont {D.}~\bibnamefont
  {Bhowmik}}\ and\ \bibinfo {author} {\bibfnamefont {M.}~\bibnamefont
  {Shanahan}},\ }\bibfield  {title} {\enquote {\bibinfo {title} {How well do
  oscillator models capture the behaviour of biological neurons?}}\ }in\
  \href@noop {} {\emph {\bibinfo {booktitle} {The 2012 International Joint
  Conference on Neural Networks (IJCNN)}}}\ (\bibinfo {year} {2012})\ pp.\
  \bibinfo {pages} {1--8}\BibitemShut {NoStop}%
\bibitem [{\citenamefont {Mirollo}\ and\ \citenamefont
  {Strogatz}(1990)}]{MirolloStrogatz90}%
  \BibitemOpen
  \bibfield  {author} {\bibinfo {author} {\bibfnamefont {R.}~\bibnamefont
  {Mirollo}}\ and\ \bibinfo {author} {\bibfnamefont {S.}~\bibnamefont
  {Strogatz}},\ }\bibfield  {title} {\enquote {\bibinfo {title}
  {Synchronization of pulse-coupled biological oscillators},}\ }\href {\doibase
  10.1137/0150098} {\bibfield  {journal} {\bibinfo  {journal} {SIAM Journal on
  Applied Mathematics}\ }\textbf {\bibinfo {volume} {50}},\ \bibinfo {pages}
  {1645--1662} (\bibinfo {year} {1990})},\ \Eprint
  {http://arxiv.org/abs/https://doi.org/10.1137/0150098}
  {https://doi.org/10.1137/0150098} \BibitemShut {NoStop}%
\bibitem [{\citenamefont {Filatrella}, \citenamefont {Nielsen},\ and\
  \citenamefont {Pedersen}(2008)}]{FilatrellaEtAl08}%
  \BibitemOpen
  \bibfield  {author} {\bibinfo {author} {\bibfnamefont {G.}~\bibnamefont
  {Filatrella}}, \bibinfo {author} {\bibfnamefont {A.~H.}\ \bibnamefont
  {Nielsen}}, \ and\ \bibinfo {author} {\bibfnamefont {N.~F.}\ \bibnamefont
  {Pedersen}},\ }\bibfield  {title} {\enquote {\bibinfo {title} {Analysis of a
  power grid using a {K}uramoto-like model},}\ }\href {\doibase
  10.1140/epjb/e2008-00098-8} {\bibfield  {journal} {\bibinfo  {journal} {Eur.
  Phys J. B}\ }\textbf {\bibinfo {volume} {61}},\ \bibinfo {pages} {485--491}
  (\bibinfo {year} {2008})}\BibitemShut {NoStop}%
\bibitem [{\citenamefont {Nishikawa}\ and\ \citenamefont
  {Motter}(2015)}]{NishikawaMotter15}%
  \BibitemOpen
  \bibfield  {author} {\bibinfo {author} {\bibfnamefont {T.}~\bibnamefont
  {Nishikawa}}\ and\ \bibinfo {author} {\bibfnamefont {A.~E.}\ \bibnamefont
  {Motter}},\ }\bibfield  {title} {\enquote {\bibinfo {title} {Comparative
  analysis of existing models for power-grid synchronization},}\ }\href
  {\doibase 10.1088/1367-2630/17/1/015012} {\bibfield  {journal} {\bibinfo
  {journal} {New J. Phys.}\ }\textbf {\bibinfo {volume} {17}} (\bibinfo {year}
  {2015}),\ 10.1088/1367-2630/17/1/015012}\BibitemShut {NoStop}%
\bibitem [{\citenamefont {Watanabe}\ and\ \citenamefont
  {Strogatz}(1994)}]{WatanabeStrogatz94}%
  \BibitemOpen
  \bibfield  {author} {\bibinfo {author} {\bibfnamefont {S.}~\bibnamefont
  {Watanabe}}\ and\ \bibinfo {author} {\bibfnamefont {S.~H.}\ \bibnamefont
  {Strogatz}},\ }\bibfield  {title} {\enquote {\bibinfo {title} {Constants of
  motion for superconducting {J}osephson arrays},}\ }\href@noop {} {\bibfield
  {journal} {\bibinfo  {journal} {Physica D}\ }\textbf {\bibinfo {volume}
  {74}},\ \bibinfo {pages} {197 -- 253} (\bibinfo {year} {1994})}\BibitemShut
  {NoStop}%
\bibitem [{\citenamefont {Wiesenfeld}, \citenamefont {Colet},\ and\
  \citenamefont {Strogatz}(1998)}]{WiesenfeldEtAl98}%
  \BibitemOpen
  \bibfield  {author} {\bibinfo {author} {\bibfnamefont {K.}~\bibnamefont
  {Wiesenfeld}}, \bibinfo {author} {\bibfnamefont {P.}~\bibnamefont {Colet}}, \
  and\ \bibinfo {author} {\bibfnamefont {S.~H.}\ \bibnamefont {Strogatz}},\
  }\bibfield  {title} {\enquote {\bibinfo {title} {Frequency locking in
  {J}osephson arrays: {C}onnection with the {K}uramoto model},}\ }\href
  {\doibase 10.1103/PhysRevE.57.1563} {\bibfield  {journal} {\bibinfo
  {journal} {Phys. Rev. E}\ }\textbf {\bibinfo {volume} {57}},\ \bibinfo
  {pages} {1563--1569} (\bibinfo {year} {1998})}\BibitemShut {NoStop}%
\bibitem [{\citenamefont {Kuramoto}(1984)}]{Kuramoto84}%
  \BibitemOpen
  \bibfield  {author} {\bibinfo {author} {\bibfnamefont {Y.}~\bibnamefont
  {Kuramoto}},\ }\href {\doibase 10.1007/978-3-642-69689-3} {\emph {\bibinfo
  {title} {Chemical {O}scillations, {W}aves, and {T}urbulence}}},\ \bibinfo
  {series} {Springer Series in Synergetics}, Vol.~\bibinfo {volume} {19}\
  (\bibinfo  {publisher} {Springer-Verlag},\ \bibinfo {address} {Berlin},\
  \bibinfo {year} {1984})\ pp.\ \bibinfo {pages} {viii+156}\BibitemShut
  {NoStop}%
\bibitem [{\citenamefont {Sakaguchi}\ and\ \citenamefont
  {Kuramoto}(1986)}]{SakaguchiKuramoto86}%
  \BibitemOpen
  \bibfield  {author} {\bibinfo {author} {\bibfnamefont {H.}~\bibnamefont
  {Sakaguchi}}\ and\ \bibinfo {author} {\bibfnamefont {Y.}~\bibnamefont
  {Kuramoto}},\ }\bibfield  {title} {\enquote {\bibinfo {title} {{A soluble
  active rotator model showing phase transitions via mutual entrainment}},}\
  }\href {\doibase 10.1143/PTP.76.576} {\bibfield  {journal} {\bibinfo
  {journal} {Prog. Theor. Phys.}\ }\textbf {\bibinfo {volume} {76}},\ \bibinfo
  {pages} {576--581} (\bibinfo {year} {1986})}\BibitemShut {NoStop}%
\bibitem [{\citenamefont {Strogatz}(2000)}]{Strogatz00}%
  \BibitemOpen
  \bibfield  {author} {\bibinfo {author} {\bibfnamefont {S.~H.}\ \bibnamefont
  {Strogatz}},\ }\bibfield  {title} {\enquote {\bibinfo {title} {From
  {K}uramoto to {C}rawford: {E}xploring the onset of synchronization in
  populations of coupled oscillators},}\ }\href {\doibase
  10.1016/S0167-2789(00)00094-4} {\bibfield  {journal} {\bibinfo  {journal}
  {Physica D}\ }\textbf {\bibinfo {volume} {143}},\ \bibinfo {pages} {1--20}
  (\bibinfo {year} {2000})}\BibitemShut {NoStop}%
\bibitem [{\citenamefont {Pikovsky}, \citenamefont {Rosenblum},\ and\
  \citenamefont {Kurths}(2001)}]{PikovskyEtAl01}%
  \BibitemOpen
  \bibfield  {author} {\bibinfo {author} {\bibfnamefont {A.}~\bibnamefont
  {Pikovsky}}, \bibinfo {author} {\bibfnamefont {M.}~\bibnamefont {Rosenblum}},
  \ and\ \bibinfo {author} {\bibfnamefont {J.}~\bibnamefont {Kurths}},\
  }\href@noop {} {\emph {\bibinfo {title} {{Synchronization: {A} {U}niversal
  {C}oncept in {N}onlinear {S}ciences}}}}\ (\bibinfo  {publisher} {Cambridge
  University Press},\ \bibinfo {address} {Cambridge},\ \bibinfo {year}
  {2001})\BibitemShut {NoStop}%
\bibitem [{\citenamefont {Acebr\'on}\ \emph {et~al.}(2005)\citenamefont
  {Acebr\'on}, \citenamefont {Bonilla}, \citenamefont {P\'erez~Vicente},
  \citenamefont {Ritort},\ and\ \citenamefont {Spigler}}]{AcebronEtAl05}%
  \BibitemOpen
  \bibfield  {author} {\bibinfo {author} {\bibfnamefont {J.~A.}\ \bibnamefont
  {Acebr\'on}}, \bibinfo {author} {\bibfnamefont {L.~L.}\ \bibnamefont
  {Bonilla}}, \bibinfo {author} {\bibfnamefont {C.~J.}\ \bibnamefont
  {P\'erez~Vicente}}, \bibinfo {author} {\bibfnamefont {F.}~\bibnamefont
  {Ritort}}, \ and\ \bibinfo {author} {\bibfnamefont {R.}~\bibnamefont
  {Spigler}},\ }\bibfield  {title} {\enquote {\bibinfo {title} {The {K}uramoto
  model: {A} simple paradigm for synchronization phenomena},}\ }\href {\doibase
  10.1103/RevModPhys.77.137} {\bibfield  {journal} {\bibinfo  {journal} {Rev.
  Mod. Phys.}\ }\textbf {\bibinfo {volume} {77}},\ \bibinfo {pages} {137--185}
  (\bibinfo {year} {2005})}\BibitemShut {NoStop}%
\bibitem [{\citenamefont {Osipov}, \citenamefont {Kurths},\ and\ \citenamefont
  {Zhou}(2007)}]{OsipovEtAl07}%
  \BibitemOpen
  \bibfield  {author} {\bibinfo {author} {\bibfnamefont {G.~V.}\ \bibnamefont
  {Osipov}}, \bibinfo {author} {\bibfnamefont {J.}~\bibnamefont {Kurths}}, \
  and\ \bibinfo {author} {\bibfnamefont {C.}~\bibnamefont {Zhou}},\ }\href
  {\doibase 10.1007/978-3-540-71269-5} {\emph {\bibinfo {title}
  {Synchronization in {O}scillatory {N}etworks}}},\ Springer Series in
  Synergetics\ (\bibinfo  {publisher} {Springer},\ \bibinfo {address}
  {Berlin},\ \bibinfo {year} {2007})\ p.\ \bibinfo {pages} {37c}\BibitemShut
  {NoStop}%
\bibitem [{\citenamefont {Arenas}\ \emph {et~al.}(2008)\citenamefont {Arenas},
  \citenamefont {Diaz-Guilera}, \citenamefont {Kurths}, \citenamefont
  {Moreno},\ and\ \citenamefont {Zhou}}]{ArenasEtAl08}%
  \BibitemOpen
  \bibfield  {author} {\bibinfo {author} {\bibfnamefont {A.}~\bibnamefont
  {Arenas}}, \bibinfo {author} {\bibfnamefont {A.}~\bibnamefont
  {Diaz-Guilera}}, \bibinfo {author} {\bibfnamefont {J.}~\bibnamefont
  {Kurths}}, \bibinfo {author} {\bibfnamefont {Y.}~\bibnamefont {Moreno}}, \
  and\ \bibinfo {author} {\bibfnamefont {C.}~\bibnamefont {Zhou}},\ }\bibfield
  {title} {\enquote {\bibinfo {title} {Synchronization in complex networks},}\
  }\href {\doibase 10.1016/j.physrep.2008.09.002} {\bibfield  {journal}
  {\bibinfo  {journal} {Phys. Rep.}\ }\textbf {\bibinfo {volume} {469}},\
  \bibinfo {pages} {93--153} (\bibinfo {year} {2008})}\BibitemShut {NoStop}%
\bibitem [{\citenamefont {D{\"o}rfler}\ and\ \citenamefont
  {Bullo}(2014)}]{DorflerBullo14}%
  \BibitemOpen
  \bibfield  {author} {\bibinfo {author} {\bibfnamefont {F.}~\bibnamefont
  {D{\"o}rfler}}\ and\ \bibinfo {author} {\bibfnamefont {F.}~\bibnamefont
  {Bullo}},\ }\bibfield  {title} {\enquote {\bibinfo {title} {Synchronization
  in complex networks of phase oscillators: {A} survey},}\ }\href@noop {}
  {\bibfield  {journal} {\bibinfo  {journal} {Automatica}\ }\textbf {\bibinfo
  {volume} {50}},\ \bibinfo {pages} {1539 -- 1564} (\bibinfo {year}
  {2014})}\BibitemShut {NoStop}%
\bibitem [{\citenamefont {Rodrigues}\ \emph {et~al.}(2016)\citenamefont
  {Rodrigues}, \citenamefont {Peron}, \citenamefont {Ji},\ and\ \citenamefont
  {Kurths}}]{RodriguesEtAl16}%
  \BibitemOpen
  \bibfield  {author} {\bibinfo {author} {\bibfnamefont {F.~A.}\ \bibnamefont
  {Rodrigues}}, \bibinfo {author} {\bibfnamefont {T.~K.~D.}\ \bibnamefont
  {Peron}}, \bibinfo {author} {\bibfnamefont {P.}~\bibnamefont {Ji}}, \ and\
  \bibinfo {author} {\bibfnamefont {J.}~\bibnamefont {Kurths}},\ }\bibfield
  {title} {\enquote {\bibinfo {title} {The {K}uramoto model in complex
  networks},}\ }\href@noop {} {\bibfield  {journal} {\bibinfo  {journal} {Phys.
  Rep.}\ }\textbf {\bibinfo {volume} {610}},\ \bibinfo {pages} {1 -- 98}
  (\bibinfo {year} {2016})}\BibitemShut {NoStop}%
\bibitem [{\citenamefont {Ott}\ and\ \citenamefont
  {Antonsen}(2008)}]{OttAntonsen08}%
  \BibitemOpen
  \bibfield  {author} {\bibinfo {author} {\bibfnamefont {E.}~\bibnamefont
  {Ott}}\ and\ \bibinfo {author} {\bibfnamefont {T.~M.}\ \bibnamefont
  {Antonsen}},\ }\bibfield  {title} {\enquote {\bibinfo {title} {Low
  dimensional behavior of large systems of globally coupled oscillators},}\
  }\href {\doibase 10.1063/1.2930766} {\bibfield  {journal} {\bibinfo
  {journal} {Chaos}\ }\textbf {\bibinfo {volume} {18}},\ \bibinfo {pages}
  {037113, 6} (\bibinfo {year} {2008})}\BibitemShut {NoStop}%
\bibitem [{\citenamefont {Delgadino}, \citenamefont {Gvalani},\ and\
  \citenamefont {Pavliotis}(2021)}]{DelgadinoEtAl21}%
  \BibitemOpen
  \bibfield  {author} {\bibinfo {author} {\bibfnamefont {M.~G.}\ \bibnamefont
  {Delgadino}}, \bibinfo {author} {\bibfnamefont {R.~S.}\ \bibnamefont
  {Gvalani}}, \ and\ \bibinfo {author} {\bibfnamefont {G.~A.}\ \bibnamefont
  {Pavliotis}},\ }\bibfield  {title} {\enquote {\bibinfo {title} {On the
  diffusive-mean field limit for weakly interacting diffusions exhibiting phase
  transitions},}\ }\href {\doibase 10.1007/s00205-021-01648-1} {\bibfield
  {journal} {\bibinfo  {journal} {Arch. Ration. Mech. Anal.}\ }\textbf
  {\bibinfo {volume} {241}},\ \bibinfo {pages} {91--148} (\bibinfo {year}
  {2021})}\BibitemShut {NoStop}%
\bibitem [{\citenamefont {Giacomin}\ and\ \citenamefont
  {Poquet}(2015)}]{GiacominPoquet15}%
  \BibitemOpen
  \bibfield  {author} {\bibinfo {author} {\bibfnamefont {G.}~\bibnamefont
  {Giacomin}}\ and\ \bibinfo {author} {\bibfnamefont {C.}~\bibnamefont
  {Poquet}},\ }\bibfield  {title} {\enquote {\bibinfo {title} {Noise,
  interaction, nonlinear dynamics and the origin of rhythmic behaviors},}\
  }\href@noop {} {\bibfield  {journal} {\bibinfo  {journal} {Braz. J. Probab.
  Stat.}\ }\textbf {\bibinfo {volume} {29}},\ \bibinfo {pages} {460--493}
  (\bibinfo {year} {2015})}\BibitemShut {NoStop}%
\bibitem [{\citenamefont {Lu\c{c}on}(2015)}]{Lucon15}%
  \BibitemOpen
  \bibfield  {author} {\bibinfo {author} {\bibfnamefont {E.}~\bibnamefont
  {Lu\c{c}on}},\ }\bibfield  {title} {\enquote {\bibinfo {title} {Large
  population asymptotics for interacting diffusions in a quenched random
  environment},}\ }in\ \href@noop {} {\emph {\bibinfo {booktitle} {From
  particle systems to partial differential equations. {II}}}},\ \bibinfo
  {series} {Springer Proc. Math. Stat.}, Vol.\ \bibinfo {volume} {129}\
  (\bibinfo  {publisher} {Springer, Cham},\ \bibinfo {year} {2015})\ pp.\
  \bibinfo {pages} {231--251}\BibitemShut {NoStop}%
\bibitem [{\citenamefont {Gottwald}(2015)}]{Gottwald15}%
  \BibitemOpen
  \bibfield  {author} {\bibinfo {author} {\bibfnamefont {G.~A.}\ \bibnamefont
  {Gottwald}},\ }\bibfield  {title} {\enquote {\bibinfo {title} {Model
  reduction for networks of coupled oscillators},}\ }\href {\doibase
  10.1063/1.4921295} {\bibfield  {journal} {\bibinfo  {journal} {Chaos}\
  }\textbf {\bibinfo {volume} {25}},\ \bibinfo {pages} {053111, 12} (\bibinfo
  {year} {2015})}\BibitemShut {NoStop}%
\bibitem [{\citenamefont {Gottwald}(2017)}]{Gottwald17}%
  \BibitemOpen
  \bibfield  {author} {\bibinfo {author} {\bibfnamefont {G.~A.}\ \bibnamefont
  {Gottwald}},\ }\bibfield  {title} {\enquote {\bibinfo {title} {Finite-size
  effects in a stochastic {K}uramoto model},}\ }\href {\doibase
  10.1063/1.5004618} {\bibfield  {journal} {\bibinfo  {journal} {Chaos}\
  }\textbf {\bibinfo {volume} {27}},\ \bibinfo {pages} {101103} (\bibinfo
  {year} {2017})}\BibitemShut {NoStop}%
\bibitem [{\citenamefont {Hildebrand}, \citenamefont {Buice},\ and\
  \citenamefont {Chow}(2007)}]{HildebrandtEtAl07}%
  \BibitemOpen
  \bibfield  {author} {\bibinfo {author} {\bibfnamefont {E.~J.}\ \bibnamefont
  {Hildebrand}}, \bibinfo {author} {\bibfnamefont {M.~A.}\ \bibnamefont
  {Buice}}, \ and\ \bibinfo {author} {\bibfnamefont {C.~C.}\ \bibnamefont
  {Chow}},\ }\bibfield  {title} {\enquote {\bibinfo {title} {Kinetic theory of
  coupled oscillators},}\ }\href {\doibase 10.1103/PhysRevLett.98.054101}
  {\bibfield  {journal} {\bibinfo  {journal} {Phys. Rev. Lett.}\ }\textbf
  {\bibinfo {volume} {98}},\ \bibinfo {pages} {054101} (\bibinfo {year}
  {2007})}\BibitemShut {NoStop}%
\bibitem [{\citenamefont {Buice}\ and\ \citenamefont
  {Chow}(2007)}]{BuiceChow07}%
  \BibitemOpen
  \bibfield  {author} {\bibinfo {author} {\bibfnamefont {M.~A.}\ \bibnamefont
  {Buice}}\ and\ \bibinfo {author} {\bibfnamefont {C.~C.}\ \bibnamefont
  {Chow}},\ }\bibfield  {title} {\enquote {\bibinfo {title} {Correlations,
  fluctuations, and stability of a finite-size network of coupled
  oscillators},}\ }\href {\doibase 10.1103/PhysRevE.76.031118} {\bibfield
  {journal} {\bibinfo  {journal} {Phys. Rev. E}\ }\textbf {\bibinfo {volume}
  {76}},\ \bibinfo {pages} {031118} (\bibinfo {year} {2007})}\BibitemShut
  {NoStop}%
\bibitem [{\citenamefont {Hong}, \citenamefont {Ha},\ and\ \citenamefont
  {Park}(2007)}]{HongEtAl07}%
  \BibitemOpen
  \bibfield  {author} {\bibinfo {author} {\bibfnamefont {H.}~\bibnamefont
  {Hong}}, \bibinfo {author} {\bibfnamefont {M.}~\bibnamefont {Ha}}, \ and\
  \bibinfo {author} {\bibfnamefont {H.}~\bibnamefont {Park}},\ }\bibfield
  {title} {\enquote {\bibinfo {title} {Finite-size scaling in complex
  networks},}\ }\href {\doibase 10.1103/PhysRevLett.98.258701} {\bibfield
  {journal} {\bibinfo  {journal} {Phys. Rev. Lett.}\ }\textbf {\bibinfo
  {volume} {98}},\ \bibinfo {pages} {258701} (\bibinfo {year}
  {2007})}\BibitemShut {NoStop}%
\bibitem [{\citenamefont {Tang}(2011)}]{Tang11}%
  \BibitemOpen
  \bibfield  {author} {\bibinfo {author} {\bibfnamefont {L.-H.}\ \bibnamefont
  {Tang}},\ }\bibfield  {title} {\enquote {\bibinfo {title} {To synchronize or
  not to synchronize, that is the question: finite-size scaling and fluctuation
  effects in the {K}uramoto model},}\ }\href
  {http://stacks.iop.org/1742-5468/2011/i=01/a=P01034} {\bibfield  {journal}
  {\bibinfo  {journal} {Journal of Statistical Mechanics: Theory and
  Experiment}\ }\textbf {\bibinfo {volume} {2011}},\ \bibinfo {pages} {P01034}
  (\bibinfo {year} {2011})}\BibitemShut {NoStop}%
\bibitem{OmelchenkoGottwald25}Omel'chenko, O. \& Gottwald, G. ``Mean-field approach to finite-size fluctuations in the Kuramoto-Sakaguchi model,'' {\em ArXiv:2511.03700 [nlin.AO]}. (2025)
\bibitem [{\citenamefont {Okuda}\ and\ \citenamefont
  {Kuramoto}(1991)}]{OkudaKuramoto91}%
  \BibitemOpen
  \bibfield  {author} {\bibinfo {author} {\bibfnamefont {K.}~\bibnamefont
  {Okuda}}\ and\ \bibinfo {author} {\bibfnamefont {Y.}~\bibnamefont
  {Kuramoto}},\ }\bibfield  {title} {\enquote {\bibinfo {title} {Mutual
  entrainment between populations of coupled oscillators},}\ }\href {\doibase
  10.1143/PTP.86.1159} {\bibfield  {journal} {\bibinfo  {journal} {Progr.
  Theoret. Phys.}\ }\textbf {\bibinfo {volume} {86}},\ \bibinfo {pages}
  {1159--1176} (\bibinfo {year} {1991})}\BibitemShut {NoStop}%
\bibitem [{\citenamefont {Montbri\'o}, \citenamefont {Kurths},\ and\
  \citenamefont {Blasius}(2004)}]{MontbrioEtAl04}%
  \BibitemOpen
  \bibfield  {author} {\bibinfo {author} {\bibfnamefont {E.}~\bibnamefont
  {Montbri\'o}}, \bibinfo {author} {\bibfnamefont {J.}~\bibnamefont {Kurths}},
  \ and\ \bibinfo {author} {\bibfnamefont {B.}~\bibnamefont {Blasius}},\
  }\bibfield  {title} {\enquote {\bibinfo {title} {Synchronization of two
  interacting populations of oscillators},}\ }\href {\doibase
  10.1103/PhysRevE.70.056125} {\bibfield  {journal} {\bibinfo  {journal} {Phys.
  Rev. E}\ }\textbf {\bibinfo {volume} {70}},\ \bibinfo {pages} {056125}
  (\bibinfo {year} {2004})}\BibitemShut {NoStop}%
\bibitem [{\citenamefont {Kuramoto}\ and\ \citenamefont
  {Battogtokh}(2002)}]{KuramotoBattogkh02}%
  \BibitemOpen
  \bibfield  {author} {\bibinfo {author} {\bibfnamefont {Y.}~\bibnamefont
  {Kuramoto}}\ and\ \bibinfo {author} {\bibfnamefont {D.}~\bibnamefont
  {Battogtokh}},\ }\bibfield  {title} {\enquote {\bibinfo {title} {Coexistence
  of coherence and incoherence in nonlocally coupled phase oscillators},}\
  }\href@noop {} {\bibfield  {journal} {\bibinfo  {journal} {Nonlinear Phenom.
  Complex Syst.}\ }\textbf {\bibinfo {volume} {5}},\ \bibinfo {pages} {380?385}
  (\bibinfo {year} {2002})}\BibitemShut {NoStop}%
\bibitem [{\citenamefont {Omel'chenko}, \citenamefont {Maistrenko},\ and\
  \citenamefont {Tass}(2008)}]{OmelchenkoEtAl08}%
  \BibitemOpen
  \bibfield  {author} {\bibinfo {author} {\bibfnamefont {O.~E.}\ \bibnamefont
  {Omel'chenko}}, \bibinfo {author} {\bibfnamefont {Y.~L.}\ \bibnamefont
  {Maistrenko}}, \ and\ \bibinfo {author} {\bibfnamefont {P.~A.}\ \bibnamefont
  {Tass}},\ }\bibfield  {title} {\enquote {\bibinfo {title} {Chimera states:
  The natural link between coherence and incoherence},}\ }\href {\doibase
  10.1103/PhysRevLett.100.044105} {\bibfield  {journal} {\bibinfo  {journal}
  {Phys. Rev. Lett.}\ }\textbf {\bibinfo {volume} {100}},\ \bibinfo {pages}
  {044105} (\bibinfo {year} {2008})}\BibitemShut {NoStop}%
\bibitem [{\citenamefont {Omel'chenko}, \citenamefont {Wolfrum},\ and\
  \citenamefont {Maistrenko}(2010)}]{OmelchenkoEtAl10}%
  \BibitemOpen
  \bibfield  {author} {\bibinfo {author} {\bibfnamefont {O.~E.}\ \bibnamefont
  {Omel'chenko}}, \bibinfo {author} {\bibfnamefont {M.}~\bibnamefont
  {Wolfrum}}, \ and\ \bibinfo {author} {\bibfnamefont {Y.~L.}\ \bibnamefont
  {Maistrenko}},\ }\bibfield  {title} {\enquote {\bibinfo {title} {Chimera
  states as chaotic spatiotemporal patterns},}\ }\href {\doibase
  10.1103/PhysRevE.81.065201} {\bibfield  {journal} {\bibinfo  {journal} {Phys.
  Rev. E}\ }\textbf {\bibinfo {volume} {81}},\ \bibinfo {pages} {065201}
  (\bibinfo {year} {2010})}\BibitemShut {NoStop}%
\bibitem [{\citenamefont {Omel'chenko}(2018)}]{Omelchenko18}%
  \BibitemOpen
  \bibfield  {author} {\bibinfo {author} {\bibfnamefont {O.~E.}\ \bibnamefont
  {Omel'chenko}},\ }\bibfield  {title} {\enquote {\bibinfo {title} {The
  mathematics behind chimera states},}\ }\href {\doibase
  10.1088/1361-6544/aaaa07} {\bibfield  {journal} {\bibinfo  {journal}
  {Nonlinearity}\ }\textbf {\bibinfo {volume} {31}},\ \bibinfo {pages} {R121}
  (\bibinfo {year} {2018})}\BibitemShut {NoStop}%
\bibitem{DeHaroGarciaTorres25}Haro García, A. \& Torres, J. ``Emergence of chimeras states in one-dimensional Ising model with long-range diffusion,'' {\em ArXiv:2510.24903 [nlin.AO]}. (2025)
\bibitem [{\citenamefont {Panaggio}\ \emph {et~al.}(2016)\citenamefont
  {Panaggio}, \citenamefont {Abrams}, \citenamefont {Ashwin},\ and\
  \citenamefont {Laing}}]{PanaggioEtAl16}%
  \BibitemOpen
  \bibfield  {author} {\bibinfo {author} {\bibfnamefont {M.~J.}\ \bibnamefont
  {Panaggio}}, \bibinfo {author} {\bibfnamefont {D.~M.}\ \bibnamefont
  {Abrams}}, \bibinfo {author} {\bibfnamefont {P.}~\bibnamefont {Ashwin}}, \
  and\ \bibinfo {author} {\bibfnamefont {C.~R.}\ \bibnamefont {Laing}},\
  }\bibfield  {title} {\enquote {\bibinfo {title} {Chimera states in networks
  of phase oscillators: {T}he case of two small populations},}\ }\href
  {\doibase 10.1103/PhysRevE.93.012218} {\bibfield  {journal} {\bibinfo
  {journal} {Phys. Rev. E}\ }\textbf {\bibinfo {volume} {93}},\ \bibinfo
  {pages} {012218} (\bibinfo {year} {2016})}\BibitemShut {NoStop}%
\bibitem [{\citenamefont {Burylko}, \citenamefont {Martens},\ and\
  \citenamefont {Bick}(2022)}]{BurylkoEtAl22}%
  \BibitemOpen
  \bibfield  {author} {\bibinfo {author} {\bibfnamefont {O.}~\bibnamefont
  {Burylko}}, \bibinfo {author} {\bibfnamefont {E.~A.}\ \bibnamefont
  {Martens}}, \ and\ \bibinfo {author} {\bibfnamefont {C.}~\bibnamefont
  {Bick}},\ }\bibfield  {title} {\enquote {\bibinfo {title} {{Symmetry breaking
  yields chimeras in two small populations of {K}uramoto-type oscillators}},}\
  }\href {\doibase 10.1063/5.0088465} {\bibfield  {journal} {\bibinfo
  {journal} {Chaos: An Interdisciplinary Journal of Nonlinear Science}\
  }\textbf {\bibinfo {volume} {32}},\ \bibinfo {pages} {093109} (\bibinfo
  {year} {2022})}\BibitemShut {NoStop}%
\bibitem [{\citenamefont {Abrams}\ \emph {et~al.}(2008)\citenamefont {Abrams},
  \citenamefont {Mirollo}, \citenamefont {Strogatz},\ and\ \citenamefont
  {Wiley}}]{AbramsEtAl08}%
  \BibitemOpen
  \bibfield  {author} {\bibinfo {author} {\bibfnamefont {D.~M.}\ \bibnamefont
  {Abrams}}, \bibinfo {author} {\bibfnamefont {R.}~\bibnamefont {Mirollo}},
  \bibinfo {author} {\bibfnamefont {S.~H.}\ \bibnamefont {Strogatz}}, \ and\
  \bibinfo {author} {\bibfnamefont {D.~A.}\ \bibnamefont {Wiley}},\ }\bibfield
  {title} {\enquote {\bibinfo {title} {Solvable model for chimera states of
  coupled oscillators},}\ }\href {\doibase 10.1103/physrevlett.101.084103}
  {\bibfield  {journal} {\bibinfo  {journal} {Physical Review Letters}\
  }\textbf {\bibinfo {volume} {101}} (\bibinfo {year} {2008}),\
  10.1103/physrevlett.101.084103}\BibitemShut {NoStop}%
\bibitem [{\citenamefont {Laing}(2009)}]{Laing09}%
  \BibitemOpen
  \bibfield  {author} {\bibinfo {author} {\bibfnamefont {C.~R.}\ \bibnamefont
  {Laing}},\ }\bibfield  {title} {\enquote {\bibinfo {title} {Chimera states in
  heterogeneous networks},}\ }\href {\doibase 10.1063/1.3068353} {\bibfield
  {journal} {\bibinfo  {journal} {Chaos: An Interdisciplinary Journal of
  Nonlinear Science}\ }\textbf {\bibinfo {volume} {19}} (\bibinfo {year}
  {2009}),\ 10.1063/1.3068353}\BibitemShut {NoStop}%
\bibitem [{\citenamefont {Martens}, \citenamefont {Bick},\ and\ \citenamefont
  {Panaggio}(2016)}]{MartensEtAl16}%
  \BibitemOpen
  \bibfield  {author} {\bibinfo {author} {\bibfnamefont {E.~A.}\ \bibnamefont
  {Martens}}, \bibinfo {author} {\bibfnamefont {C.}~\bibnamefont {Bick}}, \
  and\ \bibinfo {author} {\bibfnamefont {M.~J.}\ \bibnamefont {Panaggio}},\
  }\bibfield  {title} {\enquote {\bibinfo {title} {Chimera states in two
  populations with heterogeneous phase-lag},}\ }\href {\doibase
  10.1063/1.4958930} {\bibfield  {journal} {\bibinfo  {journal} {Chaos: An
  Interdisciplinary Journal of Nonlinear Science}\ }\textbf {\bibinfo {volume}
  {26}} (\bibinfo {year} {2016}),\ 10.1063/1.4958930}\BibitemShut {NoStop}%
\bibitem [{\citenamefont {Ma}, \citenamefont {Wang},\ and\ \citenamefont
  {Liu}(2010)}]{MaEtAl10}%
  \BibitemOpen
  \bibfield  {author} {\bibinfo {author} {\bibfnamefont {R.}~\bibnamefont
  {Ma}}, \bibinfo {author} {\bibfnamefont {J.}~\bibnamefont {Wang}}, \ and\
  \bibinfo {author} {\bibfnamefont {Z.}~\bibnamefont {Liu}},\ }\bibfield
  {title} {\enquote {\bibinfo {title} {Robust features of chimera states and
  the implementation of alternating chimera states},}\ }\href {\doibase
  10.1209/0295-5075/91/40006} {\bibfield  {journal} {\bibinfo  {journal}
  {Europhysics Letters}\ }\textbf {\bibinfo {volume} {91}},\ \bibinfo {pages}
  {40006} (\bibinfo {year} {2010})}\BibitemShut {NoStop}%
\bibitem [{\citenamefont {Laing}(2012)}]{Laing12}%
  \BibitemOpen
  \bibfield  {author} {\bibinfo {author} {\bibfnamefont {C.~R.}\ \bibnamefont
  {Laing}},\ }\bibfield  {title} {\enquote {\bibinfo {title} {{Disorder-induced
  dynamics in a pair of coupled heterogeneous phase oscillator networks}},}\
  }\href {\doibase 10.1063/1.4758814} {\bibfield  {journal} {\bibinfo
  {journal} {Chaos: An Interdisciplinary Journal of Nonlinear Science}\
  }\textbf {\bibinfo {volume} {22}},\ \bibinfo {pages} {043104} (\bibinfo
  {year} {2012})}\BibitemShut {NoStop}%
\bibitem [{\citenamefont {Buscarino}\ \emph {et~al.}(2015)\citenamefont
  {Buscarino}, \citenamefont {Frasca}, \citenamefont {Gambuzza},\ and\
  \citenamefont {H\"ovel}}]{BuscarinoEtAl15}%
  \BibitemOpen
  \bibfield  {author} {\bibinfo {author} {\bibfnamefont {A.}~\bibnamefont
  {Buscarino}}, \bibinfo {author} {\bibfnamefont {M.}~\bibnamefont {Frasca}},
  \bibinfo {author} {\bibfnamefont {L.~V.}\ \bibnamefont {Gambuzza}}, \ and\
  \bibinfo {author} {\bibfnamefont {P.}~\bibnamefont {H\"ovel}},\ }\bibfield
  {title} {\enquote {\bibinfo {title} {Chimera states in time-varying complex
  networks},}\ }\href {\doibase 10.1103/PhysRevE.91.022817} {\bibfield
  {journal} {\bibinfo  {journal} {Phys. Rev. E}\ }\textbf {\bibinfo {volume}
  {91}},\ \bibinfo {pages} {022817} (\bibinfo {year} {2015})}\BibitemShut
  {NoStop}%
\bibitem [{\citenamefont {Semenova}\ \emph {et~al.}(2016)\citenamefont
  {Semenova}, \citenamefont {Zakharova}, \citenamefont {Anishchenko},\ and\
  \citenamefont {Sch\"oll}}]{SemenovaEtAl16}%
  \BibitemOpen
  \bibfield  {author} {\bibinfo {author} {\bibfnamefont {N.}~\bibnamefont
  {Semenova}}, \bibinfo {author} {\bibfnamefont {A.}~\bibnamefont {Zakharova}},
  \bibinfo {author} {\bibfnamefont {V.}~\bibnamefont {Anishchenko}}, \ and\
  \bibinfo {author} {\bibfnamefont {E.}~\bibnamefont {Sch\"oll}},\ }\bibfield
  {title} {\enquote {\bibinfo {title} {Coherence-resonance chimeras in a
  network of excitable elements},}\ }\href {\doibase
  10.1103/PhysRevLett.117.014102} {\bibfield  {journal} {\bibinfo  {journal}
  {Phys. Rev. Lett.}\ }\textbf {\bibinfo {volume} {117}},\ \bibinfo {pages}
  {014102} (\bibinfo {year} {2016})}\BibitemShut {NoStop}%
\bibitem [{\citenamefont {Zhang}\ \emph {et~al.}(2020)\citenamefont {Zhang},
  \citenamefont {Nicolaou}, \citenamefont {Hart}, \citenamefont {Roy},\ and\
  \citenamefont {Motter}}]{ZhangEtAl20}%
  \BibitemOpen
  \bibfield  {author} {\bibinfo {author} {\bibfnamefont {Y.}~\bibnamefont
  {Zhang}}, \bibinfo {author} {\bibfnamefont {Z.~G.}\ \bibnamefont {Nicolaou}},
  \bibinfo {author} {\bibfnamefont {J.~D.}\ \bibnamefont {Hart}}, \bibinfo
  {author} {\bibfnamefont {R.}~\bibnamefont {Roy}}, \ and\ \bibinfo {author}
  {\bibfnamefont {A.~E.}\ \bibnamefont {Motter}},\ }\bibfield  {title}
  {\enquote {\bibinfo {title} {Critical switching in globally attractive
  chimeras},}\ }\href {\doibase 10.1103/PhysRevX.10.011044} {\bibfield
  {journal} {\bibinfo  {journal} {Phys. Rev. X}\ }\textbf {\bibinfo {volume}
  {10}},\ \bibinfo {pages} {011044} (\bibinfo {year} {2020})}\BibitemShut
  {NoStop}%
\bibitem [{\citenamefont {Bick}(2018)}]{Bick18}%
  \BibitemOpen
  \bibfield  {author} {\bibinfo {author} {\bibfnamefont {C.}~\bibnamefont
  {Bick}},\ }\bibfield  {title} {\enquote {\bibinfo {title} {Heteroclinic
  switching between chimeras},}\ }\href {\doibase 10.1103/PhysRevE.97.050201}
  {\bibfield  {journal} {\bibinfo  {journal} {Phys. Rev. E}\ }\textbf {\bibinfo
  {volume} {97}},\ \bibinfo {pages} {050201} (\bibinfo {year}
  {2018})}\BibitemShut {NoStop}%
\bibitem [{\citenamefont {Bick}(2019)}]{Bick19}%
  \BibitemOpen
  \bibfield  {author} {\bibinfo {author} {\bibfnamefont {C.}~\bibnamefont
  {Bick}},\ }\bibfield  {title} {\enquote {\bibinfo {title} {Heteroclinic
  dynamics of localized frequency synchrony: {H}eteroclinic cycles for small
  populations},}\ }\href@noop {} {\bibfield  {journal} {\bibinfo  {journal}
  {Journal of Nonlinear Science}\ }\textbf {\bibinfo {volume} {29}},\ \bibinfo
  {pages} {2547--2570} (\bibinfo {year} {2019})}\BibitemShut {NoStop}%
\bibitem [{\citenamefont {Bick}\ and\ \citenamefont
  {Lohse}(2019)}]{BickLohse19}%
  \BibitemOpen
  \bibfield  {author} {\bibinfo {author} {\bibfnamefont {C.}~\bibnamefont
  {Bick}}\ and\ \bibinfo {author} {\bibfnamefont {A.}~\bibnamefont {Lohse}},\
  }\bibfield  {title} {\enquote {\bibinfo {title} {Heteroclinic dynamics of
  localized frequency synchrony: Stability of heteroclinic cycles and
  networks},}\ }\href@noop {} {\bibfield  {journal} {\bibinfo  {journal}
  {Journal of Nonlinear Science}\ }\textbf {\bibinfo {volume} {29}},\ \bibinfo
  {pages} {2571--2600} (\bibinfo {year} {2019})}\BibitemShut {NoStop}%
\bibitem [{\citenamefont {Ebrahimzadeh}\ \emph {et~al.}(2020)\citenamefont
  {Ebrahimzadeh}, \citenamefont {Schiek}, \citenamefont {Jaros}, \citenamefont
  {Kapitaniak}, \citenamefont {van Waasen},\ and\ \citenamefont
  {Maistrenko}}]{EbrahimzadehEtAl20}%
  \BibitemOpen
  \bibfield  {author} {\bibinfo {author} {\bibfnamefont {P.}~\bibnamefont
  {Ebrahimzadeh}}, \bibinfo {author} {\bibfnamefont {M.}~\bibnamefont
  {Schiek}}, \bibinfo {author} {\bibfnamefont {P.}~\bibnamefont {Jaros}},
  \bibinfo {author} {\bibfnamefont {T.}~\bibnamefont {Kapitaniak}}, \bibinfo
  {author} {\bibfnamefont {S.}~\bibnamefont {van Waasen}}, \ and\ \bibinfo
  {author} {\bibfnamefont {Y.}~\bibnamefont {Maistrenko}},\ }\bibfield  {title}
  {\enquote {\bibinfo {title} {Minimal chimera states in phase-lag coupled
  mechanical oscillators},}\ }\href@noop {} {\bibfield  {journal} {\bibinfo
  {journal} {The European Physical Journal Special Topics}\ }\textbf {\bibinfo
  {volume} {229}},\ \bibinfo {pages} {2205--2214} (\bibinfo {year}
  {2020})}\BibitemShut {NoStop}%
\bibitem [{\citenamefont {Goldschmidt}, \citenamefont {Pikovsky},\ and\
  \citenamefont {Politi}(2019)}]{GoldschmidtEtAl19}%
  \BibitemOpen
  \bibfield  {author} {\bibinfo {author} {\bibfnamefont {R.~J.}\ \bibnamefont
  {Goldschmidt}}, \bibinfo {author} {\bibfnamefont {A.}~\bibnamefont
  {Pikovsky}}, \ and\ \bibinfo {author} {\bibfnamefont {A.}~\bibnamefont
  {Politi}},\ }\bibfield  {title} {\enquote {\bibinfo {title} {{Blinking
  chimeras in globally coupled rotators}},}\ }\href {\doibase
  10.1063/1.5105367} {\bibfield  {journal} {\bibinfo  {journal} {Chaos: An
  Interdisciplinary Journal of Nonlinear Science}\ }\textbf {\bibinfo {volume}
  {29}},\ \bibinfo {pages} {071101} (\bibinfo {year} {2019})}\BibitemShut
  {NoStop}%
\bibitem [{\citenamefont {Brezetsky}\ \emph {et~al.}(2021)\citenamefont
  {Brezetsky}, \citenamefont {Jaros}, \citenamefont {Levchenko}, \citenamefont
  {Kapitaniak},\ and\ \citenamefont {Maistrenko}}]{BrezetskyEtAl21}%
  \BibitemOpen
  \bibfield  {author} {\bibinfo {author} {\bibfnamefont {S.}~\bibnamefont
  {Brezetsky}}, \bibinfo {author} {\bibfnamefont {P.}~\bibnamefont {Jaros}},
  \bibinfo {author} {\bibfnamefont {R.}~\bibnamefont {Levchenko}}, \bibinfo
  {author} {\bibfnamefont {T.}~\bibnamefont {Kapitaniak}}, \ and\ \bibinfo
  {author} {\bibfnamefont {Y.}~\bibnamefont {Maistrenko}},\ }\bibfield  {title}
  {\enquote {\bibinfo {title} {Chimera complexity},}\ }\href {\doibase
  10.1103/PhysRevE.103.L050204} {\bibfield  {journal} {\bibinfo  {journal}
  {Phys. Rev. E}\ }\textbf {\bibinfo {volume} {103}},\ \bibinfo {pages}
  {L050204} (\bibinfo {year} {2021})}\BibitemShut {NoStop}%
\bibitem [{\citenamefont {Lee}\ and\ \citenamefont
  {Krischer}(2023)}]{LeeKrischer23}%
  \BibitemOpen
  \bibfield  {author} {\bibinfo {author} {\bibfnamefont {S.}~\bibnamefont
  {Lee}}\ and\ \bibinfo {author} {\bibfnamefont {K.}~\bibnamefont {Krischer}},\
  }\bibfield  {title} {\enquote {\bibinfo {title} {{Heteroclinic switching
  between chimeras in a ring of six oscillator populations}},}\ }\href
  {\doibase 10.1063/5.0147228} {\bibfield  {journal} {\bibinfo  {journal}
  {Chaos: An Interdisciplinary Journal of Nonlinear Science}\ }\textbf
  {\bibinfo {volume} {33}},\ \bibinfo {pages} {063120} (\bibinfo {year}
  {2023})}\BibitemShut {NoStop}%
\bibitem{LeeKrischer23b}Lee, S. \& Krischer, K. ``Chimera dynamics of generalized Kuramoto–Sakaguchi oscillators in two-population networks,'' {\em Journal Of Physics A: Mathematical And Theoretical}. \textbf{56}, 405001 (2023,9), https://doi.org/10.1088/1751-8121/acf4d6
\bibitem [{\citenamefont {Yue}\ and\ \citenamefont
  {Gottwald}(2024)}]{YueGottwald24}%
  \BibitemOpen
  \bibfield  {author} {\bibinfo {author} {\bibfnamefont {W.}~\bibnamefont
  {Yue}}\ and\ \bibinfo {author} {\bibfnamefont {G.~A.}\ \bibnamefont
  {Gottwald}},\ }\bibfield  {title} {\enquote {\bibinfo {title} {A stochastic
  approximation for the finite-size {K}uramoto-{S}akaguchi model},}\ }\href
  {\doibase https://doi.org/10.1016/j.physd.2024.134292} {\bibfield  {journal}
  {\bibinfo  {journal} {Physica D: Nonlinear Phenomena}\ }\textbf {\bibinfo
  {volume} {468}},\ \bibinfo {pages} {134292} (\bibinfo {year}
  {2024})}\BibitemShut {NoStop}%
\bibitem [{\citenamefont {Kramers}(1940)}]{Kramers40}%
  \BibitemOpen
  \bibfield  {author} {\bibinfo {author} {\bibfnamefont {H.}~\bibnamefont
  {Kramers}},\ }\bibfield  {title} {\enquote {\bibinfo {title} {Brownian motion
  in a field of force and the diffusion model of chemical reactions},}\ }\href
  {\doibase https://doi.org/10.1016/S0031-8914(40)90098-2} {\bibfield
  {journal} {\bibinfo  {journal} {Physica}\ }\textbf {\bibinfo {volume} {7}},\
  \bibinfo {pages} {284--304} (\bibinfo {year} {1940})}\BibitemShut {NoStop}%
\bibitem [{\citenamefont {Bick}\ \emph {et~al.}(2020)\citenamefont {Bick},
  \citenamefont {Goodfellow}, \citenamefont {Laing},\ and\ \citenamefont
  {Martens}}]{BickEtAl20}%
  \BibitemOpen
  \bibfield  {author} {\bibinfo {author} {\bibfnamefont {C.}~\bibnamefont
  {Bick}}, \bibinfo {author} {\bibfnamefont {M.}~\bibnamefont {Goodfellow}},
  \bibinfo {author} {\bibfnamefont {C.~R.}\ \bibnamefont {Laing}}, \ and\
  \bibinfo {author} {\bibfnamefont {E.~A.}\ \bibnamefont {Martens}},\
  }\bibfield  {title} {\enquote {\bibinfo {title} {Understanding the dynamics
  of biological and neural oscillator networks through exact mean-field
  reductions: a review},}\ }\href {\doibase 10.1186/s13408-020-00086-9}
  {\bibfield  {journal} {\bibinfo  {journal} {J. Math. Neurosc.}\ }\textbf
  {\bibinfo {volume} {10}},\ \bibinfo {pages} {9} (\bibinfo {year}
  {2020})}\BibitemShut {NoStop}%
\bibitem{Omelchenko12}Omel'chenko, O. \& Wolfrum, M. ``Nonuniversal transitions to synchrony in the Sakaguchi-Kuramoto model,'' {\em Phys. Rev. Lett.}. \textbf{109}, 164101 (2012,10), https://link.aps.org/doi/10.1103/PhysRevLett.109.164101
\bibitem [{\citenamefont {Peter}\ and\ \citenamefont
  {Pikovsky}(2018)}]{PeterPikovsky18}%
  \BibitemOpen
  \bibfield  {author} {\bibinfo {author} {\bibfnamefont {F.}~\bibnamefont
  {Peter}}\ and\ \bibinfo {author} {\bibfnamefont {A.}~\bibnamefont
  {Pikovsky}},\ }\bibfield  {title} {\enquote {\bibinfo {title} {Transition to
  collective oscillations in finite {K}uramoto ensembles},}\ }\href {\doibase
  10.1103/PhysRevE.97.032310} {\bibfield  {journal} {\bibinfo  {journal} {Phys.
  Rev. E}\ }\textbf {\bibinfo {volume} {97}},\ \bibinfo {pages} {032310}
  (\bibinfo {year} {2018})}\BibitemShut {NoStop}%
\bibitem [{\citenamefont {Fialkowski}\ \emph {et~al.}(2023)\citenamefont
  {Fialkowski}, \citenamefont {Yanchuk}, \citenamefont {Sokolov}, \citenamefont
  {Sch\"oll}, \citenamefont {Gottwald},\ and\ \citenamefont
  {Berner}}]{FialkowskiEtAl23}%
  \BibitemOpen
  \bibfield  {author} {\bibinfo {author} {\bibfnamefont {J.}~\bibnamefont
  {Fialkowski}}, \bibinfo {author} {\bibfnamefont {S.}~\bibnamefont {Yanchuk}},
  \bibinfo {author} {\bibfnamefont {I.~M.}\ \bibnamefont {Sokolov}}, \bibinfo
  {author} {\bibfnamefont {E.}~\bibnamefont {Sch\"oll}}, \bibinfo {author}
  {\bibfnamefont {G.~A.}\ \bibnamefont {Gottwald}}, \ and\ \bibinfo {author}
  {\bibfnamefont {R.}~\bibnamefont {Berner}},\ }\bibfield  {title} {\enquote
  {\bibinfo {title} {Heterogeneous nucleation in finite-size adaptive dynamical
  networks},}\ }\href {\doibase 10.1103/PhysRevLett.130.067402} {\bibfield
  {journal} {\bibinfo  {journal} {Phys. Rev. Lett.}\ }\textbf {\bibinfo
  {volume} {130}},\ \bibinfo {pages} {067402} (\bibinfo {year}
  {2023})}\BibitemShut {NoStop}%
\bibitem{MasseyFJ51}Massey, F. ``The Kolmogorov-Smirnov test for goodness of fit,'' {\em Journal Of The American Statistical Association}. \textbf{46}, 68-78 (1951), http://www.jstor.org/stable/2280095
\bibitem{WolfrumOmelchenko11}Wolfrum, M. \& Omel'chenko, O. ``Chimera states are chaotic transients,'' {\em Phys. Rev. E}. \textbf{84}, 015201 (2011,7), https://link.aps.org/doi/10.1103/PhysRevE.84.015201
\bibitem{WolfrumEtAl11}Wolfrum, M., Omel'chenko, O., Yanchuk, S. \& Maistrenko, Y. ``Spectral properties of chimera states,'' {\em Chaos: An Interdisciplinary Journal Of Nonlinear Science}. \textbf{21}, 013112 (2011,3), https://doi.org/10.1063/1.3563579
\bibitem [{\citenamefont {Hancock}\ and\ \citenamefont
  {Gottwald}(2018)}]{HancockGottwald18}%
  \BibitemOpen
  \bibfield  {author} {\bibinfo {author} {\bibfnamefont {E.~J.}\ \bibnamefont
  {Hancock}}\ and\ \bibinfo {author} {\bibfnamefont {G.~A.}\ \bibnamefont
  {Gottwald}},\ }\bibfield  {title} {\enquote {\bibinfo {title} {Model
  reduction for {K}uramoto models with complex topologies},}\ }\href {\doibase
  10.1103/PhysRevE.98.012307} {\bibfield  {journal} {\bibinfo  {journal} {Phys.
  Rev. E}\ }\textbf {\bibinfo {volume} {98}},\ \bibinfo {pages} {012307}
  (\bibinfo {year} {2018})}\BibitemShut {NoStop}%
\bibitem [{\citenamefont {Smith}\ and\ \citenamefont
  {Gottwald}(2019)}]{SmithGottwald19}%
  \BibitemOpen
  \bibfield  {author} {\bibinfo {author} {\bibfnamefont {L.~D.}\ \bibnamefont
  {Smith}}\ and\ \bibinfo {author} {\bibfnamefont {G.~A.}\ \bibnamefont
  {Gottwald}},\ }\bibfield  {title} {\enquote {\bibinfo {title} {Chaos in
  networks of coupled oscillators with multimodal natural frequency
  distributions},}\ }\href {\doibase 10.1063/1.5109130} {\bibfield  {journal}
  {\bibinfo  {journal} {Chaos}\ }\textbf {\bibinfo {volume} {29}},\ \bibinfo
  {pages} {093127} (\bibinfo {year} {2019})}\BibitemShut {NoStop}%
\bibitem [{\citenamefont {Smith}\ and\ \citenamefont
  {Gottwald}(2020)}]{SmithGottwald20}%
  \BibitemOpen
  \bibfield  {author} {\bibinfo {author} {\bibfnamefont {L.~D.}\ \bibnamefont
  {Smith}}\ and\ \bibinfo {author} {\bibfnamefont {G.~A.}\ \bibnamefont
  {Gottwald}},\ }\bibfield  {title} {\enquote {\bibinfo {title} {{Model
  reduction for the collective dynamics of globally coupled oscillators: {F}rom
  finite networks to the thermodynamic limit}},}\ }\href {\doibase
  10.1063/5.0009790} {\bibfield  {journal} {\bibinfo  {journal} {Chaos: An
  Interdisciplinary Journal of Nonlinear Science}\ }\textbf {\bibinfo {volume}
  {30}} (\bibinfo {year} {2020}),\ 10.1063/5.0009790}\BibitemShut {NoStop}%
\bibitem [{\citenamefont {Yue}, \citenamefont {Smith},\ and\ \citenamefont
  {Gottwald}(2020)}]{YueEtAl20}%
  \BibitemOpen
  \bibfield  {author} {\bibinfo {author} {\bibfnamefont {W.}~\bibnamefont
  {Yue}}, \bibinfo {author} {\bibfnamefont {L.~D.}\ \bibnamefont {Smith}}, \
  and\ \bibinfo {author} {\bibfnamefont {G.~A.}\ \bibnamefont {Gottwald}},\
  }\bibfield  {title} {\enquote {\bibinfo {title} {Model reduction for the
  {K}uramoto-{S}akaguchi model: The importance of nonentrained rogue
  oscillators},}\ }\href {\doibase 10.1103/PhysRevE.101.062213} {\bibfield
  {journal} {\bibinfo  {journal} {Phys. Rev. E}\ }\textbf {\bibinfo {volume}
  {101}},\ \bibinfo {pages} {062213} (\bibinfo {year} {2020})}\BibitemShut
  {NoStop}%
\bibitem [{\citenamefont {Vatiwutipong}\ and\ \citenamefont
  {Phewchean}(2019)}]{VatiwutipongPhewchean19}%
  \BibitemOpen
  \bibfield  {author} {\bibinfo {author} {\bibfnamefont {P.}~\bibnamefont
  {Vatiwutipong}}\ and\ \bibinfo {author} {\bibfnamefont {N.}~\bibnamefont
  {Phewchean}},\ }\bibfield  {title} {\enquote {\bibinfo {title} {Alternative
  way to derive the distribution of the multivariate {O}rnstein--{U}hlenbeck
  process},}\ }\href@noop {} {\bibfield  {journal} {\bibinfo  {journal}
  {Advances in Difference Equations}\ }\textbf {\bibinfo {volume} {2019}},\
  \bibinfo {pages} {276} (\bibinfo {year} {2019})}\BibitemShut {NoStop}%
\bibitem [{\citenamefont {Sachs}, \citenamefont {Leimkuhler},\ and\
  \citenamefont {Danos}(2017)}]{SachsEtAl17}%
  \BibitemOpen
  \bibfield  {author} {\bibinfo {author} {\bibfnamefont {M.}~\bibnamefont
  {Sachs}}, \bibinfo {author} {\bibfnamefont {B.}~\bibnamefont {Leimkuhler}}, \
  and\ \bibinfo {author} {\bibfnamefont {V.}~\bibnamefont {Danos}},\ }\bibfield
   {title} {\enquote {\bibinfo {title} {Langevin dynamics with variable
  coefficients and nonconservative forces: {F}rom stationary states to
  numerical methods},}\ }\href {\doibase 10.3390/e19120647} {\bibfield
  {journal} {\bibinfo  {journal} {Entropy}\ }\textbf {\bibinfo {volume} {19}}
  (\bibinfo {year} {2017}),\ 10.3390/e19120647}\BibitemShut {NoStop}%
\end{thebibliography}
\end{document}